\documentclass[12pt]{article}
\pdfoutput=1
\usepackage{jheppub}
\usepackage[utf8]{inputenc}
\usepackage{graphicx}
\usepackage{amsfonts}
\usepackage[dvipsnames]{xcolor}
\usepackage{enumerate}
\usepackage{amsmath}
\usepackage{amssymb}
\usepackage[mathscr]{eucal}
\usepackage{slashed}
\usepackage{array}
\usepackage{amsthm}
\usepackage{epsfig}
\usepackage{wasysym} 
\usepackage{tikz-cd} 
\usepgfmodule{shapes}
\usetikzlibrary{backgrounds, arrows,calc,shapes,decorations.pathreplacing, decorations.markings, automata,positioning,calligraphy}
\usepackage{ytableau} 
\usepackage{dirtytalk} 
\usepackage{cancel} 
\usepackage{changepage} 
\usepackage{float} 
\usepackage{comment} 
\usepackage{stackengine}

\tikzset{
  arrow/.pic={\path[tips,every arrow/.try,->,>=#1] (0,0) -- +(.1pt,0);},
  pics/arrow/.default={latex,very thick}
}

\stackMath
\newcommand\tsup[2][2]{%
 \def\useanchorwidth{T}%
  \ifnum#1>1%
    \stackon[-.5pt]{\tsup[\numexpr#1-1\relax]{#2}}{\scriptscriptstyle\sim}%
  \else%
    \stackon[.5pt]{#2}{\scriptscriptstyle\sim}%
  \fi%
}

\newcommand{\be}{\begin{equation}}
\newcommand{\ee}{\end{equation}}
\newcommand{\bea}{\begin{eqnarray}}
\newcommand{\eea}{\end{eqnarray}}
\newcommand{\ba}{\begin{array}}
\newcommand{\ea}{\end{array}}
\newcommand{\bpic}{\begin{tikzpicture}}
\newcommand{\epic}{\end{tikzpicture}}

\newcommand{\nn}{\nonumber}

\newcommand{\tr}{\text{tr}\,}

\newcommand{\Flip}[1]{\text{Flip#1}}
\newcommand{\Flipper}[1]{\text{Flipper#1}}
\newcommand{\Triangle}{\text{Triangle}\,}
\newcommand{\triangles}{\text{Triangles}\,}
\newcommand{\quartic}{\text{Quartic}\,}
\newcommand{\flips}{\text{Flips}\,}

\newcommand{\lra}{\leftrightarrow}
\newcommand{\llra}{\longleftrightarrow}

\newcommand\at{\tilde a}
\newcommand\bt{\tilde b}
\newcommand\ct{\tilde c}
\newcommand\dt{\tilde d}
\newcommand\kt{\tilde k}
\newcommand\lt{\tilde l}

\newcommand\nt{\tilde n}
\newcommand\ot{\tilde o}
\newcommand\pt{\tilde p}
\newcommand\qt{\tilde q}
\newcommand\rt{\tilde r}
\newcommand\st{\tilde s}
\newcommand\ttilde{\tilde t}
\newcommand\ut{\tilde u}

\newcommand\Bt{\tilde B}
\newcommand\Dt{\tilde D}
\newcommand\Ft{\tilde F}
\newcommand\Lt{\tilde L}
\newcommand\Qt{\tilde Q}

\newcommand\Rt{\tilde R}
\newcommand\Ut{\tilde U}
\newcommand\Vt{\tilde V}

\newcommand{\cN}{\mathcal{N}}
\newcommand{\cO}{\mathcal{O}}

\newcommand{\cW}{\mathcal{W}}

\newcommand{\bZ}{\mathbb{Z}}

\def\a{\alpha}

\def\Bt{\tilde{B}}

\title{$4d$ $\,\mathcal{N}\!=\!1$ dualities from $5d$ dualities}

\author[1]{Stephane Bajeot}
\author[2]{Sergio Benvenuti}

\affiliation[1]{SISSA, Via Bonomea 265, 34136 Trieste, Italy}
\affiliation[2]{INFN, Sezione di Trieste, SISSA, via Bonomea 265, 34136 Trieste, Italy}
\emailAdd{sbajeot@sissa.it, benve79@gmail.com}

\abstract{We consider $5d$ KK dualities, that is multiple $5d$ gauge theories with the same $6d$ infinite coupling limit. We provide a prescription to associate $4d$ $\mathcal{N}=1$ quivers to the $5d$ dual quivers, such that the $4d$ quivers are also dual to each other. The $4d$ dualities are infrared dualities which can be checked matching global symmetry anomalies and in certain cases can be proven using basic Seiberg dualities sequentially. We also consider dualities obtained by Higgsing in two different ways the same $5d$ theory, in some simple examples.}

\begin{document}

\maketitle

\section{Introduction and summary}  \label{Algorithm}
Gauge theories living in space-time dimension greater than $4$ are infrared free, nevertheless, in many supersymmetric cases, it is possible to argue that at strong coupling lives a ultraviolet superconformal field theory (SCFT) \cite{Seiberg:1996bd, Seiberg:1996qx, Morrison:1996xf}. In this paper we are interested in $5d$ quiver gauge theories whose UV completion is actually a $6d$ SCFT. Such models go under the name of Kaluza-Klein (KK) theories.

In many instances, there are more than one  $5d$ gauge theories with the same infinite coupling SCFT ($5d$ or $6d$). This phenomenon goes under the name of \emph{5d dualities}, even if the language is slightly improper, since the physical picture is really that the UV SCFT can be relevantly deformed  in various different ways, triggering RG flows to different IR gauge theories. 

A powerful tool to analyze the strong coupling behavior of $5d$ $\cN=1$ gauge theories is given by Hanany-Witten branes setups \cite{Hanany:1996ie}, which in this case involve webs of $5$-branes, a.k.a. pq-webs \cite{Aharony:1997ju, Aharony:1997bh, Benini:2009gi}. Pq-webs were used to study $5d$ dualities in \cite{Bergman:2013aca, Zafrir:2014ywa, Bergman:2014kza, Bergman:2015dpa}. Later, the pq-web technology to deal with KK theories  was developed: \cite{Kim:2015jba,Hayashi:2015fsa,Hayashi:2015zka,Zafrir:2015rga, Zafrir:2015ftn, Hayashi:2015vhy} discuss many examples of different $5d$ $\cN=1$ quiver gauge theories with the same $6d$ SCFT in the infinite coupling limit, described by Type IIA brane systems \cite{Brunner:1997gk, Brunner:1997gf, Hanany:1997gh}.
 
In this paper, we take  $5d$ KK dualities and associate to them $4d$ $\cN=1$ dualities. Starting from a $5d$ KK quiver with $8$ supercharges, the $4d$ quivers has the same gauge structure (but in $4d$ the nodes are $\cN=1$ $4$ supercharges nodes), the same matter fields (but in $4d$ there are chiral multiplets instead of hyper multiplets) plus for each bifundamental we add a "triangle". A "triangle" means that if in $5d$ there is a bifundamental hyper connecting node $A$ with node $B$, in $4d$ there is a chiral bifundamental going from node $A$ to node $B$, a fundamental going from node $B$ to a global $SU(2)$ node, and a fundamental going from the global $SU(2)$ node to node $A$. We also add a cubic $SU(2)$ invariant superpotential term. See eq. \ref{IntroRules1}. Such triangles are meant to reproduce the $5d$ axial symmetries (which are anomalous in $4d$ but not in $5d$) and the $5d$ instantonic symmetries (which do not exist in $4d$). With this prescription we are able to associate a $4d$ quiver to $5d$ quivers, in such a way that the rank of the global $4d$ symmetry is equal to the rank of the global $5d$ symmetry minus $2$. We only consider quivers such that this prescription yields a $4d$ quivers without gauge anomalies. The previous prescription is illustrated by the following example, where round red nodes are $SU$ gauge groups\footnote{Our notation is explained at the end of this section.}

\be \label{IntroRules1} \scalebox{0.9}{\bpic[node distance=2cm,gSUnode/.style={circle,red,draw,minimum size=8mm},gUSpnode/.style={circle,blue,draw,minimum size=8mm},fnode/.style={rectangle,red,draw,minimum size=8mm}]  
\node at (-7.5,0) {$5d:$};
\node[fnode] (F1) at (-5.5,0) {$N+2$};
\node[gSUnode] (G1) at (-3.5,0) {$N$};
\node[gSUnode] (G2) at (-2,0) {$N$};
\node[gSUnode] (G3) at (-0.5,0) {$N$};
\node[fnode] (F2) at (1.5,0) {$N+2$};
\draw (F1) -- (G1) -- (G2) -- (G3) -- (F2);
\node at (-2,-1.5) {\scalebox{1.5}{$\Downarrow$}};
\epic} \ee 
\be\nn \scalebox{0.9}{\bpic[node distance=2cm,gSUnode/.style={circle,red,draw,minimum size=8mm},gUSpnode/.style={circle,blue,draw,minimum size=8mm},fnode/.style={rectangle,red,draw,minimum size=8mm}]  
\node at (-7.5,0) {$4d:$};
\node[fnode,orange] (F1) at (-6,0) {$N$};
\node[gSUnode] (G1) at (-4,0) {$N$};
\node[gSUnode] (G2) at (-2,0) {$N$};
\node[gSUnode] (G3) at (0,0) {$N$};
\node[fnode] (F2) at (2,0) {$N$};
\node[fnode] (F3) at (-3,-1.7) {$2$};
\node[fnode] (F4) at (-1,-1.7) {$2$};
\node[fnode,orange] (F5) at (-5.5,-1.7) {$2$};
\node[fnode,violet] (F6) at (1.5,-1.7) {$2$};
\draw (F1) -- pic[pos=0.7,sloped]{arrow} (G1) -- pic[pos=0.7,sloped]{arrow} (G2) -- pic[pos=0.7,sloped]{arrow} (G3) -- pic[pos=0.7,sloped]{arrow} (F2);
\draw (G1) -- pic[pos=0.5,sloped,very thick]{arrow=latex reversed} (F3);
\draw (G2) -- pic[pos=0.5,sloped,very thick]{arrow=latex reversed} (F3);
\draw (G2) -- pic[pos=0.5,sloped,very thick]{arrow=latex reversed} (F4);
\draw (G3) -- pic[pos=0.5,sloped,very thick]{arrow=latex reversed} (F4);
\draw (G1) -- pic[pos=0.5,sloped,very thick]{arrow=latex reversed} (F5);
\draw (G3) -- pic[pos=0.5,sloped,very thick]{arrow=latex reversed} (F6);
\node[right] at (-5,-3) {$ \cW= 2 \, \triangles + Flips$};
\epic} \ee    

One remark about the prescription is that when we go from a hyper in $5d$ to a chiral in $4d$ we are free to choose it in the fundamental or anti-fundamental representation of the gauge group. The constraint on gauge anomaly cancellation fixes the choice of the representation. In the example that we show the $N+2$ hypers on the left in $5d$ have been split in $N$ fundamentals and 2 anti-fundamentals. Our prescription produces a non-anomalous $4d$ $\cN=1$ theory only if the ranks of the $SU$ nodes are constant. It would be interesting to generalize it to more general quivers, e.g. unitary tails with non-constant rank or ortho-symplectic quivers. Another remark is that the $4d$ dualities involve flipping fields, that is gauge singlets that enter the superpotential linearly, multiplying a gauge invariant mesonic or baryonic operator.

The main point of this paper is that starting from two $5d$ dual KK quivers, hence with the same $6d$ SCFT UV completion, the two $4d$ quivers constructed with the above prescription are infrared dual.

We discuss two classes of theories. In the first class, $R_{N,k}$, Sec. \ref{fam1} and \ref{fam1g}, we are able to prove the  $4d$ $\cN=1$ dualities using basic Seiberg dualities  \cite{Seiberg:1994pq, Intriligator:1995ne}, that is we use deconfinement and basic dualities sequentially, in the same spirit of \cite{Benvenuti:2017kud, Benvenuti:2017bpg, Giacomelli:2017vgk, Pasquetti:2019tix, Benvenuti:2020gvy, Bajeot:2022kwt,Bottini:2022vpy, Bajeot:2022lah, Amariti:2022tbd, Bajeot:2023gyl}.
 In the second class, Sec. \ref{fam2} and \ref{4ddualityToCheck}, we do not have such a proof and the proposed dualities are tested by matching the t'Hooft anomalies and the central charges. We conclude with section \ref{higgsing}, where we discuss a set of theories obtaining by Higgsing the class $R_{N,k}$.

\subsection*{Possible interpretation}
In this paper we provide a prescription to obtain $4d$ duality from $5d$ dualities, but we do not investigate \emph{why} our prescription works, that is why the $5d$ UV KK duality is transferred to a $4d$ IR duality. This is obviously an important question, so let us close this introduction with some speculations about a possible explanation.

There should be a connection between our prescription and the compactfication of $6d$ $(1,0)$ SCFT's on Riemann surfaces. Such compactification is usually done in two steps: first, one compactified the $6d$ brane system on a circle, getting a pq-web and the associated infrared $5d$ KK gauge theory (this is exactly what we are doing in this paper). Second, one constructs a $4d$ $\cN=1$ supersymmetric duality wall \cite{Gaiotto:2015usa, Gaiotto:2015una, Razamat:2016dpl,Bah:2017gph,Kim:2017toz,Kim:2018bpg,Kim:2018lfo,Razamat:2018gro,Razamat:2019mdt,Pasquetti:2019hxf,Razamat:2019ukg,Razamat:2020bix,Hwang:2021xyw,Sabag:2022hyw}. This second step is very similar to our prescription, the difference is that we are adding the triangle terms and we are gauging the $5d$ gauge groups also in $4d$. Gauging such puncture symmetry should be related to gluing the two boundaries of the tube into a torus.

This suggests that our $4d$ gauge theories are related to their \emph{mother} $6d$ SCFT on a Riemann surface with flux, but no punctures (a puncture would reveal itself as some global symmetry descending from a $5d$ gauge symmetry). More precisely, since the rank of the $4d$ global symmetry for our theories is the rank of the $6d$ global symmetry minus one, one can expect them to be a relevant superpotential deformation of the $4d$ SCFTS obtained by $6d$ SCFT on a Riemann surface with flux (which instead have the rank of the $4d$ global symmetry equal to the rank of the $6d$ global symmetry).

We leave an investigation of these issues to future work.

\subsection*{Notations} \label{notationQuiver}
In this paper we use the quiver notation to denote the theories we are studying. The $4d$ quivers that we are going to study denote theories with 4-supercharges. Let us summarize here the notation that we will use.
\paragraph{Quiver diagrams}
\begin{itemize}
\item a circle node denotes a gauge group and the colour will specify which kind 
\begin{itemize}
\item a red node $\bpic \node[circle,draw,red] at(0,-0.3) {$\tiny{\!\!N\!\!}$}; \epic$ denotes $SU(N)$
\item a blue node $\bpic \node[circle,draw,blue] at(0,-0.3) {$\tiny{\!\!2N\!\!}$}; \epic$ denotes $USp(2N)$
\end{itemize}
\item a square node $\bpic \node[rectangle,draw, red] at(0,-0.3) {$\,\tiny{N}\,$}; \epic$ denotes a $SU(N)$ flavor group\footnote{Sometimes we will use a different colour for the flavor group. It happens when inside a quiver we have two (or more) identical nodes and we want an easier way to distinguish what symmetry we are talking about.}
\item An oriented link between two nodes {\tiny$\bpic \node[circle,draw,red](x) at(0,-0.3) {$\tiny{\!\!N_1\!\!}$}; \node[circle,draw,red](y) at(1.1,-0.3) {$\tiny{\!\!N_2\!\!}$}; \draw (x) -- pic[pos=0.7,sloped]{arrow} (y);\epic$} denotes a chiral field in the fundamental representation of $SU(N_2)$ and in the anti-fundamental representation of $SU(N_1)$
\item An arc on a node $\bpic \node[circle,draw,red] at(0,0) {$\tiny{\!\!N\!\!}$}; \draw (0.3,0.2) to[out=90,in=0] pic[pos=0.1,sloped]{arrow} (0,0.6) to[out=180,in=90] pic[pos=0.5,sloped,very thick]{arrow=latex reversed} (-0.3,0.2); \epic$ denotes a chiral field in the antisymmetric representation
\end{itemize}

\paragraph{Flips}
In this paper, an important role will be played by a class of gauge singlet chiral field $\sigma$ called flippers. We say that $\sigma$ \emph{flips an operator} $\cO$ when it enters the superpotential through the term $\sigma\cdot \cO$. Most of the time, we will not draw these flippers in the quiver. Their presence can be inferred looking at the superpotential.

\paragraph{Superpotential}
In theories with 4-supercharges, the holomorphic function $\cW$ called the superpotential plays a really important role in the dynamics. 
\begin{itemize}
\item A term in the superpotential is represented by a closed loop in the quiver notation. Often we will denote these terms by the geometrical shape and not by the actual names of the fields. For example, for a cubic term represented by the following quiver {\tiny$\bpic \node[circle,draw,red](x) at(0,0.3) {$\tiny{\!\!N_1\!\!}$}; \node[circle,draw,red](y) at(1.1,0.3) {$\tiny{\!\!N_2\!\!}$}; \node[rectangle, draw, red] (z) at(0.5, -0.5){$2$}; \draw (x) -- pic[pos=0.7,sloped]{arrow} (y);\draw (y) -- pic[pos=0.3,sloped,very thick]{arrow=latex reversed} (z); \draw (z) -- pic[pos=0.4,sloped,very thick]{arrow=latex reversed} (x); \node at(0.5,0.5) {$a$}; \node at(1,-0.2) {$b$}; \node at(0,-0.2) {$c$}; \epic$} we will either write $\cW= a b c $ or $\cW = \Triangle$
\item Concerning the flippers interaction, instead of writing $\cW = \sigma \cdot \cO$ we will often use the following notation $\cW = \Flip[\cO]$. Using this notation, we could avoid giving a name to the flipper $\sigma$. When we want to refer to a specific flipper we will use the notation $\Flipper[\cO]$ (or an explicit name if we gave one). 
\end{itemize}

\section{A simple class: $R_{N, 2}$ and its two duals}\label{fam1}
In this section we consider a simple class of theories, which are special cases of the more general class studied in Section \ref{fam1g}.

\subsection{5d triality} \label{5dDualityI}
\begin{figure}[H]
\begin{minipage}{0.3\textwidth}
\centering
\includegraphics[height=0.24\textheight]{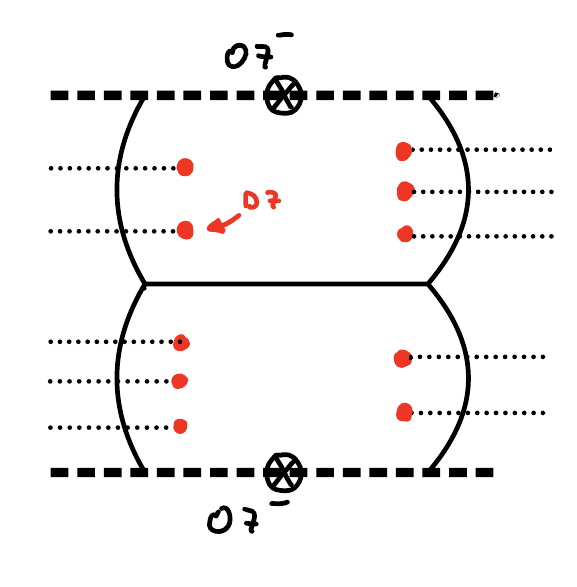}
\end{minipage}%
\begin{minipage}{0.4\textwidth}
\centering
\includegraphics[height=0.24\textheight]{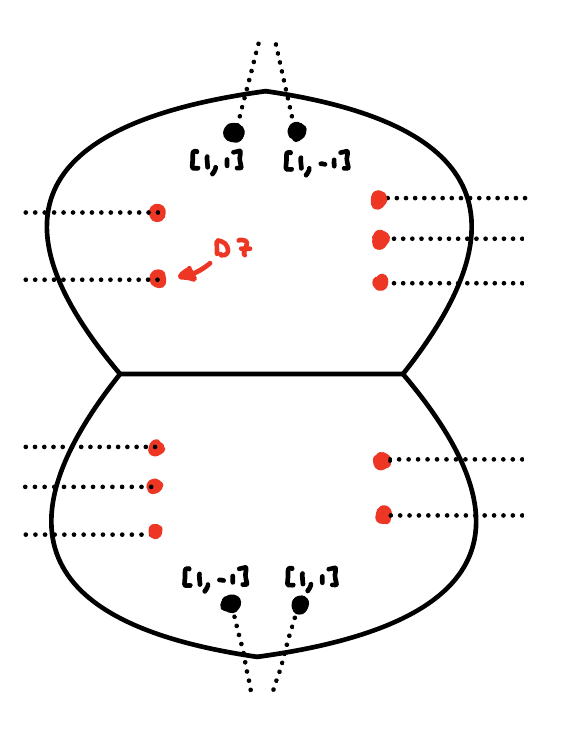}
\end{minipage}%
\begin{minipage}{0.3\textwidth}
\centering
\includegraphics[height=0.24\textheight]{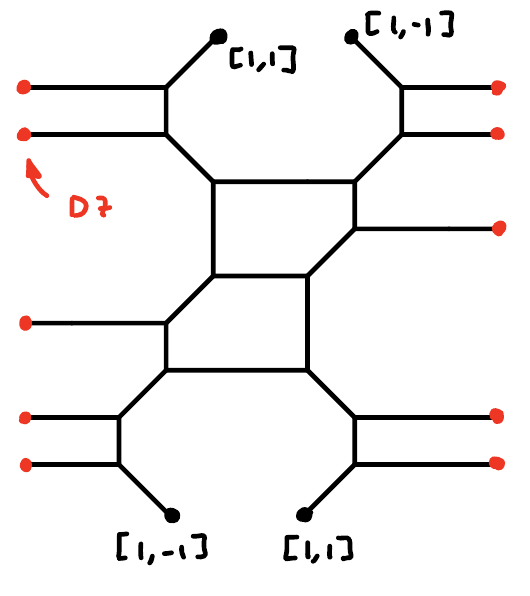}
\end{minipage}
\caption{Resolution of the two $O7^-$ planes leading to $R_{3,2}$: a $SU(3)$ gauge theory with 10 fundamentals.}
\label{SU3+10F1}
\end{figure}

\begin{figure}[H]
\begin{minipage}{0.3\textwidth}
\centering
\includegraphics[height=0.2\textheight]{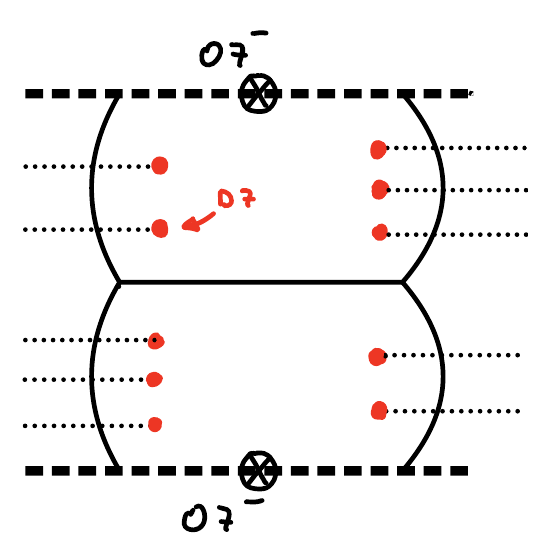}
\end{minipage}%
\begin{minipage}{0.4\textwidth}
\centering
\includegraphics[height=0.2\textheight]{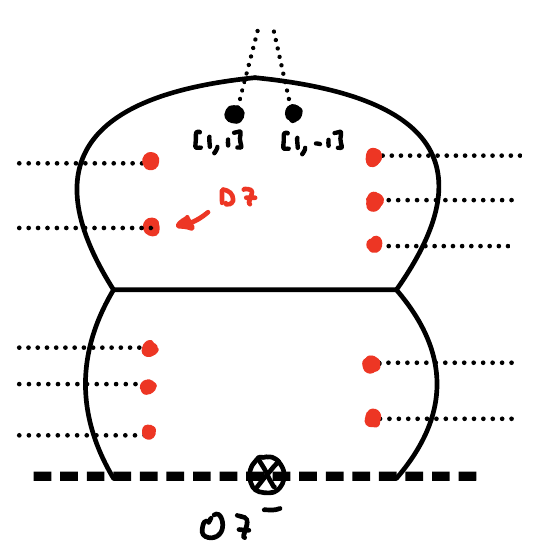}
\end{minipage}%
\begin{minipage}{0.3\textwidth}
\centering
\includegraphics[height=0.18\textheight]{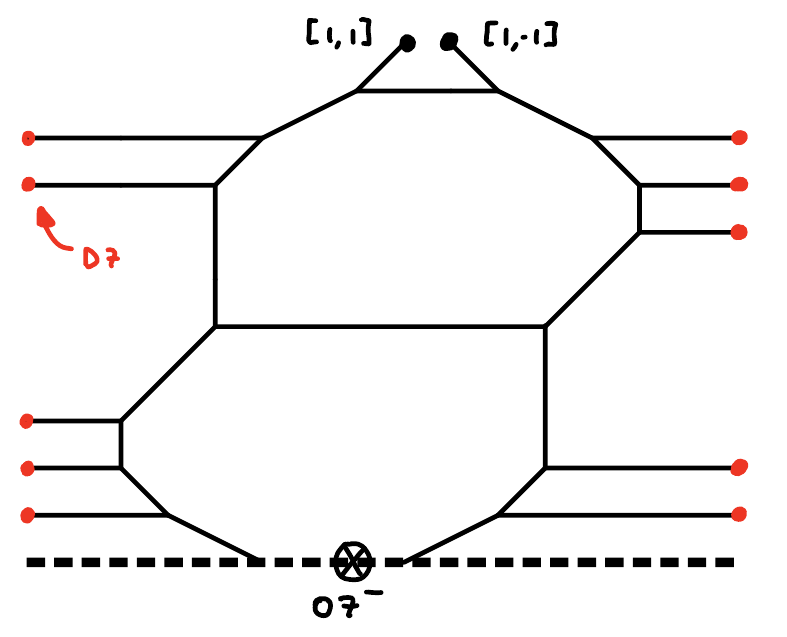}
\end{minipage}
\caption{Resolution of one of the two $O7^-$ planes leading to the $USp(4)$ with 10 fundamental hypers gauge theory.}
\label{USp4+10F1}
\end{figure}

\begin{figure}[H]
\centering
\includegraphics[height=0.22\textheight]{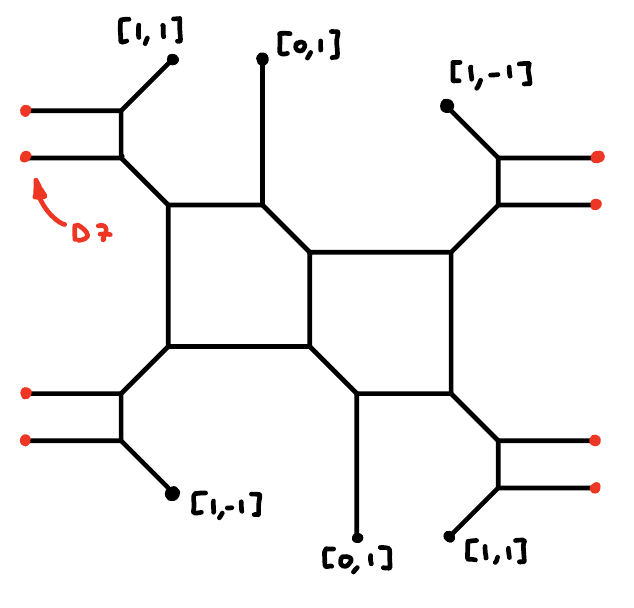}
\caption{$4F + SU(2)-SU(2)+4F$}
\label{SU2Quiver1}
\end{figure}

The first $5d$ dualities that we are studying combine into a triality:
\be \label{UVdualitiesFamilyI} \scalebox{0.9}{\bpic[node distance=2cm,gSUnode/.style={circle,red,draw,minimum size=8mm},gUSpnode/.style={circle,blue,draw,minimum size=8mm},fnode/.style={rectangle,red,draw,minimum size=8mm}]
\begin{scope}[shift={(-3,0)}]  
\node at (-0.8,1.2) {$\star_1)$};
\node[gSUnode] (G1) at (0,0) {$N$};
\node[fnode] (F1) at (2.3,0) {$2N+4$};
\draw (G1) --  (F1);
\end{scope}
\begin{scope}[shift={(2,0)}]
\node at (-0.8,1.2) {$\star_2)$};
\node[gUSpnode] (G2) at (0,0) {\scalebox{0.8}{$2N-2$}};
\node[fnode] (F2) at (2.2,0) {$2N+4$};
\draw (G2) --  (F2);
\end{scope}
\begin{scope}[shift={(6.7,0)}]
\node at (-0.8,1.2) {$\star_3)$};
\node[fnode] (F3) at (0,0) {$4$};
\node[gSUnode] (G3) at (1.5,0) {$2$};
\node (G4) at (2.5,0) {$\dots$};
\node[gSUnode] (G5) at (3.5,0) {$2$};
\node[fnode] (F4) at (5,0) {$4$};
\draw (F3) -- (G3) -- (G4) -- (G5) -- (F4);
\draw[decorate,decoration={calligraphic brace,mirror,amplitude=7pt}] (1.1,-0.5) -- (3.9,-0.5) node[pos=0.5,below=9pt,black] {$N-1$};
\end{scope}
\epic} \ee 
Throughout the paper, we denote $SU$ nodes with red circles, $Usp$ nodes with blue circles and global $SU$ symmetries with square.
To understand why the three theories in \eqref{UVdualitiesFamilyI} are dual to each other, we recall the analysis done in \cite{Hayashi:2015fsa, Hayashi:2015zka,Zafrir:2015rga}. We start from the $6d$ Type IIA brane setup Figure \ref{TypeIIA6dCompletionFamilyIFig}. Then, we do a circle compactification and perfom T-duality along the compactified direction. We obtain a Type IIB brane setup. The $O8^-$ plane becomes two $O7^{-}$ planes and the $D8$ become $D7$. The resulting brane web, for $N=3$, is shown on the left in Figure \ref{SU3+10F1}. Then in order to read the gauge theory we have to resolve the $O7^{-}$ plane by 7-branes \cite{Sen:1996vd}. We have the choice to resolve the two $O7^{-}$ or just one. If we resolve the two $O7^{-}$ we get the brane web in the middle of Figure \ref{SU3+10F1}. After pulling-out the 7-branes we obtain the $SU(3)$ gauge theory with $10$ fundamental hypers shown in the right of Figure \ref{SU3+10F1}. The general $N$ case corresponds to the left theory in \eqref{UVdualitiesFamilyI} and we call it $R_{N,2}$. This name will be clear when we consider the generalization in the next section.
Now, if we resolve only one $O7^-$ plane we obtain, after pulling out the 7-branes, the $USp(4)$ gauge theory on the right of Figure \ref{USp4+10F1}. It correponds to the middle theory in \eqref{UVdualitiesFamilyI}.
If we perform an S-duality on the right figure of Figure \ref{SU3+10F1} (which amounts to a $90^{\circ}$ rotation of the pq-web), we obtain Figure \ref{SU2Quiver1} which describes $4F+SU(2)-SU(2)+4F$\footnote{The notation means that the first and the second $SU(2)$ are coupled to 4 fundamental hypers and the horizontal bar represents a bifundamental hyper between the two gauge nodes.} quiver theory. It corresponds to the right theory in \eqref{UVdualitiesFamilyI}.
Since the theories in \eqref{UVdualitiesFamilyI} are either coming from the same brane system or  are related by S-duality, it is clear that they are UV dual in the sense of completed by the same theory.

\subsection{6d UV completion: $(D_{N+2},D_{N+2})$ Minimal Conformal Matter}
The $6d$ UV completion of the $5d$ theories in \eqref{UVdualitiesFamilyI} is given by the following Type IIA brane setup \cite{Hayashi:2015fsa, Hayashi:2015zka,Zafrir:2015rga}: 
\begin{figure}[H]
\centering
\includegraphics[height=0.25\textheight]{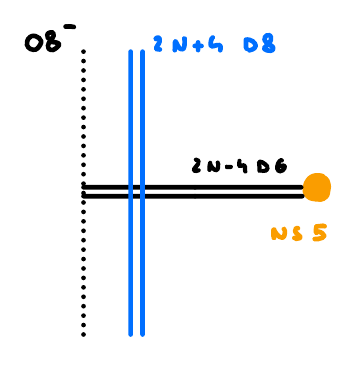}
\caption{Type IIA brane setup corresponding to the $6d$ UV completion of $R_{N,2}$.}
\label{TypeIIA6dCompletionFamilyIFig}
\end{figure}
This theory is called the $(D_{N+2},D_{N+2})$ Minimal Conformal Matter. On the tensor branch, the system flows to the following gauge theory:
\be \label{TypeIIA6dCompletionFamilyI} \scalebox{0.9}{\bpic[node distance=2cm,gSUnode/.style={circle,red,draw,minimum size=8mm},gUSpnode/.style={circle,blue,draw,minimum size=8mm},fnode/.style={rectangle,red,draw,minimum size=8mm}]
\node[gUSpnode] (G1) at (-4,0) {\scalebox{0.9}{$2N-4$}};
\node[fnode] (F1) at (-1.5,0) {\scalebox{0.8}{$2N+4$}};
\draw (G1) -- (F1);
\epic} \ee

\subsection{4d triality} \label{4ddualitiesFamilyI}
We can now apply our prescription, already described in section \ref{Algorithm}. Starting from the left theory in \eqref{UVdualitiesFamilyI}, we replace the hypers in $5d$ by a chiral field in $4d$. We also have to split the $2N+4$ hypers into $N+2$ chirals and $N+2$ anti-chirals for the theory to be non-anomalous. We get the theory $\star_1)$ in \eqref{4dFamilyI1}. We have also added a gauge singlet in the bifundamental of the $SU(N+2)_Q \times SU(N+2)_{\Qt}$ flavor symmetry and a flipping type superpotential. The role of this flipper is essential for the duality to be true as we will see in the following. Our procedure applied to the theory in the middle of \eqref{UVdualitiesFamilyI} produces the theory $\star_2)$ in \eqref{4dFamilyI1}. Finally, we focus to the right theory of \eqref{UVdualitiesFamilyI}. In this case, since we have a quiver in $5d$ our procedure tells us that for each bifundamental we have to associate a \say{triangle} with an explicit $SU(2)$ symmetry. We obtain the following $4d$ quiver $\star_3)$ in \eqref{4dFamilyI2} with the correct set of flippers. 
\be \label{4dFamilyI1} \scalebox{0.9}{\bpic[node distance=2cm,gSUnode/.style={circle,red,draw,minimum size=8mm},gUSpnode/.style={circle,blue,draw,minimum size=8mm},fnode/.style={rectangle,red,draw,minimum size=8mm}] 
\begin{scope}[shift={(-3,0)}] 
\node at (-1.5,1) {$\star_1)$};
\node[gSUnode] (G1) at (0,0) {$N$};
\node[fnode] (F1) at (-1,-2) {$N+2$};
\node[fnode] (F2) at (1,-2) {$N+2$};
\draw (G1) -- pic[pos=0.2,sloped]{arrow} (F1);
\draw (G1) -- pic[pos=0.5,sloped]{arrow} (F2);
\node[right] at (-1.3,-3) {$ \cW= \Flip[Q \, \Qt]$};
\node at (-0.9,-0.8) {$Q$};
\node at (0.9,-0.8) {$\Qt$};
\end{scope}
\node at (0,0) {$\longleftrightarrow$};
\begin{scope}[shift={(2.5,0)}]
\node at (-1,1) {$\star_2)$};
\node[gUSpnode] (G2) at (0,0) {\scalebox{0.9}{$2N-2$}};
\node[fnode] (F2) at (2.5,0) {$2N+4$};
\draw (G2) -- (F2);
\node[right] at (-0.2,-1.2) {$ \cW= \Flip[Q \, Q]$};
\node at (1.2,0.3) {$Q$};
\end{scope}
\epic} \ee 
\be \label{4dFamilyI2} \scalebox{0.9}{\bpic[node distance=2cm,gSUnode/.style={circle,red,draw,minimum size=8mm},gUSpnode/.style={circle,blue,draw,minimum size=8mm},fnode/.style={rectangle,red,draw,minimum size=8mm}]  
\node at (-6.5,0) {$\longleftrightarrow$};
\node at (-5.5,1) {$\star_3)$};
\node[fnode,orange] (F1) at (-4,0) {$4$};
\node[gSUnode] (G1) at (-2.5,0) {$2$};
\node[gSUnode] (G2) at (-1,0) {$2$};
\node (G3) at (0.5,0) {$\dots$};
\node (G4) at (1.5,0) {$\dots$};
\node[gSUnode] (G5) at (3,0) {$2$};
\node[fnode] (F2) at (4.5,0) {$4$};
\node[fnode,orange] (F3) at (-1.8,-1.5) {$2$};
\node[fnode] (F4) at (-0.2,-1.5) {$2$};
\node[fnode,violet] (F5) at (2.4,-1.5) {$2$};
\draw (F1) -- pic[pos=0.7,sloped]{arrow} (G1) -- (G2) -- (G3);
\draw (G4) -- (G5) -- pic[pos=0.7,sloped]{arrow} (F2);
\draw (G1) -- (F3);
\draw (G2) -- (F3);
\draw (G2) -- (F4);
\draw (G3) -- (F4);
\draw (G4) -- (F5);
\draw (G5) -- (F5);
\node[right] at (-5.5,-3) {$ \cW= (N-2) \, \triangles + \Flip[L \, L] + \Flip[R \, R] + \displaystyle\sum_{i=1}^{N-2} \Flip[B_i \, B_i]$};
\node at (-3.2,0.4) {$L$};
\node at (-1.8,0.4) {$B_1$};
\node at (-0.3,0.4) {$B_2$};
\node at (2.1,0.4) {$B_{N-2}$};
\node at (3.7,0.4) {$R$};
\node at (-2.4,-0.8) {$V_1$};
\node at (-1.1,-0.8) {$D_1$};
\node at (-0.3,-0.8) {$V_2$};
\node at (0.4,-0.8) {$D_2$};
\node at (1.5,-0.8) {$V_{N-2}$};
\node at (3.2,-0.8) {$D_{N-2}$};
\epic} \ee 
The mapping of the chiral ring generators between the different frames is
\vspace{-2.5cm}
\be \label{mappingFamilyI}
\scalebox{0.9}{$
\ba[t]{c}\star_1) \\
\\
\\[6pt]
\begin{cases}
\Flipper[Q \, \Qt] \\
Q^N \\
\Qt^N
\end{cases}
\ea
\ba{c} \\
\\
\\
\\
\\
\\
\\
\\
\\
\\[-10pt]
\Longleftrightarrow
\ea
\ba[t]{c} \star_2) \\
\\
\\
\\
\\[-5pt]
\Flipper[Q \, Q]
\ea
\ba{c} \\
\\
\\
\\
\\
\\
\\
\\
\\
\\[-10pt]
\Longleftrightarrow
\ea
\ba[t]{c}\star_3) \\
\begin{cases}
\Flipper[L \, L] \\
\Flipper[R \, R] \\
L B_1 \dots B_{N-2} R \\
L B_1 \dots B_i V_{i+1} \\
D_j B_{j+1} \dots B_{N-2} R \\
\Flipper[B_k \, B_k] \\
D_i B_{i+1} \dots B_j V_j
\end{cases}
\ea
\qquad
\ba[t]{l}\\
\\
\\
\\
\\[-4pt]
i=0, \dots, N-3 \\[4pt]
j=1, \dots, N-2 \\[4pt]
k=1, \dots N-2 \\[4pt]
i=1, \dots, N-3 \, \& \\
j=i+1, \dots, N-2
\ea
$}
\ee
We have to understand the mapping \eqref{mappingFamilyI} in the following way. In the UV, the manifest global symmetries in $\star_1)$, $\star_2)$ and $\star_3)$ are different. In the IR, there is the emergence of the global symmetry. Therefore some operators in the UV will combined into an operator transforming into the bigger symmetry group. In our case the global symmetry group\footnote{In this article, we don't pay attention to the global structure of the global symmetry.} in the IR is $SU(2N+4)$. Then we claim that in the frame $\star_1)$ the three operators $\Flipper[Q \, \Qt], \, Q^N$ and $\Qt^N$ will combine into an operator that transforms into an antisymmetric representation of the emergent $SU(2N+4)$ global symmetry group. One necessary condition to make sense is that the number of degrees of freedom (d.o.f) corresponds to the dimension of the representation. In this case $\Flipper[Q \, \Qt]$ contains $N^2+4N+4$ d.o.f, $Q^N$ and $\Qt^N$ $\frac{1}{2}(N+2)(N+1)$ each. The sum equals to $2N^2 + 7N + 6$ which indeed correspond to the dimension of the antisymmetric representation of $SU(2N+4)$. The same kind of counting works for the frame $\star_3)$. 

\subsection{Proof of the 4d dualities} \label{Proof4ddualitiesFamilyI}
In this subsection, we provide a \say{proof} of the $4d$ triality \eqref{4dFamilyI1}-\eqref{4dFamilyI2}. By \say{proof}, we mean the use of a sequence of well-established dualities as in \cite{Bajeot:2022kwt,Bajeot:2022lah}. Starting from $\star_1)$ and apply the Seiberg duality \cite{Seiberg:1994pq}, we obtain
\be \label{4dFamilyI3} \scalebox{0.9}{\bpic[node distance=2cm,gSUnode/.style={circle,red,draw,minimum size=8mm},gUSpnode/.style={circle,blue,draw,minimum size=8mm},fnode/.style={rectangle,red,draw,minimum size=8mm}]
\node[gSUnode] (G1) at (-4,0) {$2$};
\node[fnode] (F1) at (-2,0) {\scalebox{0.8}{$2N+4$}};
\draw (G1) -- (F1);
\node at (0,0) {$\cW = 0$};
\epic} \ee
We see that the role of the flipper in $\star_1)$ in \eqref{4dFamilyI1} is to give a mass to the singlet present in the Seiberg duality and therefore get $\cW=0$ in \eqref{4dFamilyI3}. 

Now starting from $\star_2)$ in \eqref{4dFamilyI1} and applying the Intriligator-Pouliot (IP) duality \cite{Intriligator:1995ne}, we once again get \eqref{4dFamilyI3}. This implies that also $\star_1)$ and $\star_2)$ are dual.

More work has to be done in order to prove that also $\star_3)$ is dual. It goes as follows. We first apply the Csaki-Schmaltz-Skiba-Terning (CSST) duality \cite{Csaki:1997cu} to the left $SU(2)$. The form of this duality that is useful for our purpose is the following
\be \label{CSST} \scalebox{0.9}{\bpic[node distance=2cm,gSUnode/.style={circle,red,draw,minimum size=8mm},gUSpnode/.style={circle,blue,draw,minimum size=8mm},fnode/.style={rectangle,red,draw,minimum size=8mm}] 
\begin{scope}[shift={(-3,0)}] 
\node[gSUnode] (G1) at (0,0) {$2$};
\node[fnode,orange] (F1) at (-1.5,0) {$4$};
\node[fnode,orange] (F2) at (0.8,-1.5) {$2$};
\node[fnode] (F3) at (1.5,0) {$2$};
\draw (G1) -- pic[pos=0.3,sloped]{arrow} (F1);
\draw (G1) -- (F2);
\draw (G1) -- (F3);
\node[right] at (-1,-2.4) {$ \cW= 0$};
\node at (-0.7,0.3) {$L$};
\node at (0.8,0.3) {$B_1$};
\node at (0.1,-0.8) {$V_1$};
\end{scope}
\node at (1,0) {$\longleftrightarrow$};
\begin{scope}[shift={(5.5,0)}]
\node[gSUnode] (G1) at (0,0) {$2$};
\node[fnode,orange] (F1) at (-1.5,0) {$4$};
\node[fnode,orange] (F2) at (0.8,-1.5) {$2$};
\node[fnode] (F3) at (1.5,0) {$2$};
\draw (G1) -- pic[pos=0.5,sloped,very thick]{arrow=latex reversed} (F1);
\draw (G1) -- (F2);
\draw (G1) -- (F3);
\draw (F2) -- (F3);
\node at (-0.7,0.3) {$l$};
\node at (0.8,0.3) {$b_1$};
\node at (0.1,-0.8) {$v_1$};
\node[right] at (-3.3,-2.4) {$ \cW= \Flip[l \, l] + \Flip[b_1 b_1] + \Flip[v_1 v_1] + \Triangle$};
\end{scope}
\epic} \ee 
The important effect of this duality is to give a mass to the field $D_1$ in \eqref{4dFamilyI2}. Indeed, we are left with
\be \label{Proof4dFamilyI1} \scalebox{0.9}{\bpic[node distance=2cm,gSUnode/.style={circle,red,draw,minimum size=8mm},gUSpnode/.style={circle,blue,draw,minimum size=8mm},fnode/.style={rectangle,red,draw,minimum size=8mm}]  
\node[fnode,orange] (F1) at (-4,0) {$4$};
\node[gSUnode] (G1) at (-2.5,0) {$2$};
\node[gSUnode] (G2) at (-1,0) {$2$};
\node (G3) at (0.5,0) {$\dots$};
\node (G4) at (1.5,0) {$\dots$};
\node[gSUnode] (G5) at (3,0) {$2$};
\node[fnode] (F2) at (4.5,0) {$4$};
\node[fnode,orange] (F3) at (-1.8,-1.5) {$2$};
\node[fnode] (F4) at (-0.2,-1.5) {$2$};
\node[fnode,violet] (F5) at (2.4,-1.5) {$2$};
\draw (F1) -- pic[pos=0.7,sloped]{arrow} (G1) -- (G2) -- (G3);
\draw (G4) -- (G5) -- pic[pos=0.7,sloped]{arrow} (F2);
\draw (G1) -- (F3);
\draw (G2) -- (F4);
\draw (G3) -- (F4);
\draw (G4) -- (F5);
\draw (G5) -- (F5);
\node[right] at (-5.5,-3) {$ \cW= (N-3) \, \triangles + \Flip[R \, R] + \displaystyle\sum_{i=2}^{N-2} \Flip[B_i \, B_i]$};
\node at (-3.2,0.4) {$l$};
\node at (-1.8,0.4) {$b_1$};
\node at (-0.3,0.4) {$B_2$};
\node at (2.1,0.4) {$B_{N-2}$};
\node at (3.7,0.4) {$R$};
\node at (-2.4,-0.8) {$v_1$};
\node at (-0.3,-0.8) {$V_2$};
\node at (0.4,-0.8) {$D_2$};
\node at (1.5,-0.8) {$V_{N-2}$};
\node at (3.2,-0.8) {$D_{N-2}$};
\epic} \ee
Now we realize that the second $SU(2)$ is coupled to $6$ chirals and therefore we can use the IP confinement for this $SU(2)$ \cite{Intriligator:1995ne}. The form useful of this confinement is the following
\be \label{ConfinementSU} \scalebox{0.9}{\bpic[node distance=2cm,gSUnode/.style={circle,red,draw,minimum size=8mm},gUSpnode/.style={circle,blue,draw,minimum size=8mm},fnode/.style={rectangle,red,draw,minimum size=8mm}] 
\begin{scope}[shift={(-3,0)}] 
\node[gSUnode] (G1) at (0,0) {$2$};
\node[gSUnode] (G2) at (1.5,0) {$2$};
\node[gSUnode] (G3) at (-1.5,0) {$2$};
\node[fnode] (F2) at (0.8,-1.5) {$2$};
\draw (G1) -- (G2);
\draw (G1) -- (G3);
\draw (G1) -- (F2);
\draw (G2) -- (F2);
\node[right] at (-2,-2.4) {$ \cW= \Flip[b_1 \, b_1] + \Flip[B_2 \, B_2]$};
\node at (-0.7,0.3) {$b_1$};
\node at (0.8,0.3) {$B_2$};
\node at (0.1,-0.8) {$V_2$};
\node at (1.5,-0.8) {$D_2$};
\end{scope}
\node at (1,0) {$\longleftrightarrow$};
\begin{scope}[shift={(3.5,0)}]
\node[gSUnode] (G1) at (0,0) {$2$};
\node[gSUnode] (G2) at (1.5,0) {$2$};
\node[fnode] (F2) at (0.8,-1.5) {$2$};
\draw (G1) -- (G2);
\draw (G1) -- (F2);
\node[right] at (-0.5,-2.4) {$ \cW= \Flip[b_2 \, b_2]$};
\node at (0.8,0.4) {$b_2$};
\end{scope}
\epic} \ee
After the confinement of the second $SU(2)$ we can see that the next one on the right is also coupled to $6$ chirals and therefore we can iterate the use of \eqref{ConfinementSU}. We can do $(N-4)$ more s-confining \eqref{ConfinementSU}. We get     
\be \label{Proof4dFamilyI2} \scalebox{0.9}{\bpic[node distance=2cm,gSUnode/.style={circle,red,draw,minimum size=8mm},gUSpnode/.style={circle,blue,draw,minimum size=8mm},fnode/.style={rectangle,red,draw,minimum size=8mm}]  
\node[gSUnode] (G1) at (0,0) {$2$};
\node[gSUnode] (G2) at (2,0) {$2$};
\node[fnode,orange] (F1) at (-1.5,0) {$4$};
\node[fnode] (F2) at (3.5,0) {$4$};
\node[fnode,orange] (F3) at (-1.4,-1.6) {$2$};
\node[fnode,violet] (F4) at (1.6,-1.6) {$2$};
\node[fnode] (F5) at (-0.2,-1.6) {$2$};
\draw (F1) -- (G1) -- (G2) -- pic[pos=0.7,sloped]{arrow} (F2);
\draw (F3) -- (G1);
\draw (F4) -- (G1);
\draw (F5) -- (G1);
\node[right] at (-2,-2.6) {$ \cW= \Flip[b_{N-2} \, b_{N-2}] + \Flip[R \, R]$};
\node at (-0.7,0.4) {$l$};
\node at (-1,-0.8) {$v_1$};
\node at (0.2,-0.8) {$v_2$};
\node at (0.7,-1.6) {$\dots$};
\node at (1.3,-0.8) {$v_{N-2}$};
\node at (1,0.4) {$b_{N-2}$};
\node at (2.7,0.4) {$R$};
\epic} \ee
The last $SU(2)$ is once again coupled to $6$ chirals and therefore we can use for the last time the confinement \cite{Seiberg:1994bz}. We end up with
\be \label{Proof4dFamilyI3} \scalebox{0.9}{\bpic[node distance=2cm,gSUnode/.style={circle,red,draw,minimum size=8mm},gUSpnode/.style={circle,blue,draw,minimum size=8mm},fnode/.style={rectangle,red,draw,minimum size=8mm}]
\begin{scope}[shift={(-3,0)}]   
\node[gSUnode] (G1) at (0,0) {$2$};
\node[fnode,orange] (F1) at (-1.5,0) {$4$};
\node[fnode] (F2) at (1.5,0) {$4$};
\node[fnode,orange] (F3) at (-1.4,-1.6) {$2$};
\node[fnode,violet] (F4) at (1.6,-1.6) {$2$};
\node[fnode] (F5) at (-0.2,-1.6) {$2$};
\draw (F1) -- (G1) -- (F2);
\draw (F3) -- (G1);
\draw (F4) -- (G1);
\draw (F5) -- (G1);
\node[right] at (-0.5,-2.4) {$ \cW= 0$};
\node at (-0.7,0.4) {$l$};
\node at (-1,-0.8) {$v_1$};
\node at (0.2,-0.8) {$v_2$};
\node at (0.7,-1.6) {$\dots$};
\node at (1.3,-0.8) {$v_{N-2}$};
\node at (0.7,0.4) {$r$};
\end{scope}
\node at (0,0) {$\equiv$};
\begin{scope}[shift={(1.3,0)}]
\node[gSUnode] (G1) at (0,0) {$2$};
\node[fnode] (F1) at (2,0) {$2N+4$};
\draw (G1) -- (F1);
\node[right] at (0.2,-1) {$ \cW= 0$};
\node at (0.8,0.4) {$ p$};
\end{scope}
\epic} \ee
To summarize, starting from $\star_3)$ in \eqref{4dFamilyI2} and doing the CSST duality followed by $(N-2)$ s-confining duality we get \eqref{4dFamilyI3} which proves the $4d$ triality \eqref{4dFamilyI1}-\eqref{4dFamilyI2}.  

Notice that the duality between $\star_3)$ and \eqref{Proof4dFamilyI3} is one of the simplest instances of the $4d$ \emph{mirror symmetry} of \cite{Pasquetti:2019uop, Bottini:2021vms, Hwang:2021ulb}, and it uplifts the $3d$ mirror symmetry between $U(1)$ with $N$ flavors and the linear Abelian quiver $U(1)^{N-1}$.

Now using the proof we can justify the mapping \eqref{mappingFamilyI}. Indeed we can obtain the mapping from the frame \eqref{4dFamilyI2} to the frame \eqref{Proof4dFamilyI3} by following the mapping of the basic dualities (CSST and the IP confinement). We get
\vspace{-2.5cm}
\be \label{mappingFamilyI2}
\scalebox{0.9}{$
\ba[t]{c}\eqref{4dFamilyI2} \\
\begin{cases}
\Flipper[L \, L] \\
\Flipper[R \, R] \\
L B_1 \dots B_{N-2} R \\
L B_1 \dots B_i V_{i+1} \\
D_j B_{j+1} \dots B_{N-2} R \\
\Flipper[B_k \, B_k] \\
D_i B_{i+1} \dots B_j V_j
\end{cases}
\ea
\ba{c} \\
\\
\\
\\
\\
\\
\\
\\
\\
\\[-10pt]
\Longleftrightarrow
\ea
\ba[t]{c}\eqref{Proof4dFamilyI3} \\
\begin{cases}
l \, l \\
r \, r \\
l \, r \\
l \, v_{i+1} \\
r \, v_j \\
v_k^2 \\
v_i \, v_j
\end{cases}
\ea
\qquad
\ba[t]{l}\\
\\
\\
\\
\\[-4pt]
i=0, \dots, N-3 \\[4pt]
j=1, \dots, N-2 \\[4pt]
k=1, \dots N-2 \\[4pt]
i=1, \dots, N-3 \, \& \\
j=i+1, \dots, N-2
\ea
$}
\ee
Then since there is no superpotential in \eqref{Proof4dFamilyI3} all the operators in the RHS of the mapping \eqref{mappingFamilyI2} combine into $pp$ which transforms in the antisymmetric representation of the $SU(2N+4)$ global symmetry as previously claimed.

\section{Rectangular pq-webs: the $R_{N,k}$ theories}\label{fam1g}
\subsection{5d theories and duality $R_{N,k} \leftrightarrow R_{k,N}$}
\subsubsection*{$R_{N,k}$ theories}
\begin{figure}[H]
\begin{minipage}{0.6\textwidth}
\centering
\includegraphics[height=0.25\textheight]{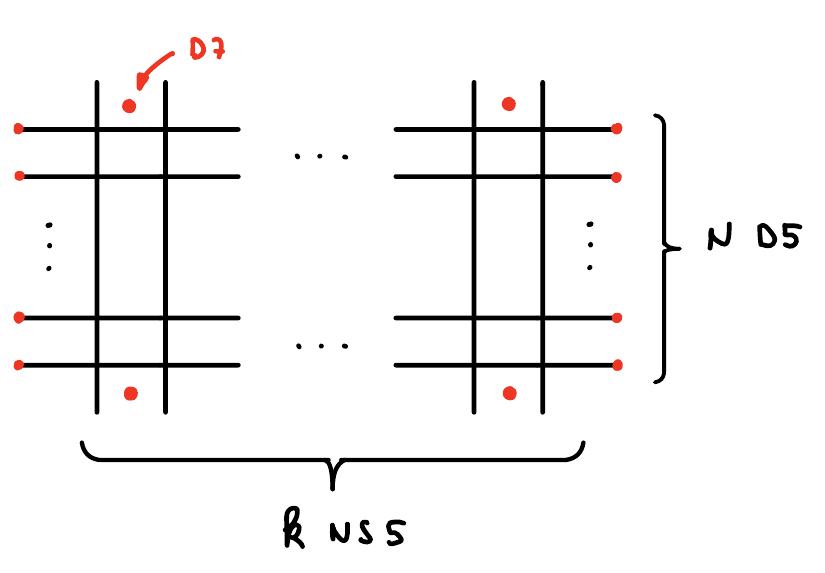}
\end{minipage}%
\begin{minipage}{0.4\textwidth}
\centering
\includegraphics[height=0.3\textheight]{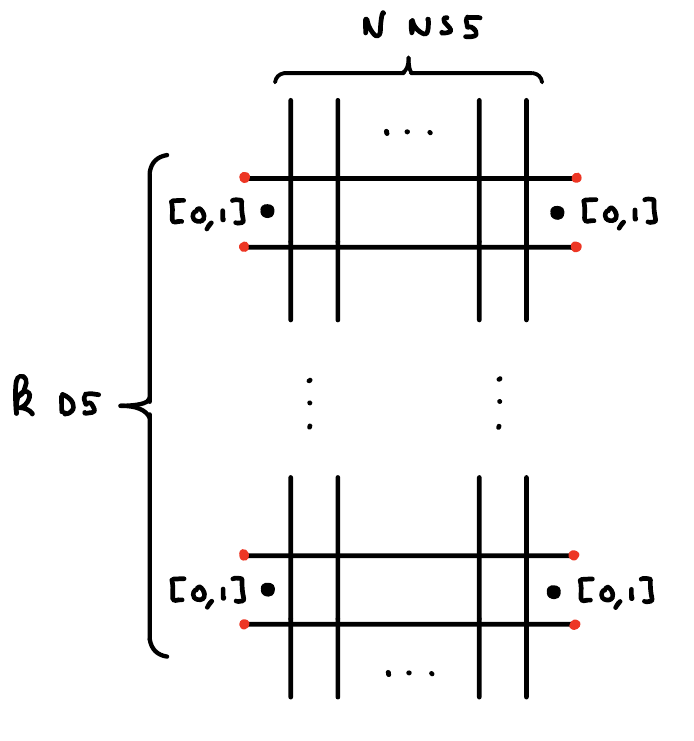}
\end{minipage}
\caption{Brane setup for $(N+2)F+SU(N)^{k-1}+(N+2)F$ on the left and for $(k+2)F+SU(k)^{N-1}+(k+2)F$ for the right. The two brane systems are S-dual.}
\label{TypeIIB5dGeneralFamilyI1}
\end{figure}

In this section, we generalize the discussion of Sec. \ref{fam1} by considering the following two-parameter family of $5d$ theories, that we call $R_{N,k}$:
\be \label{RNk} \scalebox{0.9}{\bpic[node distance=2cm,gSUnode/.style={circle,red,draw,minimum size=8mm},gUSpnode/.style={circle,blue,draw,minimum size=8mm},fnode/.style={rectangle,red,draw,minimum size=8mm}]
\node at (-6.5,1) {$R_{N,k}:$};
\node[fnode] (F1) at (-5,0) {$N+2$};
\node[gSUnode] (G1) at (-3,0) {$N$};
\node[gSUnode] (G2) at (-1.5,0) {$N$};
\node (G3) at (0,0) {$\dots$};
\node[gSUnode] (G4) at (1.5,0) {$N$};
\node[fnode] (F2) at (3.5,0) {$N+2$};
\draw (F1) -- (G1) -- (G2) -- (G3) -- (G4) -- (F2);
\draw[decorate,decoration={calligraphic brace,mirror,amplitude=7pt}] (-3.4,-0.5) -- (1.9,-0.5) node[pos=0.5,below=9pt,black] {$k-1$};
\epic} \ee 
The brane web associated to the $R_{N,k}$ is shown on the left of Figure \ref{TypeIIB5dGeneralFamilyI1}. We can perform S-duality on this brane system and we obtain the web on the right of Figure \ref{TypeIIB5dGeneralFamilyI1}.
It is not completely obvious how to read off the gauge theory for the S-dual theory. Let us illustrate  the case of $N=2$ and $k=3$, Figure \ref{TypeIIB5dGeneralFamilyISdual1}:
\begin{figure}[H]
\centering
\includegraphics[height=0.25\textheight]{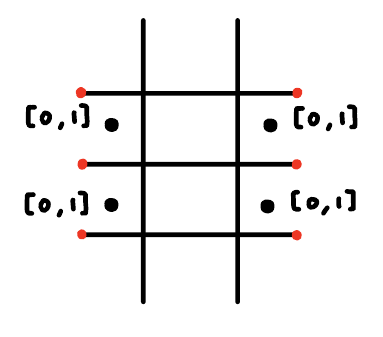}
\caption{The S-dual brane system of $4F+SU(2)^2+4F$.}
\label{TypeIIB5dGeneralFamilyISdual1}
\end{figure}
First, we  pull out the $[0,1]$ 7-branes through the D5 branes. Due to the Hanany-Witten effect, we get 
\begin{figure}[H]
\centering
\includegraphics[height=0.25\textheight]{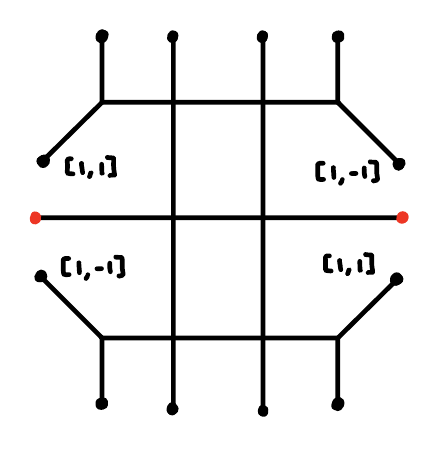}
\caption{Brane system after pulling out the $[0,1]$ 7-branes of Figure \ref{TypeIIB5dGeneralFamilyISdual1}.}
\label{TypeIIB5dGeneralFamilyISdual2}
\end{figure}
The second step is to pull out the $[1,1]$ and $[1,-1]$ 7-branes through the D5 branes. We get
\begin{figure}[H]
\centering
\includegraphics[height=0.25\textheight]{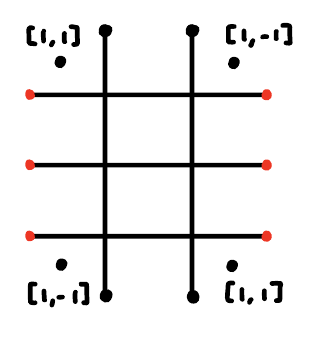}
\caption{Brane system after pulling out the $[1,1]$ and $[1,-1]$ 7-branes of Figure \ref{TypeIIB5dGeneralFamilyISdual2} through the D5.}
\label{TypeIIB5dGeneralFamilyISdual3}
\end{figure}
The final step is to pull out the $[1,1]$ and $[1,-1]$ 7-branes through the NS5 brane. We get
\begin{figure}[H]
\centering
\includegraphics[height=0.25\textheight]{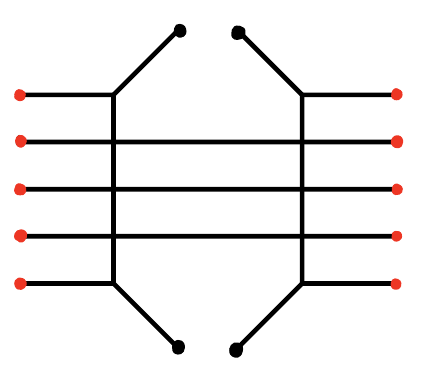}
\caption{Brane system after pulling out the $[1,1]$ and $[1,-1]$ 7-branes of Figure \ref{TypeIIB5dGeneralFamilyISdual3} through the NS5. In this frame, it is easy to read off the gauge theory that is $SU(3)+10F$.}
\label{TypeIIB5dGeneralFamilyISdual4}
\end{figure}
It is easy to generalize the previous discussion and we find that the brane system on the right of Figure \ref{TypeIIB5dGeneralFamilyI1} describes $(k+2)F+SU(k)^{N-1}+(k+2)F$ gauge theory which corresponds to $R_{k,N}$. This result is valid for arbitrary $N$ and $k$. Therefore we have shown that $R_{N,k}$ and $R_{k,N}$ are UV duals.

As in the previous section, for general $k$ and $N$, there is a third dual frame involving an $Usp$ gauge group or an antisymmetric field. While this dual frame will not play a role in $4d$, let us discuss it for completeness. In order to get this $5d$ UV dual, we assume $N \ge k$ and distinguish between the case $k$ even and $k$ odd (as we will discuss later, also the $6d$ UV completion depends on the parity of this parameter).

\subsection*{$k$ even: $k=2l$}
The $5d$ triality reads
\be \label{5dGeneralFamilyIkEven1} \scalebox{0.9}{\bpic[node distance=2cm,gSUnode/.style={circle,red,draw,minimum size=8mm},gUSpnode/.style={circle,blue,draw,minimum size=8mm},fnode/.style={rectangle,red,draw,minimum size=8mm}]
\node at (-6.5,1) {$\star_1)$};
\node[fnode] (F1) at (-5,0) {$N+2$};
\node[gSUnode] (G1) at (-3,0) {$N$};
\node[gSUnode] (G2) at (-1.5,0) {$N$};
\node (G3) at (0,0) {$\dots$};
\node[gSUnode] (G4) at (1.5,0) {$N$};
\node[fnode] (F2) at (3.5,0) {$N+2$};
\draw (F1) -- (G1) -- (G2) -- (G3) -- (G4) -- (F2);
\draw[decorate,decoration={calligraphic brace,mirror,amplitude=7pt}] (-3.4,-0.5) -- (1.9,-0.5) node[pos=0.5,below=9pt,black] {$2l-1$};
\epic} \ee
\be \label{5dGeneralFamilyIkEven2} \scalebox{0.9}{\bpic[node distance=2cm,gSUnode/.style={circle,red,draw,minimum size=8mm},gUSpnode/.style={circle,blue,draw,minimum size=8mm},fnode/.style={rectangle,red,draw,minimum size=8mm}]
\node at (-6.5,1) {$\star_2)$};
\node[gUSpnode] (G1) at (-5,0) {\scalebox{0.9}{$2N-2l$}};
\node[gSUnode] (G2) at (-2.5,0) {\scalebox{0.8}{$2N-2l+4$}};
\node[gSUnode] (G3) at (0,0) {\scalebox{0.8}{$2N-2l+8$}};
\node (G4) at (2,0) {$\dots$};
\node[gSUnode] (G5) at (4,0) {\scalebox{0.8}{$2N+2l-4$}};
\node[fnode] (F1) at (7,0) {\scalebox{0.9}{$2N+2l+2$}};
\draw (G1) -- (G2) -- (G3) -- (G4) -- (G5) -- (F1);
\draw[decorate,decoration={calligraphic brace,mirror,amplitude=7pt}] (-3.5,-1.2) -- (5,-1.2) node[pos=0.5,below=9pt,black] {$l-1$};
\epic} \ee
\be \label{5dGeneralFamilyIkEven3} \scalebox{0.9}{\bpic[node distance=2cm,gSUnode/.style={circle,red,draw,minimum size=8mm},gUSpnode/.style={circle,blue,draw,minimum size=8mm},fnode/.style={rectangle,red,draw,minimum size=8mm}]
\node at (-6.5,1) {$\star_3)$};
\node[fnode] (F1) at (-5,0) {$2l+2$};
\node[gSUnode] (G1) at (-3,0) {$2l$};
\node[gSUnode] (G2) at (-1.5,0) {$2l$};
\node (G3) at (0,0) {$\dots$};
\node[gSUnode] (G4) at (1.5,0) {$2l$};
\node[fnode] (F2) at (3.5,0) {$2l+2$};
\draw (F1) -- (G1) -- (G2) -- (G3) -- (G4) -- (F2);
\draw[decorate,decoration={calligraphic brace,mirror,amplitude=7pt}] (-3.4,-0.5) -- (1.9,-0.5) node[pos=0.5,below=9pt,black] {$N-1$};
\epic} \ee 
First remark, if we put $l=1$ we recover the triality studied in section \ref{5dDualityI}. The logic to understand why these theories are UV duals is the same as before. We start from the brane system in Figure \ref{TypeIIA6dCompletionGeneralFamilyIkEvenFig} describing a $6d$ theory. Then we compactify this system into an $S^1$ and we perform T-duality along the compactified dimension. The $O8^-$ plane becomes two $O7^-$. Then we have the choice to resolve one or two $O7$'s. If we resolve two, we get the theory $\star_1)$ and if we resolve only one, we get $\star_2)$. Finally, as we have seen, $\star_1)$ and $\star_3)$ are S-dual one to each other. We have been very brief about the derivation because all the details can be found in \cite{Hayashi:2015zka, Zafrir:2015rga}.
\subsection*{$k$ odd: $k=2l+1$}
Similar arguments \cite{Hayashi:2015zka, Zafrir:2015rga} show that the triality reads
\be \label{5dGeneralFamilyIkOdd1} \scalebox{0.9}{\bpic[node distance=2cm,gSUnode/.style={circle,red,draw,minimum size=8mm},gUSpnode/.style={circle,blue,draw,minimum size=8mm},fnode/.style={rectangle,red,draw,minimum size=8mm}] 
\node at (-6.5,1) {$\star_1)$};
\node[fnode] (F1) at (-5,0) {$N+2$};
\node[gSUnode] (G1) at (-3,0) {$N$};
\node[gSUnode] (G2) at (-1.5,0) {$N$};
\node (G3) at (0,0) {$\dots$};
\node[gSUnode] (G4) at (1.5,0) {$N$};
\node[fnode] (F2) at (3.5,0) {$N+2$};
\draw (F1) -- (G1) -- (G2) -- (G3) -- (G4) -- (F2);
\draw[decorate,decoration={calligraphic brace,mirror,amplitude=7pt}] (-3.4,-0.5) -- (1.9,-0.5) node[pos=0.5,below=9pt,black] {$2l$};
\epic} \ee
\be \label{5dGeneralFamilyIkOdd2} \scalebox{0.9}{\bpic[node distance=2cm,gSUnode/.style={circle,red,draw,minimum size=8mm},gUSpnode/.style={circle,blue,draw,minimum size=8mm},fnode/.style={rectangle,red,draw,minimum size=8mm}]
\node at (-6.5,1.8) {$\star_2)$};
\node[gSUnode] (G1) at (-5,0) {\scalebox{0.8}{$2N-2l+1$}};
\node[gSUnode] (G2) at (-2.5,0) {\scalebox{0.8}{$2N-2l+5$}};
\node[gSUnode] (G3) at (0,0) {\scalebox{0.8}{$2N-2l+9$}};
\node (G4) at (2,0) {$\dots$};
\node[gSUnode] (G5) at (4,0) {\scalebox{0.8}{$2N+2l-3$}};
\node[fnode] (F1) at (7,0) {\scalebox{0.9}{$2N+2l+3$}};
\draw (G1) -- (G2) -- (G3) -- (G4) -- (G5) -- (F1);
\draw (-4.5,0.9) to[out=90,in=0]  (-5,1.5) to[out=180,in=90] (-5.5,0.9);
\draw[decorate,decoration={calligraphic brace,mirror,amplitude=7pt}] (-3.5,-1.2) -- (5,-1.2) node[pos=0.5,below=9pt,black] {$l-1$};
\epic} \ee
\be \label{5dGeneralFamilyIkOdd3} \scalebox{0.9}{\bpic[node distance=2cm,gSUnode/.style={circle,red,draw,minimum size=8mm},gUSpnode/.style={circle,blue,draw,minimum size=8mm},fnode/.style={rectangle,red,draw,minimum size=8mm}]
\node at (-6.5,1) {$\star_3)$};
\node[fnode] (F1) at (-5,0) {$2l+3$};
\node[gSUnode] (G1) at (-3,0) {\scalebox{0.9}{$2l+1$}};
\node[gSUnode] (G2) at (-1,0) {\scalebox{0.9}{$2l+1$}};
\node (G3) at (0.5,0) {$\dots$};
\node[gSUnode] (G4) at (2,0) {\scalebox{0.9}{$2l+1$}};
\node[fnode] (F2) at (4,0) {$2l+3$};
\draw (F1) -- (G1) -- (G2) -- (G3) -- (G4) -- (F2);
\draw[decorate,decoration={calligraphic brace,mirror,amplitude=7pt}] (-3.7,-0.8) -- (2.7,-0.8) node[pos=0.5,below=9pt,black] {$N-1$};
\epic} \ee

\subsection{6d UV completion}
The $6d$ UV completion of the theory $R_{N,k}$, $N \ge k$, theory depends on the parity of $k$.
\subsection*{$k$ even: $k=2l$}
The $6d$ completion is given by the following Type IIA brane setup \cite{Hayashi:2015zka,Zafrir:2015rga}: 
\begin{figure}[H]
\centering
\includegraphics[height=0.25\textheight]{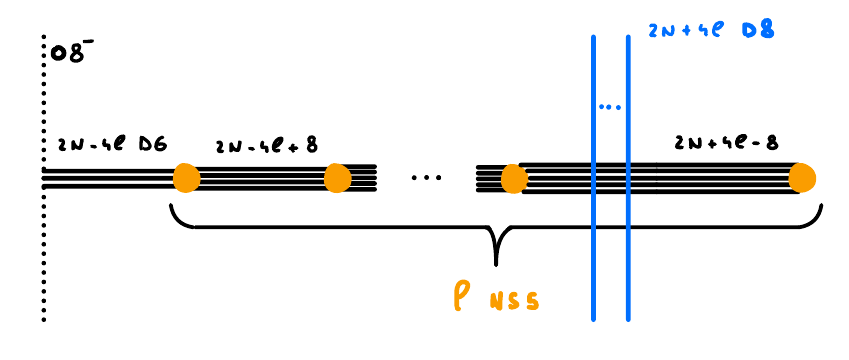}
\caption{Type IIA brane setup corresponding to the $6d$ UV completion of $R_{N,2l}$ theory.}
\label{TypeIIA6dCompletionGeneralFamilyIkEvenFig}
\end{figure}
The gauge theory corresponding to this brane system is a linear quiver with one $USp$ gauge node and $l-1$ $SU$ gauge nodes: 
\be \label{TypeIIA6dCompletionGeneralFamilyIkEven} \scalebox{0.9}{\bpic[node distance=2cm,gSUnode/.style={circle,red,draw,minimum size=8mm},gUSpnode/.style={circle,blue,draw,minimum size=8mm},fnode/.style={rectangle,red,draw,minimum size=8mm}]
\node[gUSpnode] (G1) at (-4,0) {\scalebox{0.9}{$2N-4l$}};
\node[gSUnode] (G2) at (-1,0) {\scalebox{0.8}{$2N-4l+8$}};
\node[gSUnode] (G3) at (2,0) {\scalebox{0.8}{$2N-4l+16$}};
\node (G4) at (4,0) {$\dots$};
\node[gSUnode] (G5) at (6,0) {\scalebox{0.8}{$2N+4l-8$}};
\node[fnode] (F1) at (9,0) {\scalebox{0.8}{$2N+4l$}};
\draw (G1) -- (G2) -- (G3) -- (G4) -- (G5) -- (F1);
\epic} \ee
\subsection*{$k$ odd: $k=2l+1$}
The $6d$ completion is given by the following Type IIA brane setup \cite{Hayashi:2015zka,Zafrir:2015rga}: 
\begin{figure}[H]
\centering
\includegraphics[height=0.25\textheight]{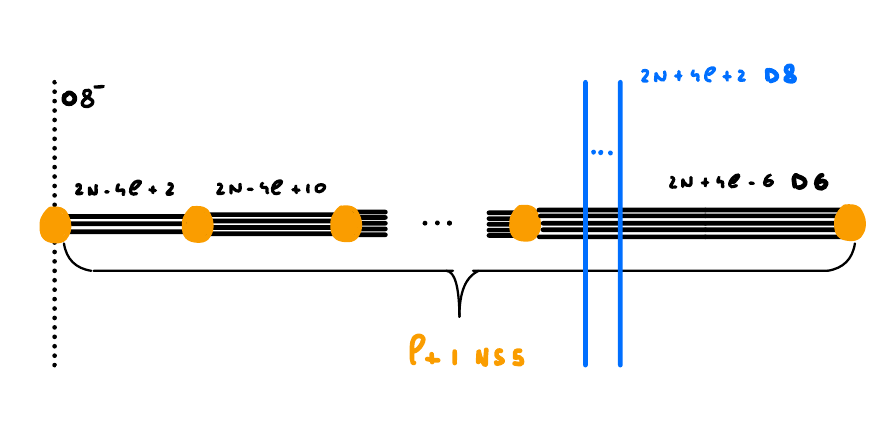}
\caption{Type IIA brane setup corresponding to the $6d$ UV completion of $R_{N,2l+1}$ theory.}
\label{TypeIIA6dCompletionGeneralFamilyIkOddFig}
\end{figure}
The gauge theory corresponding to this brane system is a linear quiver with $l$ $SU$ gauge nodes and an antisymmetric hyper attached to the first node: 
\be \label{TypeIIA6dCompletionGeneralFamilyIkOdd} \scalebox{0.9}{\bpic[node distance=2cm,gSUnode/.style={circle,red,draw,minimum size=8mm},gUSpnode/.style={circle,blue,draw,minimum size=8mm},fnode/.style={rectangle,red,draw,minimum size=8mm}] 
\node[gSUnode] (G1) at (-4,0) {\scalebox{0.8}{$2N-4l+2$}};
\node[gSUnode] (G2) at (-1,0) {\scalebox{0.8}{$2N-4l+10$}};
\node[gSUnode] (G3) at (2,0) {\scalebox{0.8}{$2N-4l+18$}};
\node (G4) at (4,0) {$\dots$};
\node[gSUnode] (G5) at (6,0) {\scalebox{0.8}{$2N+4l-6$}};
\node[fnode] (F1) at (9,0) {\scalebox{0.8}{$2N+4l+2$}};
\draw (G1) -- (G2) -- (G3) -- (G4) -- (G5) -- (F1);
\draw (-3.5,0.9) to[out=90,in=0]  (-4,1.5) to[out=180,in=90] (-4.5,0.9);
\epic} \ee

\subsection{4d duality} \label{4dGeneralFamilyI}
Having recalled the two $5d$ UV trialities \eqref{5dGeneralFamilyIkEven1}-\eqref{5dGeneralFamilyIkEven3} and \eqref{5dGeneralFamilyIkOdd1}-\eqref{5dGeneralFamilyIkOdd3} we can run our prescription of Sec. \ref{Algorithm}. We quickly realize that for generic $l$ the theories $\star_2)$ (\eqref{5dGeneralFamilyIkEven2} and \eqref{5dGeneralFamilyIkOdd2}), that is the ones involving an $Usp$ node or an antisymmetric, cannot be made non-anomalous in $4d$, this is because the ranks of the chain of $SU$ nodes are not constant\footnote{It is an interesting question if the prescription can be generalized to include quivers with non constant ranks for the $SU$ nodes. See Sec. \ref{higgsing} for a first step in this direction.}. Therefore, we do not consider these theories and treat uniformly the case $k$ even and $k$ odd. The proposed $4d$ IR duality that we obtain using our prescription is the following: 
\be \label{4dGeneralFamilyI1} \scalebox{0.9}{\bpic[node distance=2cm,gSUnode/.style={circle,red,draw,minimum size=8mm},gUSpnode/.style={circle,blue,draw,minimum size=8mm},fnode/.style={rectangle,red,draw,minimum size=8mm}]  
\node at (-6.5,1.5) {$\star_1)$};
\node[fnode,orange] (F1) at (-6,0) {$N$};
\node[gSUnode] (G1) at (-4,0) {$N$};
\node (G2) at (-2,0) {$\dots$};
\node[gSUnode] (G3) at (0,0) {$N$};
\node (G4) at (2,0) {$\dots$};
\node[gSUnode] (G5) at (4,0) {$N$};
\node[fnode] (F2) at (6,0) {$N$};
\node[fnode] (F3) at (-3,-1.7) {$2$};
\node[fnode] (F4) at (-1,-1.7) {$2$};
\node[fnode] (F5) at (1,-1.7) {$2$};
\node[fnode] (F6) at (3,-1.7) {$2$};
\node[fnode,orange] (F7) at (-5.5,-1.7) {$2$};
\node[fnode,violet] (F8) at (5.5,-1.7) {$2$};
\draw (F1) -- pic[pos=0.7,sloped]{arrow} (G1) -- pic[pos=0.7,sloped]{arrow} (G2) -- pic[pos=0.7,sloped]{arrow} (G3) -- pic[pos=0.7,sloped]{arrow} (G4) -- pic[pos=0.7,sloped]{arrow} (G5) -- pic[pos=0.7,sloped]{arrow} (F2);
\draw (G1) -- pic[pos=0.5,sloped,very thick]{arrow=latex reversed} (F3);
\draw (G2) -- pic[pos=0.5,sloped,very thick]{arrow=latex reversed} (F3);
\draw (G2) -- pic[pos=0.5,sloped,very thick]{arrow=latex reversed} (F4);
\draw (G3) -- pic[pos=0.5,sloped,very thick]{arrow=latex reversed} (F4);
\draw (G3) -- pic[pos=0.5,sloped,very thick]{arrow=latex reversed} (F5);
\draw (G4) -- pic[pos=0.5,sloped,very thick]{arrow=latex reversed} (F5);
\draw (G4) -- pic[pos=0.5,sloped,very thick]{arrow=latex reversed} (F6);
\draw (G5) -- pic[pos=0.5,sloped,very thick]{arrow=latex reversed} (F6);
\draw (G1) -- pic[pos=0.5,sloped,very thick]{arrow=latex reversed} (F7);
\draw (G5) -- pic[pos=0.5,sloped,very thick]{arrow=latex reversed} (F8);
\draw[decorate,decoration={calligraphic brace,amplitude=7pt}] (-4.4,0.8) -- (4.4,0.8) node[pos=0.5,above=9pt,black] {$k-1$};
\node[right] at (-7.5,-3) {$ \cW= (k-2) \, \triangles + \Flip[L \, l; R \, r; L^N; R^N; L \, B_1 \dots B_{k-2} \,R; l^2 \, B_1^{N-2} \dots B_{k-2}^{N-2} \, r^2] + \displaystyle\sum_{i=1}^{k-2} \Flip[B_i^N]$};
\node at (-5,0.4) {$L$};
\node at (-5,-0.7) {$l$};
\node at (-3,0.4) {$B_1$};
\node at (3,0.4) {$B_{k-2}$};
\node at (5,0.4) {$R$};
\node at (5,-0.7) {$r$};
\node at (-3.8,-1) {$V_1$};
\node at (-2.8,-0.8) {$D_1$};
\node at (-1.8,-1) {$V_2$};
\node at (4,-1) {$D_{k-2}$};
\epic} \ee  
\be \label{4dGeneralFamilyI2} \scalebox{0.9}{\bpic[node distance=2cm,gSUnode/.style={circle,red,draw,minimum size=8mm},gUSpnode/.style={circle,blue,draw,minimum size=8mm},fnode/.style={rectangle,red,draw,minimum size=8mm}]  
\node at (-9.5,0) {$\llra$}; 
\node at (-8,1.5) {$\star_3)$};
\node[fnode,orange] (F1) at (-7,0) {$k$};
\node[gSUnode] (G1) at (-5,0) {$k$};
\node (G2) at (-3,0) {$\dots$};
\node[gSUnode] (G3) at (-1,0) {$k$};
\node (G4) at (1,0) {$\dots$};
\node[gSUnode] (G5) at (3,0) {$k$};
\node[fnode] (F2) at (5,0) {$k$};
\node[fnode] (F3) at (-4,-1.7) {$2$};
\node[fnode] (F4) at (-2,-1.7) {$2$};
\node[fnode] (F5) at (0,-1.7) {$2$};
\node[fnode] (F6) at (2,-1.7) {$2$};
\node[fnode,orange] (F7) at (-6.3,-1.7) {$2$};
\node[fnode,violet] (F8) at (4.3,-1.7) {$2$};
\draw (F1) -- pic[pos=0.7,sloped]{arrow} (G1) -- pic[pos=0.7,sloped]{arrow} (G2) -- pic[pos=0.7,sloped]{arrow} (G3) -- pic[pos=0.7,sloped]{arrow} (G4) -- pic[pos=0.7,sloped]{arrow} (G5) -- pic[pos=0.7,sloped]{arrow} (F2);
\draw (G1) -- pic[pos=0.5,sloped,very thick]{arrow=latex reversed} (F3);
\draw (G2) -- pic[pos=0.5,sloped,very thick]{arrow=latex reversed} (F3);
\draw (G2) -- pic[pos=0.5,sloped,very thick]{arrow=latex reversed} (F4);
\draw (G3) -- pic[pos=0.5,sloped,very thick]{arrow=latex reversed} (F4);
\draw (G3) -- pic[pos=0.5,sloped,very thick]{arrow=latex reversed} (F5);
\draw (G4) -- pic[pos=0.5,sloped,very thick]{arrow=latex reversed} (F5);
\draw (G4) -- pic[pos=0.5,sloped,very thick]{arrow=latex reversed} (F6);
\draw (G5) -- pic[pos=0.5,sloped,very thick]{arrow=latex reversed} (F6);
\draw (G1) -- pic[pos=0.5,sloped,very thick]{arrow=latex reversed} (F7);
\draw (G5) -- pic[pos=0.5,sloped,very thick]{arrow=latex reversed} (F8);
\draw[decorate,decoration={calligraphic brace,amplitude=7pt}] (-5.4,0.8) -- (3.4,0.8) node[pos=0.5,above=9pt,black] {$N-1$};
\node[right] at (-10.5,-3) {$ \cW= (N-2) \, \triangles + \Flip[\Lt \, \lt; \Rt \, \rt; \Lt^k; \Rt^k; \Lt \, \Bt_1 \dots \Bt_{N-2} \,\Rt; \lt^2 \, \Bt_1^{k-2} \dots \Bt_{N-2}^{k-2} \, \rt^2] + \displaystyle\sum_{i=1}^{N-2} \Flip[\Bt_i^k]$};
\epic} \ee
We have denoted the fields appearing in \eqref{4dGeneralFamilyI2} with a tilde. We remark that in order to get a non-anomalous $4d$ quiver we have to split the flavor symmetries. For example, $SU(N+2)$ is split into $SU(2)$ and $SU(N)$. The expression of the superpotential in \eqref{4dGeneralFamilyI1} and \eqref{4dGeneralFamilyI2} will be justified in the next section. The mapping of the chiral ring generators is
\vspace{-2.5cm}
\be \label{mappingGeneralFamilyI1} 
\scalebox{0.9}{$
\ba[t]{c}\star_1) \\
\\
\\[6pt]
\begin{cases}
\Flipper[L B_1 \dots B_{k-2} R] \\
L^{N-2} \, B_{1}^{N-2} \dots B_{k-2}^{N-2} \, r^2 \\
l^2 \, B_{1}^{N-2} \dots B_{k-2}^{N-2} \, R^{N-2}
\end{cases}
\ea
\ba{c} \\
\\
\\
\\
\\
\\
\\
\\
\\
\\[-10pt]
\Longleftrightarrow
\ea
\ba[t]{c}\star_3) \\
\begin{cases}
\Dt_i \, \Bt_{i+1}^{k-1} \dots \Bt_{j}^{k-1} \, \Vt_{j+1} \quad i=1, \dots, N-4 \, \& \, j=i+1, \dots N-3 \\
\lt \, \Bt_{1}^{k-1} \dots \Bt_i^{k-1} \, \Vt_{i+1} \quad i=1, \dots, N-3 \\
\Dt_{j+1} \, \Bt_{i+2}^{k-1} \dots \Bt_{N-2}^{k-1} \, \rt \quad j=0, \dots, N-4 \\
\lt \Vt_1; \, \Dt_1 \Vt_2; \, \Dt_2 \Vt_3; \dots; \, \Dt_{N-3} \Vt_{N-2}; \, \Dt_{N-2} \rt \\
\Flipper[\Bt_i^k] \quad i=1, \dots, N-2 \\
\lt \, \Bt_{1}^{k-1} \dots \Bt_{N-2}^{k-1} \, \rt \\
\Flipper[\Lt^k]; \, \Flipper[\Rt^k]
\end{cases}
\ea
$}
\ee
The total number of operators is $2N^2-N$ on both sides.
\vspace{-1.5cm}
\be \label{mappingGeneralFamilyI2} 
\scalebox{0.9}{$
\ba[t]{c}\\
\begin{cases}
D_i \, B_{i+1}^{N-1} \dots B_{j}^{N-1} \, V_{j+1} \quad i=1, \dots, k-4 \, \& \, j=i+1, \dots k-3 \\
l \, B_{1}^{N-1} \dots B_i^{N-1} \, V_{i+1} \quad i=1, \dots, k-3 \\
D_{j+1} \, B_{i+2}^{N-1} \dots B_{k-2}^{N-1} \, r \quad j=0, \dots, k-4 \\
l V_1; \, D_1 V_2; \, D_2 V_3; \dots; \, D_{k-3} V_{k-2}; \, D_{k-2} r \\
\Flipper[B_i^N] \quad i=1, \dots, k-2 \\
l \, B_{1}^{N-1} \dots B_{N-2}^{N-1} \, r \\
\Flipper[L^N]; \, \Flipper[R^N]
\end{cases}
\ea
\ba{c} \\
\\
\\
\\
\\
\\
\\
\\
\\
\\[-10pt]
\Longleftrightarrow
\ea
\ba[t]{c}\\
\\
\\[6pt]
\begin{cases}
\Flipper[\Lt \Bt_1 \dots \Bt_{N-2} \Rt] \\
\Lt^{k-2} \, \Bt_{1}^{k-2} \dots \Bt_{N-2}^{k-2} \, \rt^2 \\
\lt^2 \, \Bt_{1}^{k-2} \dots \Bt_{N-2}^{k-2} \, \Rt^{k-2}
\end{cases}
\ea
$}
\ee
The total number of operators is $2k^2-k$ on both sides.
\vspace{-2cm}
\be \label{mappingGeneralFamilyI3} 
\scalebox{0.9}{$
\ba[t]{c} \\
\begin{cases}
\Flipper[L \, l]; \, \Flipper[R \, r] \\
L^{N-1} \, B_1^{N-1} \dots B_{i}^{N-1} \, V_{i+1} \quad i=0, \dots, k-3 \\
D_{k-2-i} \, B_{k-2-i+1}^{N-1} \dots B_{k-2}^{N-1} \, R^{N-1} \quad i=0, \dots, k-3 \\
L^{N-1} \, B_1^{N-1} \dots B_{k-2}^{N-1} \, r \\
l \, B_{1}^{N-1} \dots B_{k-2}^{N-1} \, R^{N-1}
\end{cases}
\ea
\ba{c} \\
\\
\\
\\
\\
\\
\\
\Longleftrightarrow
\ea
\ba[t]{c} \\
\begin{cases}
\Flipper[\Lt \, \lt]; \, \Flipper[\Rt \, \rt] \\
\Lt^{k-1} \, \Bt_1^{k-1} \dots \Bt_{i}^{k-1} \, \Vt_{i+1} \quad i=0, \dots, N-3 \\
\Dt_{N-2-i} \, \Bt_{N-2-i+1}^{k-1} \dots \Bt_{N-2}^{k-1} \, \Rt^{k-1} \quad i=0, \dots, N-3 \\
\Lt^{k-1} \, \Bt_1^{k-1} \dots \Bt_{N-2}^{k-1} \, \rt \\
\lt \, \Bt_{1}^{k-1} \dots \Bt_{N-2}^{k-1} \, \Rt^{k-1}
\end{cases}
\ea
$}
\ee
The total number of operators is $4kN$ on both sides.
\vspace{-1cm}
\be \nn
\scalebox{0.9}{$
\ba[t]{c} \\
\Flipper[l^2 \, B_1^{N-1} \dots B_{k-2}^{N-1} \, r^2] 
\ea
\ba{c} \\
\\
\hspace{2cm}\Longleftrightarrow
\ea
\ba[t]{c} \\
\hspace{2cm} \Flipper[\lt^2 \, \Bt_1^{k-1} \dots \Bt_{N-2}^{k-1} \, \rt^2]
\ea
$}
\ee
\be 
\hspace{-7.7cm} \text{The total number of operators is $1$ on both sides.} \label{mappingGeneralFamilyI4}
\ee

\noindent The way to read this mapping is the same as in \eqref{mappingFamilyI}. In the IR there is an enhancement of the global symmetry. The claim is that all the operators inside a bracket will combine, in the IR, into an operator transforming in a specific representation of the emergent global symmetry. The justification on the mapping will be clearer with the proof of the duality.

\subsection{Proof of the 4d duality} \label{Proof4dGeneralFamilyI}
Start with $\star_1)$ and do the following operations:
\begin{itemize}
\item k-1 Seiberg dualities on the SU nodes from left to right. This step transforms all the $SU(N)$ nodes into $SU(2)$ and the flavor \textcolor{orange}{$SU(N)$} is moved to the right. We get \eqref{Proof4dGeneralFamilyI1}.
\item CSST duality \eqref{CSST} on the left $SU(2)$ that will give a mass to the adjacent vertical field as in \eqref{Proof4dFamilyI1}, we obtain \eqref{Proof4dGeneralFamilyI2}.
\item $k-3$ confinements \eqref{ConfinementSU}. We end up with \eqref{Proof4dGeneralFamilyI3}.
\end{itemize}
In terms of the quiver, we get the following sequence
\be \label{Proof4dGeneralFamilyI1} \scalebox{0.9}{\bpic[node distance=2cm,gSUnode/.style={circle,red,draw,minimum size=8mm},gUSpnode/.style={circle,blue,draw,minimum size=8mm},fnode/.style={rectangle,red,draw,minimum size=8mm}]
\node at (-6.5,1.5) {$\star_1)$};
\node[gSUnode] (G1) at (-4,0) {$2$};
\node[gSUnode] (G2) at (-2,0) {$2$};
\node[gSUnode] (G3) at (0,0) {$2$};
\node (G4) at (2,0) {$\dots$};
\node[gSUnode] (G5) at (4,0) {$2$};
\node[gSUnode] (G6) at (6,0) {$2$};
\node[fnode,orange] (F1) at (8,1) {$N$};
\node[fnode] (F2) at (8,-1) {$N$};
\node[fnode] (F3) at (-4,-1.7) {$2$};
\node[fnode] (F4) at (-2,-1.7) {$2$};
\node[fnode] (F5) at (0,-1.7) {$2$};
\node[fnode] (F6) at (4,-1.7) {$2$};
\node[fnode,orange] (F7) at (-5.5,-1.7) {$2$};
\node[fnode,violet] (F8) at (6,-1.7) {$2$};
\draw (G1) -- (G2) -- (G3) -- (G4) -- (G5) -- (G6) -- (F1);
\draw (G1) -- (F7);
\draw (G6) -- (F2);
\draw (G1) -- (F3);
\draw (G1) -- (F4);
\draw (G2) -- (F4);
\draw (G2) -- (F5);
\draw (G3) -- (F5);
\draw (G3) -- (1.2,-1);
\draw (G4) -- (F6);
\draw (G5) -- (F6);
\draw (G5) -- (F8);
\draw (G6) -- (F8);
\draw[decorate,decoration={calligraphic brace,amplitude=7pt}] (-4.4,0.8) -- (6.4,0.8) node[pos=0.5,above=9pt,black] {$k-1$};
\epic} \ee 

\be \label{Proof4dGeneralFamilyI2} \scalebox{0.9}{\bpic[node distance=2cm,gSUnode/.style={circle,red,draw,minimum size=8mm},gUSpnode/.style={circle,blue,draw,minimum size=8mm},fnode/.style={rectangle,red,draw,minimum size=8mm}]
\node at (-6.5,1.5) {$\star_1)$};
\node[gSUnode] (G1) at (-4,0) {$2$};
\node[gSUnode] (G2) at (-2,0) {$2$};
\node[gSUnode] (G3) at (0,0) {$2$};
\node (G4) at (2,0) {$\dots$};
\node[gSUnode] (G5) at (4,0) {$2$};
\node[gSUnode] (G6) at (6,0) {$2$};
\node[fnode,orange] (F1) at (8,1) {$N$};
\node[fnode] (F2) at (8,-1) {$N$};
\node[fnode] (F3) at (-4,-1.7) {$2$};
\node[fnode] (F4) at (-2,-1.7) {$2$};
\node[fnode] (F5) at (0,-1.7) {$2$};
\node[fnode] (F6) at (4,-1.7) {$2$};
\node[fnode,orange] (F7) at (-5.5,-1.7) {$2$};
\node[fnode,violet] (F8) at (6,-1.7) {$2$};
\draw (G1) -- (G2) -- (G3) -- (G4) -- (G5) -- (G6) -- (F1);
\draw (G1) -- (F7);
\draw (G6) -- (F2);
\draw (G1) -- (F3);
\draw (G1) -- (F4);
\draw (G2) -- (F5);
\draw (G3) -- (F5);
\draw (G3) -- (1.2,-1);
\draw (G4) -- (F6);
\draw (G5) -- (F6);
\draw (G5) -- (F8);
\draw (G6) -- (F8);
\draw[decorate,decoration={calligraphic brace,amplitude=7pt}] (-4.4,0.8) -- (6.4,0.8) node[pos=0.5,above=9pt,black] {$k-1$};
\epic} \ee 

\be \label{Proof4dGeneralFamilyI3} \scalebox{0.9}{\bpic[node distance=2cm,gSUnode/.style={circle,red,draw,minimum size=8mm},gUSpnode/.style={circle,blue,draw,minimum size=8mm},fnode/.style={rectangle,red,draw,minimum size=8mm}]
\begin{scope}[shift={(-3,0)}]     
\node at (-5.8,2.2) {$\star_1)$};
\node[gSUnode] (G1) at (-2.5,0) {$2$};
\node[gSUnode] (G2) at (0,0) {$2$};
\node[fnode,orange] (F1) at (1.8,1.4) {$N$};
\node[fnode] (F2) at (1.8,-1.4) {$N$};
\node[fnode] (F3) at (-4.2,1.2) {$2$};
\node[fnode] (F4) at (-4.2,-1.2) {$2$};
\node[fnode,orange] (F7) at (-1,1.4) {$2$};
\node[fnode,violet] (F8) at (-1,-1.4) {$2$};
\draw (G1) -- (G2);
\draw (G2) -- (F1);
\draw (G2) -- (F2); 
\draw (G1) -- (F7);
\draw (G1) -- (F8);
\draw (G1) -- (F3.east);
\draw (G1) -- (F4.east);
\node at (-4.2,0.1) {\scalebox{1.5}{$\vdots$}};
\node[right] at (-2,-2.5) {$ \cW= 0$};
\draw[decorate,decoration={brace,amplitude=7pt,mirror}] (-5,1.6) -- (-5,-1.6) node[pos=0.5,left=9pt,black] {$k-2$};
\end{scope}
\node at (0,0) {$\equiv$};
\begin{scope}[shift={(4,0)}] 
\node[gSUnode] (G1) at (-2.5,0) {$2$};
\node[gSUnode] (G2) at (0,0) {$2$};
\node[fnode] (F1) at (0,-1.5) {$2N$};
\node[fnode] (F2) at (-2.5,-1.5) {$2k$};
\draw (G1) -- (G2);
\draw (G2) -- (F1);
\draw (G1) -- (F2);
\node[right] at (-2,-2.5) {$ \cW= 0$};
\node at (-2.8,-0.8) {$k$};
\node at (-1.2,0.3) {$b$};
\node at (0.3,-0.8) {$n$};
\end{scope}
\epic} \ee
Once again following the mapping of the basic dualities we can see that the operators in the LHS of \eqref{mappingGeneralFamilyI1}-\eqref{mappingGeneralFamilyI4} are mapped in the frame \eqref{Proof4dGeneralFamilyI3} in the following way
\vspace{-1cm}
\be  
\scalebox{0.9}{$
\ba[t]{c}\star_1) \\
\begin{cases}
\Flipper[L B_1 \dots B_{k-2} R] \\
L^{N-2} \, B_{1}^{N-2} \dots B_{k-2}^{N-2} \, r^2 \\
l^2 \, B_{1}^{N-2} \dots B_{k-2}^{N-2} \, R^{N-2}
\end{cases}
\ea
\ba{c} \\
\\
\\
\\
\\[-6pt]
\Longleftrightarrow
\ea
\ba[t]{c}\eqref{Proof4dGeneralFamilyI3} \\
\\
\\[-10pt]
\quad nn
\ea
$}
\ee
The total number of operators on both sides is $2N^2-N$.
\vspace{-3cm}
\be  
\scalebox{0.9}{$
\ba[t]{c}\\
\begin{cases}
D_i \, B_{i+1}^{N-1} \dots B_{j}^{N-1} \, V_{j+1} \quad i=1, \dots, k-4 \, \& \, j=i+1, \dots k-3 \\
l \, B_{1}^{N-1} \dots B_i^{N-1} \, V_{i+1} \quad i=1, \dots, k-3 \\
D_{j+1} \, B_{i+2}^{N-1} \dots B_{k-2}^{N-1} \, r \quad j=0, \dots, k-4 \\
l V_1; \, D_1 V_2; \, D_2 V_3; \dots; \, D_{k-3} V_{k-2}; \, D_{k-2} r \\
\Flipper[B_i^N] \quad i=1, \dots, k-2 \\
l \, B_{1}^{N-1} \dots B_{N-2}^{N-1} \, r \\
\Flipper[L^N]; \, \Flipper[R^N]
\end{cases}
\ea
\ba{c} \\
\\
\\
\\
\\
\\
\\
\\
\\
\\[-10pt]
\Longleftrightarrow
\ea
\ba[t]{c}\\
\\
\\
\\
\\[-4pt]
\quad kk
\ea
$}
\ee
The total number of operators on both sides is $2k^2-k$.
\vspace{-2.5cm}
\be  
\scalebox{0.9}{$
\ba[t]{c} \\
\begin{cases}
\Flipper[L \, l]; \, \Flipper[R \, r] \\
L^{N-1} \, B_1^{N-1} \dots B_{i}^{N-1} \, V_{i+1} \quad i=0, \dots, k-3 \\
D_{k-2-i} \, B_{k-2-i+1}^{N-1} \dots B_{k-2}^{N-1} \, R^{N-1} \quad i=0, \dots, k-3 \\
L^{N-1} \, B_1^{N-1} \dots B_{k-2}^{N-1} \, r \\
l \, B_{1}^{N-1} \dots B_{k-2}^{N-1} \, R^{N-1}
\end{cases}
\ea
\ba{c} \\
\\
\\
\\
\\
\\
\\
\Longleftrightarrow
\ea
\ba[t]{c} \\
\\
\\[10pt]
\quad kbn
\ea
$}
\ee
The total number of operators on both sides is $4kN$.
\vspace{-1cm}
\be \nn
\scalebox{0.9}{$
\ba[t]{c} \\
\Flipper[l^2 \, B_1^{N-1} \dots B_{k-2}^{N-1} \, r^2] 
\ea
\ba{c} \\
\\
\hspace{2cm}\Longleftrightarrow
\ea
\ba[t]{c} \\
\quad b b
\ea
$}
\ee
\be 
\hspace{-7.7cm} \text{The total number of operators on both sides is $1$.} 
\ee
Since in \eqref{Proof4dGeneralFamilyI3} we reach a frame where $N$ and $k$ enter symmetrically, it proves the duality $T_{N,k} \leftrightarrow T_{k,N}$ in $4d$, that is \eqref{4dGeneralFamilyI1} $\leftrightarrow$ \eqref{4dGeneralFamilyI2} and the mapping \eqref{mappingGeneralFamilyI1}-\eqref{mappingGeneralFamilyI4}.

\section{Systems with two $O7$ planes, a simple class: $A_{n,1}$}\label{fam2}
\subsection{5d duality}
In this section and the next one, we want now to test our prescription with another family of theories. We consider theories which involve two $O7^-$ planes in the Type IIB brane setup. For each $O7$, the $5d$ quivers contain either an $SU$ gauge group with antisymmetric or an $Usp$ gauge node, depending on whether a $NS5$ is stuck at the orientifold plane or not. We are going to see that also in this case our prescription works and leads to $4d$ dualities. Contrary to the previous family, we are able to prove the $4d$ dualities using basic Seiberg dualities. The $4d$ dualities that we obtain are more complicated, but are still a rather non-trivial check of the $5d$-to-$4d$ prescription.

Concretely, in this section we study the following $5d$ KK theory, that we call $A_{n,1}$.
\be \scalebox{0.9}{\bpic[node distance=2cm,gSUnode/.style={circle,red,draw,minimum size=8mm},gUSpnode/.style={circle,blue,draw,minimum size=8mm},fnode/.style={rectangle,red,draw,minimum size=8mm}]   
\begin{scope}[shift={(0,0)}]
\node[gSUnode] (G1) at (0,0) {$n$};
\node[fnode] (F1) at (0,-1.5) {$8$};
\draw (G1) -- (F1);
\draw (-0.4,0.2) to[out=135,in=-90] (-0.6,0.6) to[out=90,in=180] (-0.4,0.9) to[out=0,in=90] (-0.1,0.4);  
\draw (0.4,0.2) to[out=45,in=-90] (0.6,0.6) to[out=90,in=0] (0.4,0.9) to[out=180,in=90] (0.1,0.4);
\end{scope}
\epic} \ee
$A_{n,1}$ has a $5d$ dual theory. The form of the dual depends on the parity of $n$.
\subsection*{$n$ odd: 5d duality $A_{2N+1,1} \lra U_{2N+1,1}$}
\begin{figure}[H]
\begin{minipage}{0.5\textwidth}
\centering
\includegraphics[height=0.25\textheight]{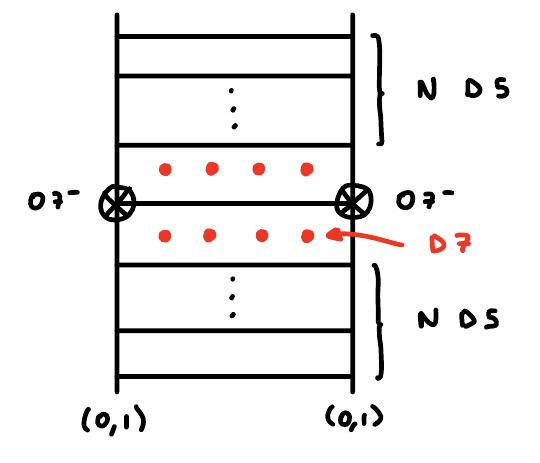}
\end{minipage}%
\begin{minipage}{0.5\textwidth}
\centering
\includegraphics[height=0.25\textheight]{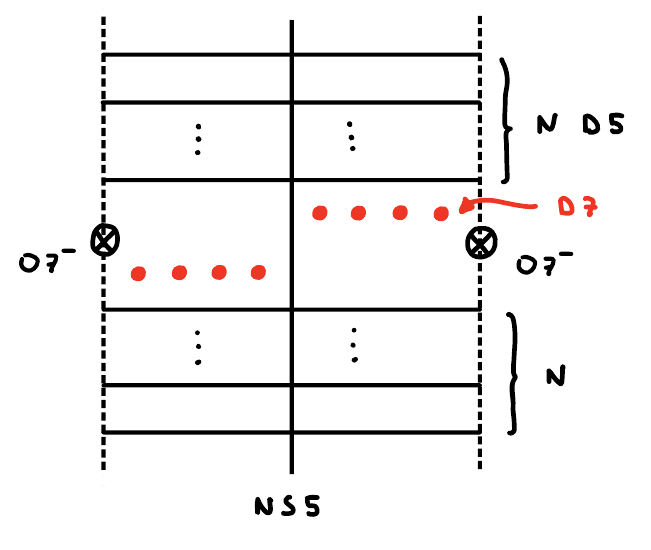}
\end{minipage}
\caption{Brane setup for $2A+SU(2N+1)+8F$ on the left with an $NS5$ stuck on each $O7^-$ plane and for $4F+USp(2N)^{2}+4F$ on the right.}
\label{TypeIIB5dFamilyIIOdd1}
\end{figure}

We call $U_{2N+1,1}$ the dual of $A_{2N+1,1}$, the statement is that the following two theories are UV dual
\be \label{UVdualitiesFamilyIIOdd} \scalebox{0.9}{\bpic[node distance=2cm,gSUnode/.style={circle,red,draw,minimum size=8mm},gUSpnode/.style={circle,blue,draw,minimum size=8mm},fnode/.style={rectangle,red,draw,minimum size=8mm}]    
\begin{scope}[shift={(0,0)}]
\node at (-2.2,1.6) {$\star_1)$};
\node[gSUnode] (G1) at (0,0) {$2N+1$};
\node[fnode] (F1) at (0,-2) {$8$};
\draw (G1) -- (F1);
\draw (-0.8,0.3) to[out=135,in=-90] (-1,0.8) to[out=90,in=180] (-0.7,1.3) to[out=0,in=90] (-0.2,0.8);  
\draw (0.8,0.3) to[out=45,in=-90] (1,0.8) to[out=90,in=0] (0.7,1.3) to[out=180,in=90] (0.2,0.8);
\end{scope}
\begin{scope}[shift={(5,0)}]
\node at (-0.7,1.6) {$\star_2)$};
\node[fnode] (F3) at (0,0) {$4$};
\node[gUSpnode] (G3) at (1.5,0) {$2N$};
\node[gUSpnode] (G4) at (3.2,0) {$2N$};
\node[fnode] (F4) at (4.7,0) {$4$};
\draw (F3) -- (G3) -- (G4) -- (F4);
\end{scope}
\epic} \ee
The analysis showing the UV duality \eqref{UVdualitiesFamilyIIOdd} is morally the same as in the previous family. We have to start with the $6d$ type IIA brane setup shown in Figure \ref{TypeIIA6dCompletionFamilyIIOddFig}, do the circle reduction, T-duality and the resolution of the $O7$-planes. Then, we have a choice on how to resolve the $O7$'s. Depending on this choice, we get two different Type IIB brane setups, see Figure \ref{TypeIIB5dFamilyIIOdd1}, which justifies the duality \eqref{UVdualitiesFamilyIIOdd}. The details can be found in \cite{Hayashi:2015zka} and will not be reproduced here. 

\subsection*{$n$ even: 5d duality $A_{2N,1} \lra U_{2N,1}$}
We call $U_{2N,1}$ the dual of $A_{2N,1}$. This duality appears in \cite{Zafrir:2018hkr} and corresponds to
\be \label{UVdualitiesFamilyIIEven} \scalebox{0.9}{\bpic[node distance=2cm,gSUnode/.style={circle,red,draw,minimum size=8mm},gUSpnode/.style={circle,blue,draw,minimum size=8mm},fnode/.style={rectangle,red,draw,minimum size=8mm}]    
\begin{scope}[shift={(0,0)}]
\node at (-2.2,1.4) {$\star_1)$};
\node[gSUnode] (G1) at (0,0) {$2N$};
\node[fnode] (F1) at (0,-2) {$8$};
\draw (G1) -- (F1);
\draw (-0.5,0.2) to[out=135,in=-90] (-0.7,0.6) to[out=90,in=180] (-0.5,0.9) to[out=0,in=90] (-0.2,0.5);  
\draw (0.5,0.2) to[out=45,in=-90] (0.7,0.6) to[out=90,in=0] (0.5,0.9) to[out=180,in=90] (0.2,0.5);
\end{scope}
\begin{scope}[shift={(5,0)}]
\node at (-0.7,1.4) {$\star_2)$};
\node[fnode] (F3) at (0,0) {$6$};
\node[gUSpnode] (G3) at (1.5,0) {$2N$};
\node[gUSpnode] (G4) at (3.2,0) {\scalebox{0.73}{$2N-2$}};
\node[fnode] (F4) at (4.8,0) {$2$};
\draw (F3) -- (G3) -- (G4) -- (F4);
\end{scope}
\epic} \ee
\begin{figure}[H]
\begin{minipage}{0.5\textwidth}
\centering
\includegraphics[height=0.25\textheight]{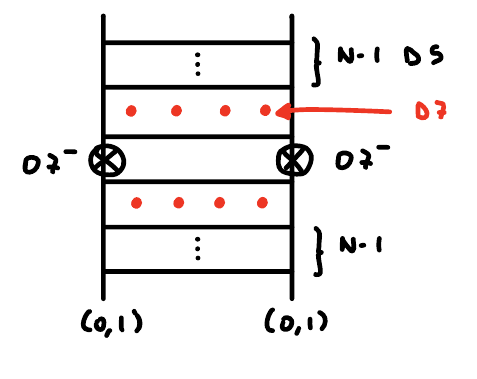}
\end{minipage}%
\begin{minipage}{0.5\textwidth}
\centering
\includegraphics[height=0.25\textheight]{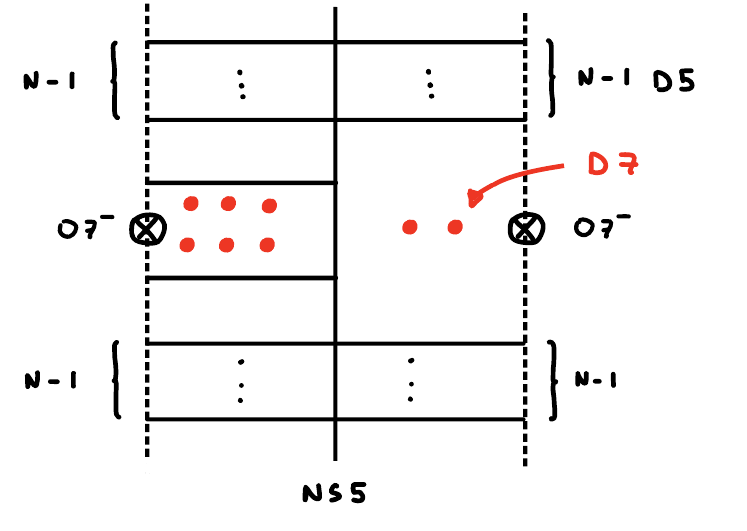}
\end{minipage}
\caption{Brane setup for $2A+SU(2N)+8F$ on the left with an $NS5$ stuck on each $O7^-$ plane and for $4F+USp(2N)-Usp(2N-2)+2F$ on the right.}
\label{TypeIIB5dFamilyIIEven1}
\end{figure}

\subsection{6d UV completion}
\subsection*{$n$ odd: $n=2N+1$}
The UV completion of the $5d$ theories in \eqref{UVdualitiesFamilyIIOdd} is a $6d$ given by the following Type IIA brane setup \cite{Hayashi:2015zka,Zafrir:2015rga}: 
\begin{figure}[H]
\centering
\includegraphics[height=0.25\textheight]{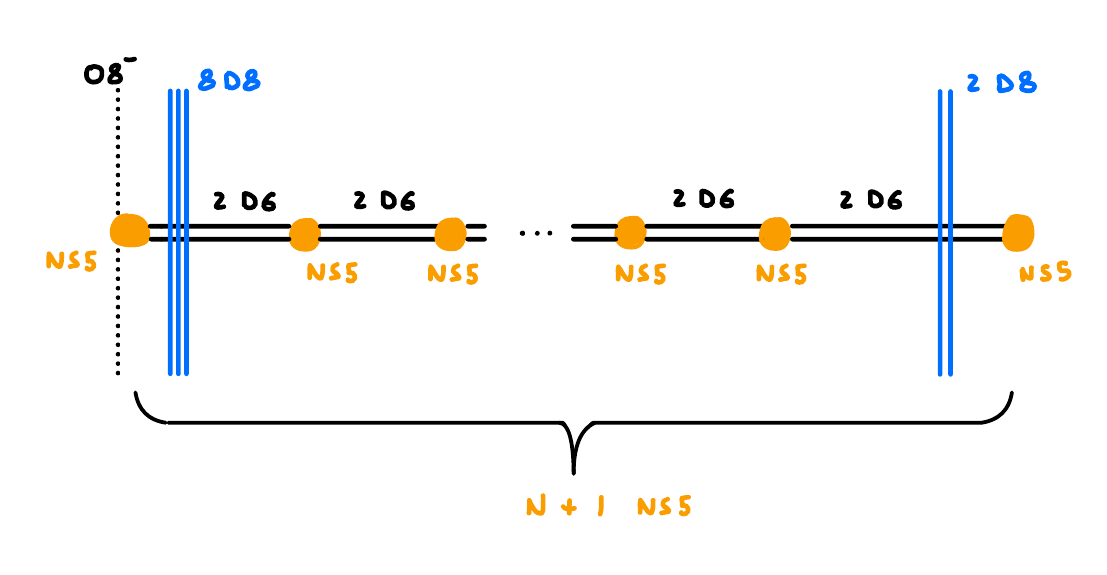}
\caption{Type IIA brane setup corresponding to the $6d$ UV completion of $A_{2N+1,1}$.}
\label{TypeIIA6dCompletionFamilyIIOddFig}
\end{figure}
On the tensor branch, the theory is given by the following gauge theory:

\be \label{TypeIIA6dCompletionFamilyIIOdd} \scalebox{0.9}{\bpic[node distance=2cm,gSUnode/.style={circle,red,draw,minimum size=8mm},gUSpnode/.style={circle,blue,draw,minimum size=8mm},fnode/.style={rectangle,red,draw,minimum size=8mm}]    
\node[fnode] (F1) at (-4.5,0) {$8$};
\node[gSUnode] (G1) at (-3,0) {$2$};
\node[gSUnode] (G2) at (-1.5,0) {$2$};
\node (G3) at (0,0) {$\dots$};
\node[gSUnode] (G4) at (1.5,0) {$2$};
\node[fnode] (F2) at (3,0) {$2$};
\draw (F1) -- (G1) -- (G2) -- (G3) -- (G4) -- (F2);
\draw[decorate,decoration={calligraphic brace,mirror,amplitude=7pt}] (-3.4,-0.5) -- (1.9,-0.5) node[pos=0.5,below=9pt,black] {$N$};
\epic} \ee

\subsection*{$n$ even: $n=2N$}
The UV completion of the $5d$ theories in \eqref{UVdualitiesFamilyIIEven} is a $6d$ given by the following Type IIA brane setup \cite{Hayashi:2015zka,Zafrir:2015rga}: 
\begin{figure}[H]
\centering
\includegraphics[height=0.25\textheight]{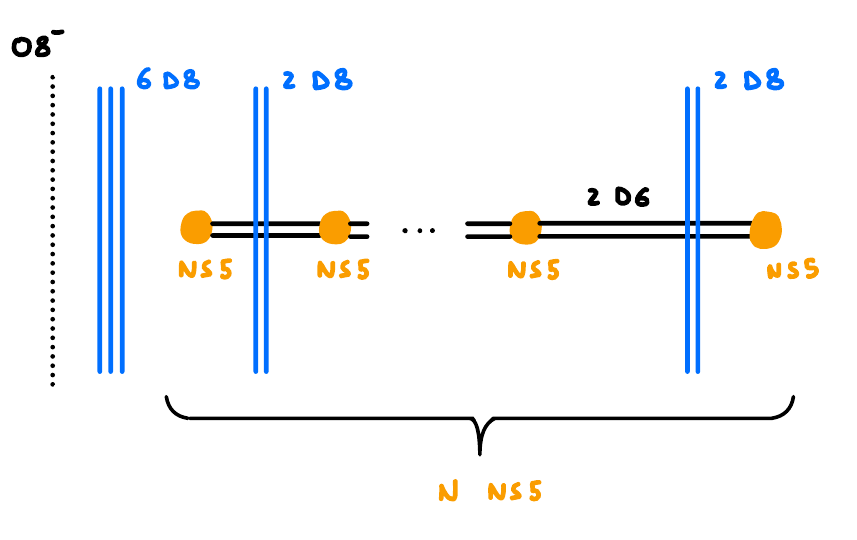}
\caption{Type IIA brane setup corresponding to the $6d$ UV completion of $A_{2N,1}$.}
\label{TypeIIA6dCompletionFamilyIIEvenFig}
\end{figure}
On the tensor branch, the theory is given by the following gauge theory:

\be \label{TypeIIA6dCompletionFamilyIIEven} \scalebox{0.9}{\bpic[node distance=2cm,gSUnode/.style={circle,red,draw,minimum size=8mm},gUSpnode/.style={circle,blue,draw,minimum size=8mm},fnode/.style={rectangle,red,draw,minimum size=8mm}] 
\node[fnode] (F1) at (-4.5,0) {$6$};
\node[gUSpnode] (G1) at (-3,0) {$0$};
\node[gSUnode] (G2) at (-1.5,0) {$2$};
\node[gSUnode] (G3) at (0,0) {$2$};
\node (G4) at (1.5,0) {$\dots$};
\node[gSUnode] (G5) at (3,0) {$2$};
\node[fnode] (F2) at (-1.5,1.5) {$2$};
\node[fnode] (F3) at (4.5,0) {$2$};
\draw (F1) -- (G1) -- (G2) -- (G3) -- (G4) -- (G5) -- (F3);
\draw (G2) -- (F2);
\draw[decorate,decoration={calligraphic brace,mirror,amplitude=7pt}] (-3.4,-0.5) -- (3.4,-0.5) node[pos=0.5,below=9pt,black] {$N$};
\epic} \ee

\subsection{4d duality}
\subsection*{$n$ odd: $n = 2N+1$}
Applying our prescription of Sec. \ref{Algorithm} to the KK duality \eqref{UVdualitiesFamilyIIOdd} leads to the following $4d$ theories that we claim are IR dual
\be \label{4dFamilyIIOdd} \scalebox{0.9}{\bpic[node distance=2cm,gSUnode/.style={circle,red,draw,minimum size=8mm},gUSpnode/.style={circle,blue,draw,minimum size=8mm},fnode/.style={rectangle,red,draw,minimum size=8mm}]     
\begin{scope}[shift={(0,0)}]
\node at (-2.2,1.6) {$\star_1)$};
\node[gSUnode] (G1) at (0,0) {$2N+1$};
\node[fnode] (F1) at (1.3,-2.3) {$4$};
\node[fnode] (F2) at (-1.3,-2.3) {$4$};
\draw (G1) -- pic[pos=0.6,sloped]{arrow} (F1);
\draw (G1) -- pic[pos=0.4,sloped]{arrow} (F2);
\draw (-0.8,0.3) to[out=135,in=-90] pic[pos=0.1,sloped]{arrow} (-1,0.8) to[out=90,in=180] (-0.7,1.3) to[out=0,in=90] pic[pos=0.9,sloped]{arrow} (-0.2,0.8);  
\draw (0.8,0.3) to[out=45,in=-90] pic[pos=0.9,sloped]{arrow} (1,0.8) to[out=90,in=0] (0.7,1.3) to[out=180,in=90] pic[pos=0.6,sloped]{arrow} (0.2,0.8);
\node[right] at (-2,-3.5) {$ \cW= \, \Flip[a^N \, q; \at^N \, \qt] $};
\node at (-1.2,1.2) {$a$};
\node at (1.2,1.2) {$\at$};
\node at (-1,-1.2) {$q$};
\node at (1,-1.2) {$\qt$};
\end{scope}
\begin{scope}[shift={(6.2,0)}]
\node at (-0.7,1.6) {$\star_2)$};
\node[fnode] (F3) at (-0.1,0) {$4$};
\node[gUSpnode] (G3) at (1.5,0) {$2N$};
\node[gUSpnode] (G4) at (3.2,0) {$2N$};
\node[fnode] (F4) at (4.8,0) {$4$};
\node[fnode] (F5) at (2.4,-2) {$2$};
\draw (F3) -- pic[pos=0.4,sloped,very thick]{arrow=latex reversed} (G3) -- (G4) -- pic[pos=0.8,sloped,very thick]{arrow=latex reversed} (F4);
\draw (G3) -- (F5) -- (G4);
\node[right] at (-3,-3.5) {$ \cW= 1 \, \Triangle + \displaystyle\sum_{i=0}^{N-1}\Flip[q_L (b b)^i q_L; q_R (b b)^i q_R; q_L b (b b)^i q_R] $};
\node at (0.7,0.3) {$q_L$};
\node at (2.3,0.3) {$b$};
\node at (4,0.3) {$q_R$};
\node at (1.6,-1.1) {$V_L$};
\node at (3.1,-1.1) {$D_R$};
\end{scope}
\epic} \ee
Of course, our prescription does not tell us the precise flippers, which are crucial in order for the duality to be correct. In section \ref{flippers} we provide a strategy to obtain such flippers, and we apply it to a quiver duality that generalizes \eqref{4dFamilyIIOdd}. 

The mapping of the chiral ring generators is given by
\vspace{-2.5cm}
\be \label{mappingFamilyIIOdd}
\scalebox{0.9}{$
\ba[t]{c}\star_1) \\
q \, \qt \, (a \, \at)^i \\
q \, q \, \at \, (a \, \at)^i \\
\qt \, \qt \, a \, (a \, \at)^i \\
(a \, \at)^j \\
\begin{cases}
\Flipper[a^{N} \, q] \\
a^{N-1} \, q^3
\end{cases}
\\
\\[-4pt]
\begin{cases}
\Flipper[\at^{N} \, \qt] \\
\at^{N-1} \, \qt^3 \\
\end{cases}
\ea
\ba{c} \\
\\
\\
\\
\\
\\
\\
\\
\\[-10pt]
\Longleftrightarrow
\ea
\ba[t]{c}\star_2) \\
\Flipper[q_L \, b (b b)^{N-1-i} \, q_R] \\
\Flipper[q_L \, (b b)^{N-1-i} \, q_L] \\
\Flipper[q_R \, (b b)^{N-1-i} \, q_R] \\
(b b)^{j} \\
\\[-4pt]
q_L \, V_L \\
\\
\\
q_R \, D_R
\ea
\qquad
\ba[t]{l} \\
i=0, \dots, N-1 \\
i=0, \dots, N-1 \\
i=0, \dots, N-1 \\
j=1, \dots, N \\
\ea
$}
\ee

For $N=1$, we go back to the situation \eqref{4dFamilyI1}-\eqref{4dFamilyI2}. For generic $N$, we don't have a proof of the duality \eqref{4dFamilyIIOdd} involving more basic Seiberg dualities. The non-trivial check of this duality is the matching of the 't Hooft anomalies and of the central charges with a-maximization \cite{Anselmi:1997am,Anselmi:1997ys,Intriligator:2003jj}.
\subsection*{$n$ even: $n = 2N$}
In this case the $4d$ duality constructed from the $5d$ duality \eqref{UVdualitiesFamilyIIEven} is
\be \label{4dFamilyIIEven} \scalebox{0.9}{\bpic[node distance=2cm,gSUnode/.style={circle,red,draw,minimum size=8mm},gUSpnode/.style={circle,blue,draw,minimum size=8mm},fnode/.style={rectangle,red,draw,minimum size=8mm}]     
\begin{scope}[shift={(0,0)}]
\node at (-2.2,1.6) {$\star_1)$};
\node[gSUnode] (G1) at (0,0) {$2N$};
\node[fnode] (F1) at (1,-2) {$4$};
\node[fnode] (F2) at (-1,-2) {$4$};
\draw (G1) -- pic[pos=0.6,sloped]{arrow} (F1);
\draw (G1) -- pic[pos=0.4,sloped]{arrow} (F2);
\draw (-0.5,0.2) to[out=135,in=-90] pic[pos=0.1,sloped]{arrow} (-0.7,0.6) to[out=90,in=180] (-0.5,0.9) to[out=0,in=90] pic[pos=0.9,sloped]{arrow} (-0.2,0.5);  
\draw (0.5,0.2) to[out=45,in=-90] pic[pos=0.9,sloped]{arrow} (0.7,0.6) to[out=90,in=0] (0.5,0.9) to[out=180,in=90] pic[pos=0.6,sloped]{arrow} (0.2,0.5);
\node[right] at (-2.8,-3.5) {$\cW= \, \Flip[a^N; \at^N] + \displaystyle\sum_{j=0}^{\left\lfloor \frac{N-1}{2} \right\rfloor} \Flip[q \, (a \at)^j \, \qt]$};
\node[right] at (-1.7,-5) {$+ \displaystyle\sum_{i=0}^{\left\lfloor \frac{N-2}{2} \right\rfloor} \Flip[\at \, (a \at)^i \, q^2; a \, (a \at)^i \, \qt^2]$};
\node at (-0.9,1) {$a$};
\node at (0.9,1) {$\at$};
\node at (-0.8,-1) {$q$};
\node at (0.8,-1) {$\qt$};
\end{scope}
\begin{scope}[shift={(7,0)}]
\node at (-0.7,1.6) {$\star_2)$};
\node[fnode] (F3) at (-0.1,0) {$6$};
\node[gUSpnode] (G3) at (1.5,0) {$2N$};
\node[gUSpnode] (G4) at (3.5,0) {\scalebox{0.73}{$2N-2$}};
\node[fnode] (F4) at (5.3,0) {$2$};
\node[fnode] (F5) at (2.4,-2) {$2$};
\draw (F3) -- (G3) -- (G4) -- (F4);
\draw (G3) -- (F5) -- (G4);
\node[right] at (-1,-3.5) {$ \cW= 1 \, \Triangle + \displaystyle\sum_{i=0}^{\left\lfloor \frac{N-2}{2} \right\rfloor} \Flip[q_L \, (b b)^i \, q_R;$};
\node[right] at (2,-5) {$q_R (b b)^i q_R; q_L b (b b)^i q_R]$};
\node at (0.7,0.3) {$q_L$};
\node at (2.4,0.3) {$b$};
\node at (4.5,0.3) {$q_R$};
\node at (1.6,-1.1) {$V_l$};
\node at (3.3,-1.1) {$D_r$};
\end{scope}
\epic} \ee
To write the mapping of the chiral ring generators we have to distinguish between $N$ even and odd.\\
\noindent \textbf{$N$ even:}
\vspace{-2cm}
\be \label{mappingFamilyIIEvenNEven1} 
\scalebox{0.9}{$
\ba[t]{c}\star_1) \\
\\
\begin{cases}
\Flipper[q \qt] \\
q \, (a \at)^{N-1} \, \qt 
\end{cases}
\ea
\,
\ba{c} \\
\\
\\
\\
\\
\\[-10pt]
\Longleftrightarrow 
\ea
\quad
\ba[t]{c}\star_2) \\
\begin{cases}
\Flipper[q_L q_L] \\
q_L \, (b b)^{N-1} \, q_L  \\
\Flipper[q_R q_R] \\
V_l^2
\end{cases}
\ea
$}
\ee
The total number of operators on both sides is $32$.
\vspace{-1.8cm}
\be \label{mappingFamilyIIEvenNEven2} 
\scalebox{0.9}{$
\ba[t]{c} \\
\\
\begin{cases}
\Flipper[q \, (a \at)^{i} \, \qt] \\
q \, (a \at)^{N-i-1} \, \qt 
\end{cases}
\ea
\,
\ba{c} \\
\\
\\
\\
\\
\\[-10pt]
\Longleftrightarrow 
\ea
\quad
\ba[t]{c} \\
\begin{cases}
\Flipper[q_L \, (b b)^{i} \, q_L] \\
q_L \, (b b)^{N-i-1} \, q_L  \\
\Flipper[q_R \, (b b)^{i} \, q_R] \\
q_R \, (b b)^{N-i-1} \, q_R
\end{cases}
\ea
\qquad
\ba[t]{l} \\
\\
\\
i=1, \dots, \frac{N-2}{2} \\
\ea
$}
\ee
The total number of operators on both sides is $16(N-2)$.
\vspace{-1.8cm}
\be \label{mappingFamilyIIEvenNEven3} 
\scalebox{0.9}{$
\ba[t]{c} \\
\begin{cases}
\Flipper[\at \, (a \at)^{j} \, q^2] \\
\at \, (a \at)^{N-j-2} \, q^2 \\
\Flipper[a \, (a \at)^{j} \, \qt^2] \\
a \, (a \at)^{N-j-2} \, \qt^2
\end{cases}
\ea
\,
\ba{c} \\
\\
\\
\\
\\
\\[-10pt]
\Longleftrightarrow 
\ea
\quad
\ba[t]{c} \\
\\
\begin{cases}
\Flipper[q_L \, b \, (b b)^{j} \, q_R] \\
q_L \, b \, (b b)^{N-j-2} \, q_R 
\end{cases}
\ea
\qquad
\ba[t]{l} \\
\\
\\
j=1, \dots, \frac{N-2}{2}-1 \\
\ea
$}
\ee
The total number of operators on both sides is $12(N-2)-24$.
\vspace{-1.8cm}
\be \label{mappingFamilyIIEvenNEven4} 
\scalebox{0.9}{$
\ba[t]{c} \\
\begin{cases}
\Flipper[\at \, (a \at)^{(N-2)/2} \, q^2] \\
a^{N-1} \, q^2 \\
\Flipper[a \, (a \at)^{(N-2)/2} \, \qt^2] \\
\at^{N-1} \, \qt^2
\end{cases}
\ea
\,
\ba{c} \\
\\
\\
\\
\\
\\[-10pt]
\Longleftrightarrow 
\ea
\quad
\ba[t]{c} \\
\\
\begin{cases}
\Flipper[q_L \, b \, (b b)^{(N-2)/2} \, q_R] \\
q_L \, V_l
\end{cases}
\ea
$}
\ee
The total number of operators on both sides is $24$.
\vspace{-1.8cm}
\be \label{mappingFamilyIIEvenNEven5} 
\scalebox{0.9}{$
\ba[t]{c} \\
\begin{cases}
\Flipper[a^N] \\
\Flipper[\at^N] \\
a^{N-2} \, q^4 \\
\at^{N-2} \, \qt^4
\end{cases}
\ea
\,
\ba{c} \\
\\
\\
\\
\\
\\[-5pt]
\Longleftrightarrow 
\ea
\quad
\ba[t]{c} \\
\\
\\
D_r \, q_R
\ea
$}
\ee
The total number of operators on both sides is $4$.
\vspace{-1cm}
\be \label{mappingFamilyIIEvenNOdd6} 
\scalebox{0.9}{$
\ba[t]{c} \\
(a \at)^{m} 
\ea
\quad
\ba{c} \\
\\
\Longleftrightarrow 
\ea
\quad
\ba[t]{c} \\
(b b)^{m} 
\ea
\qquad
\ba[t]{l} \\
m=1, \dots, N-1 \\
\ea
$}
\ee
The total number of operators on both sides is $N-1$.

\noindent \textbf{$N$ odd:}
\vspace{-1.8cm}
\be \label{mappingFamilyIIEvenNOdd1} 
\scalebox{0.9}{$
\ba[t]{c}\star_1) \\
\\
\begin{cases}
\Flipper[q \qt] \\
q \, (a \at)^{N-1} \, \qt 
\end{cases}
\ea
\,
\ba{c} \\
\\
\\
\\
\\
\\[-10pt]
\Longleftrightarrow 
\ea
\quad
\ba[t]{c}\star_2) \\
\begin{cases}
\Flipper[q_L q_L] \\
q_L \, (b b)^{N-1} \, q_L  \\
\Flipper[q_R q_R] \\
V_l^2
\end{cases}
\ea
$}
\ee
The total number of operators on both sides is $32$.
\vspace{-1.8cm}
\be \label{mappingFamilyIIEvenNOdd2} 
\scalebox{0.9}{$
\ba[t]{c} \\
\\
\begin{cases}
\Flipper[q \, (a \at)^{i} \, \qt] \\
q \, (a \at)^{N-i-1} \, \qt 
\end{cases}
\ea
\,
\ba{c} \\
\\
\\
\\
\\
\\[-10pt]
\Longleftrightarrow 
\ea
\quad
\ba[t]{c} \\
\begin{cases}
\Flipper[q_L \, (b b)^{i} \, q_L] \\
q_L \, (b b)^{N-i-1} \, q_L  \\
\Flipper[q_R \, (b b)^{i} \, q_R] \\
q_R \, (b b)^{N-i-1} \, q_R
\end{cases}
\ea
\qquad
\ba[t]{l} \\
\\
\\
i=1, \dots, \frac{N-3}{2} \\
\ea
$}
\ee
The total number of operators on both sides is $16(N-3)$.
\vspace{-1.8cm}
\be \label{mappingFamilyIIEvenNOdd3} 
\scalebox{0.9}{$
\ba[t]{c} \\
\\
\Flipper[q \, (a \at)^{(N-1)/2} \, \qt] \\
\ea
\quad
\ba{c} \\
\\
\\
\\[-5pt]
\Longleftrightarrow 
\ea
\quad
\ba[t]{c} \\
\begin{cases}
q_L \, (b b)^{(N-1)/2} \, q_L  \\
q_R \, (b b)^{(N-1)/2} \, q_R  
\end{cases}
\ea
$}
\ee
The total number of operators on both sides is $16$.
\vspace{-1.8cm}
\be \label{mappingFamilyIIEvenNOdd4} 
\scalebox{0.9}{$
\ba[t]{c} \\
\begin{cases}
\Flipper[\at \, (a \at)^{j} \, q^2] \\
\at \, (a \at)^{N-j-2} \, q^2 \\
\Flipper[a \, (a \at)^{j} \, \qt^2] \\
a \, (a \at)^{N-j-2} \, \qt^2
\end{cases}
\ea
\,
\ba{c} \\
\\
\\
\\
\\
\\[-10pt]
\Longleftrightarrow 
\ea
\quad
\ba[t]{c} \\
\\
\begin{cases}
\Flipper[q_L \, b \, (b b)^{j} \, q_R] \\
q_L \, b \, (b b)^{N-j-2} \, q_R 
\end{cases}
\ea
\qquad
\ba[t]{l} \\
\\
\\
j=0, \dots, \frac{N-3}{2} \\
\ea
$}
\ee
The total number of operators on both sides is $12(N-1)$.
\vspace{-1.4cm}
\be \label{mappingFamilyIIEvenNOdd5} 
\scalebox{0.9}{$
\ba[t]{c} \\
\begin{cases}
a^{N-1} \, q^2 \\
\at^{N-1} \, \qt^2
\end{cases}
\ea
\,
\ba{c} \\
\\
\\
\\[-10pt]
\Longleftrightarrow 
\ea
\quad
\ba[t]{c} \\
\\[-2pt]
q_L \, V_l
\ea
$}
\ee
The total number of operators on both sides is $12$.
\vspace{-1.8cm}
\be \label{mappingFamilyIIEvenNOdd6} 
\scalebox{0.9}{$
\ba[t]{c} \\
\begin{cases}
\Flipper[a^N] \\
\Flipper[\at^N] \\
a^{N-2} \, q^4 \\
\at^{N-2} \, \qt^4
\end{cases}
\ea
\,
\ba{c} \\
\\
\\
\\
\\
\\[-5pt]
\Longleftrightarrow 
\ea
\quad
\ba[t]{c} \\
\\
\\
D_r \, q_R
\ea
$}
\ee
The total number of operators on both sides is $4$.
\vspace{-1cm}
\be \label{mappingFamilyIIEvenNOdd7} 
\scalebox{0.9}{$
\ba[t]{c} \\
(a \at)^{m} 
\ea
\quad
\ba{c} \\
\\
\Longleftrightarrow 
\ea
\quad
\ba[t]{c} \\
(b b)^{m} 
\ea
\qquad
\ba[t]{l} \\
m=1, \dots, N-1 \\
\ea
$}
\ee
The total number of operators on both sides is $N-1$.

Also for this case we don't have a proof of this duality \eqref{4dFamilyIIEven}. The non-trivial check of this duality is the matching of the central charges with a-maximization.

\section{Systems with two $O7$ planes: $A_{n,m}$ and its dual} \label{4ddualityToCheck}
\noindent \textbf{$A_{n,m}$ theories:} In this section, we generalize the discussion of section \ref{fam2} considering the following two-parameter family of $5d$ theories, that we call $A_{n,m}$:
\be \label{Anm} \scalebox{0.9}{\bpic[node distance=2cm,gSUnode/.style={circle,red,draw,minimum size=8mm},gUSpnode/.style={circle,blue,draw,minimum size=8mm},fnode/.style={rectangle,red,draw,minimum size=8mm}]  
\begin{scope}[shift={(0,0)}]
\node at (-1.2,1) {$A_{n,m}:$};
\node[fnode] (F1) at (0,0) {$4$};
\node[gSUnode] (G1) at (1.5,0) {$n$};
\node[gSUnode] (G2) at (3,0) {$n$};
\node (G3) at (4.5,0) {$\dots$};
\node[gSUnode] (G4) at (6,0) {$n$};
\node[gSUnode] (G5) at (7.5,0) {$n$};
\node[fnode] (F2) at (9,0) {$4$};
\draw (F1) -- (G1) -- (G2) -- (G3) -- (G4) -- (G5) -- (F2);
\draw (1.8,0.3) to[out=90,in=0]  (1.5,0.8) to[out=180,in=90] (1.2,0.3);
\draw (7.8,0.3) to[out=90,in=0]  (7.5,0.8) to[out=180,in=90] (7.2,0.3);
\draw[decorate,decoration={calligraphic brace,mirror,amplitude=7pt}] (0.9,-0.6) -- (8.1,-0.6) node[pos=0.5,below=9pt,black] {$m$};
\end{scope}
\epic} \ee
The duality statement will depend on the parity of the parameter $n$ as in the last subsection. 
\subsection{$n$ odd: 5d duality $A_{2N+1,m} \lra U_{2N+1,m}$}
\begin{figure}[H]
\scalebox{0.9}{\bpic[node distance=2cm,gSUnode/.style={circle,red,draw,minimum size=8mm},gUSpnode/.style={circle,blue,draw,minimum size=8mm},fnode/.style={rectangle,draw,minimum size=8mm}]
\begin{scope}[shift={(-5,0)}] 
\node[inner sep=0pt] at (0,0) {\includegraphics[height=.25\textheight]{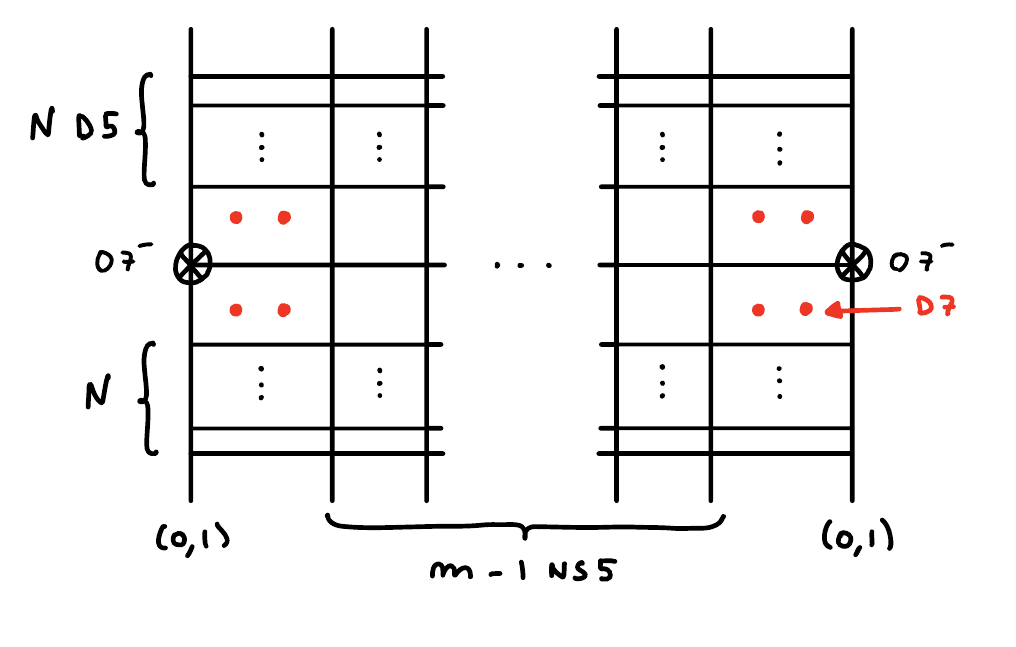}};
\end{scope}
\begin{scope}[shift={(4.5,0)}] 
\node[inner sep=0pt] at (0,0) {\includegraphics[height=.25\textheight]{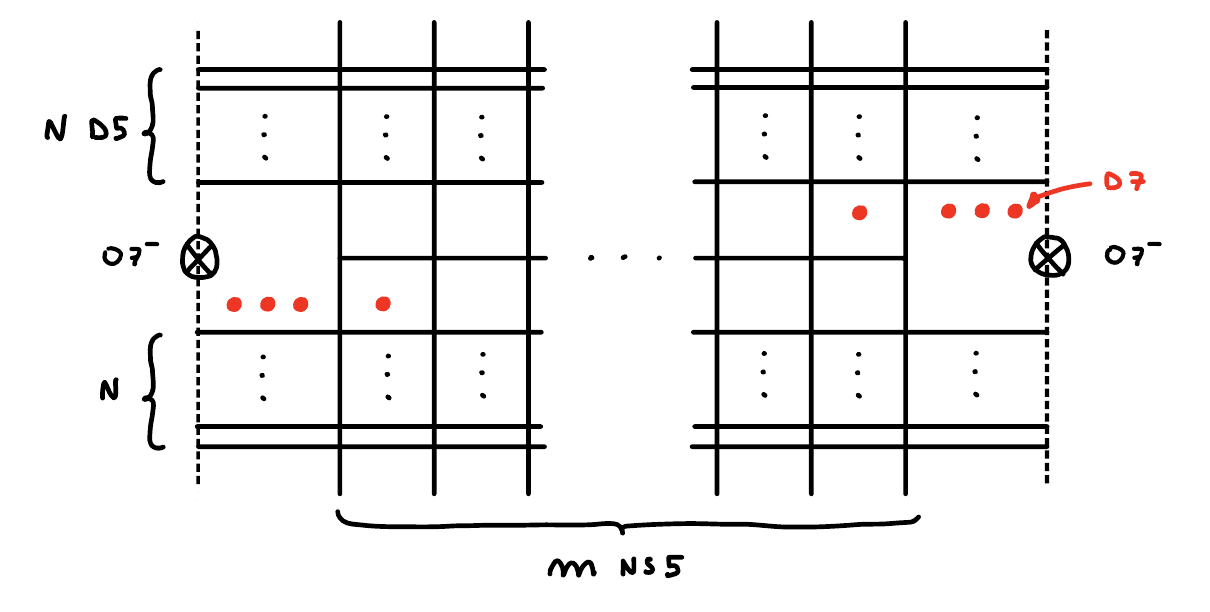}};
\end{scope}
\epic}
\caption{Brane setup for $A_{2N+1,m}$ on the left with an $NS5$ stuck on each $O7^-$ plane and for $U_{2N+1,m}$ on the right.}
\label{TypeIIB5dGeneralFamilyIIOdd1}
\end{figure}

In this section, we generalize the duality \eqref{UVdualitiesFamilyIIOdd}. We call the dual of $A_{2N+1,m}$, $U_{2N+1,m}$. $A_{2N+1,m}$ respectively $U_{2N+1,m}$ contains a hyper in the antisymmetric representation of the gauge group respectively a $USp(2N)$ gauge node at each end of the quiver. The quiver for $A_{2N+1,m}$/$U_{2N+1,m}$ is shown in \eqref{UVdualitiesGeneralFamilyIIOdd1}/\eqref{UVdualitiesGeneralFamilyIIOdd2}. We have also depicted the brane systems in Figure \ref{TypeIIB5dGeneralFamilyIIOdd1}. The claim is that $A_{2N+1,m}$ and $U_{2N+1,m}$ are UV dual. The analysis of the brane systems that lead to this duality can be found in \cite{Hayashi:2015zka}.
\be \label{UVdualitiesGeneralFamilyIIOdd1} \scalebox{0.9}{\bpic[node distance=2cm,gSUnode/.style={circle,red,draw,minimum size=8mm},gUSpnode/.style={circle,blue,draw,minimum size=8mm},fnode/.style={rectangle,red,draw,minimum size=8mm}]  
\begin{scope}[shift={(0,0)}]
\node at (-1.8,1) {$\star_1) \, \, A_{2N+1,m}:$};
\node[fnode] (F1) at (0,0) {$4$};
\node[gSUnode] (G1) at (1.5,0) {\scalebox{0.8}{$2N+1$}};
\node[gSUnode] (G2) at (3.2,0) {\scalebox{0.8}{$2N+1$}};
\node (G3) at (4.7,0) {$\dots$};
\node[gSUnode] (G4) at (6.2,0) {\scalebox{0.8}{$2N+1$}};
\node[fnode] (F2) at (7.7,0) {$4$};
\draw (F1) -- (G1) -- (G2) -- (G3) -- (G4) -- (F2);
\draw (2,0.5) to[out=90,in=0]  (1.5,1) to[out=180,in=90] (1,0.5);
\draw (6.7,0.5) to[out=90,in=0]  (6.2,1) to[out=180,in=90] (5.7,0.5);
\draw[decorate,decoration={calligraphic brace,mirror,amplitude=7pt}] (0.7,-0.8) -- (6.9,-0.8) node[pos=0.5,below=9pt,black] {$m$};
\end{scope}
\epic} \ee
\be \label{UVdualitiesGeneralFamilyIIOdd2} \scalebox{0.9}{\bpic[node distance=2cm,gSUnode/.style={circle,red,draw,minimum size=8mm},gUSpnode/.style={circle,blue,draw,minimum size=8mm},fnode/.style={rectangle,red,draw,minimum size=8mm}]  
\begin{scope}[shift={(0,0)}]
\node at (-1.3,2) {$\star_2) \, \, U_{2N+1,m}:$};
\node[fnode] (F1) at (0,0) {$3$};
\node[gUSpnode] (G1) at (1.5,0) {$2N$};
\node[gSUnode] (G2) at (3.2,0) {\scalebox{0.8}{$2N+1$}};
\node[fnode] (F2) at (3.2,1.7) {$1$};
\node[gSUnode] (G3) at (5.2,0) {\scalebox{0.8}{$2N+1$}};
\node (G4) at (7.2,0) {$\dots$};
\node[gSUnode] (G5) at (9.2,0) {\scalebox{0.8}{$2N+1$}};
\node[gSUnode] (G6) at (11.2,0) {\scalebox{0.8}{$2N+1$}};
\node[fnode] (F3) at (11.2,1.7) {$1$};
\node[gUSpnode] (G7) at (12.9,0) {$2N$};
\node[fnode] (F4) at (14.4,0) {$3$};
\draw (F1) -- (G1) -- (G2) -- (G3) -- (G4) -- (G5) -- (G6) -- (G7) -- (F4);
\draw (G2) -- (F2);
\draw (G6) -- (F3);
\draw[decorate,decoration={calligraphic brace,mirror,amplitude=7pt}] (1,-0.8) -- (13.5,-0.8) node[pos=0.5,below=9pt,black] {$m+1$};
\end{scope}
\epic} \ee 

\subsection{6d UV completion}
The UV completion of the $5d$ theories in \eqref{UVdualitiesGeneralFamilyIIOdd1}-\eqref{UVdualitiesGeneralFamilyIIOdd2} is a $6d$ given by the following Type IIA brane setup \cite{Hayashi:2015zka,Zafrir:2015rga}: 
\begin{figure}[H]
\centering
\includegraphics[height=0.25\textheight]{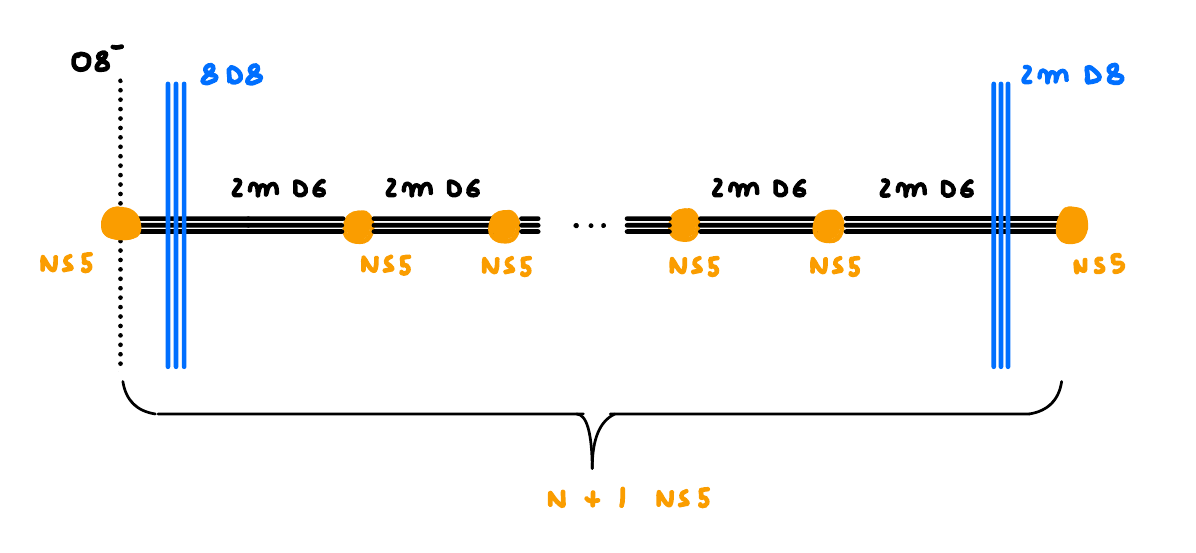}
\caption{Type IIA brane setup corresponding to the $6d$ UV completion of $A_{2N+1,m}$.}
\label{TypeIIA6dCompletionGeneralFamilyIIOddFig}
\end{figure}
On the tensor branch, the system flows to the following gauge theory:

\be \label{TypeIIA6dCompletionGeneralFamilyIIOdd} \scalebox{0.9}{\bpic[node distance=2cm,gSUnode/.style={circle,red,draw,minimum size=8mm},gUSpnode/.style={circle,blue,draw,minimum size=8mm},fnode/.style={rectangle,red,draw,minimum size=8mm}]  
\node[fnode] (F1) at (-4.5,0) {$8$};
\node[gSUnode] (G1) at (-3,0) {$2m$};
\node[gSUnode] (G2) at (-1.5,0) {$2m$};
\node (G3) at (0,0) {$\dots$};
\node[gSUnode] (G4) at (1.5,0) {$2m$};
\node[fnode] (F2) at (3,0) {$2m$};
\draw (F1) -- (G1) -- (G2) -- (G3) -- (G4) -- (F2);
\draw[decorate,decoration={calligraphic brace,mirror,amplitude=7pt}] (-3.5,-0.6) -- (2,-0.6) node[pos=0.5,below=9pt,black] {$N$};
\draw (-2.6,0.3) to[out=90,in=0]  (-3,0.8) to[out=180,in=90] (-3.4,0.3);
\epic} \ee

\subsection{4d duality}
Our prescription of section \ref{Algorithm} applied to the $5d$ duality \eqref{UVdualitiesGeneralFamilyIIOdd1}-\eqref{UVdualitiesGeneralFamilyIIOdd2} gives the following $4d$ duality
\be \label{4dGeneralFamilyIIOdd1} \scalebox{0.85}{\bpic[node distance=2cm,gSUnode/.style={circle,red,draw,minimum size=8mm},gUSpnode/.style={circle,blue,draw,minimum size=8mm},fnode/.style={rectangle,red,draw,minimum size=8mm}]
\begin{scope}[shift={(0,0)}]
\node at (-1.5,1) {$\tsup[1]{\star}_1)$};
\node[fnode] (F1) at (-0.5,0) {$3$};
\node[fnode,magenta] (F2) at (0.2,-1.7) {$1$};
\node[gSUnode] (G1) at (1.5,0) {\scalebox{0.8}{$2N+1$}};
\node[fnode,ForestGreen] (F3) at (2.4,-1.7) {$2$};
\node[gSUnode] (G2) at (3.4,0) {\scalebox{0.8}{$2N+1$}};
\node[fnode] (F4) at (4.2,-1.7) {$2$};
\node (G3) at (5.1,0) {$\dots$};
\node[fnode,YellowGreen] (F5) at (5.8,-1.7) {$2$};
\node[gSUnode] (G4) at (6.8,0) {\scalebox{0.8}{$2N+1$}};
\node[fnode,gray] (F6) at (8.1,-1.7) {$1$};
\node[fnode,Orchid] (F7) at (8.8,0) {$3$};
\draw (F1) -- pic[pos=0.7,sloped]{arrow} (G1) -- pic[pos=0.7,sloped]{arrow} (G2) -- pic[pos=0.7,sloped]{arrow} (G3) -- pic[pos=0.7,sloped]{arrow} (G4) -- pic[pos=0.7,sloped]{arrow} (F7);
\draw (G1) -- pic[pos=0.4,sloped,very thick]{arrow=latex reversed} (F2);
\draw (G1) -- pic[pos=0.5,sloped,very thick]{arrow=latex reversed} (F3);
\draw (G2) -- pic[pos=0.5,sloped,very thick]{arrow=latex reversed} (F3);
\draw (G2) -- pic[pos=0.5,sloped,very thick]{arrow=latex reversed} (F4);
\draw (G3) -- pic[pos=0.4,sloped,very thick]{arrow=latex reversed} (F4);
\draw (G3) -- pic[pos=0.5,sloped,very thick]{arrow=latex reversed} (F5);
\draw (G4) -- pic[pos=0.5,sloped,very thick]{arrow=latex reversed} (F5);
\draw (G4) -- pic[pos=0.5,sloped,very thick]{arrow=latex reversed} (F6);
\draw (2,0.5) to[out=90,in=0] pic[pos=0.2,sloped]{arrow} (1.5,1) to[out=180,in=90] pic[pos=0.5,sloped,very thick]{arrow=latex reversed} (1,0.5);
\draw (7.3,0.5) to[out=90,in=0] pic[pos=0.4,sloped,very thick]{arrow=latex reversed} (6.8,1) to[out=180,in=90] pic[pos=0.3,sloped]{arrow} (6.3,0.5);
\node at (0.3,0.4) {$L$};
\node at (0.4,-0.8) {$F_l$};
\node at (2,1.2) {$A_l$};
\node at (2.5,0.4) {$B_1$};
\node at (4.4,0.4) {$B_2$};
\node at (5.6,0.4) {$B_{m-1}$};
\node at (7.3,1.2) {$A_r$};
\node at (8,0.4) {$R$};
\node at (7.9,-0.8) {$F_r$};
\node at (1.7,-1.1) {$V_1$};
\node at (3.2,-1.1) {$D_1$};
\node at (6.8,-1.1) {$D_{m-1}$};
\end{scope}
\epic} \ee
\be \label{4dGeneralFamilyIIOdd2} \scalebox{0.85}{\bpic[node distance=2cm,gSUnode/.style={circle,red,draw,minimum size=8mm},gUSpnode/.style={circle,blue,draw,minimum size=8mm},fnode/.style={rectangle,red,draw,minimum size=8mm}]
\begin{scope}[shift={(0,0)}]
\node at (-0.5,2.2) {$\tsup[1]{\star}_2)$};
\node[fnode] (F1) at (0,0) {$3$};
\node[gUSpnode,MidnightBlue] (G1) at (1.5,0) {$2N$};
\node[fnode,Violet] (F2) at (2.3,-1.7) {$2$};
\node[gSUnode] (G2) at (3.2,0) {\scalebox{0.8}{$2N+1$}};
\node[fnode,magenta] (F3) at (3.2,1.9) {$1$};
\node[fnode,ForestGreen] (F4) at (4.2,-1.7) {$2$};
\node[gSUnode] (G3) at (5.2,0) {\scalebox{0.8}{$2N+1$}};
\node[fnode] (F5) at (6.2,-1.7) {$2$};
\node (G4) at (7.2,0) {$\dots$};
\node[fnode] (F6) at (8.2,-1.7) {$2$};
\node[gSUnode] (G5) at (9.2,0) {\scalebox{0.8}{$2N+1$}};
\node[fnode,YellowGreen] (F7) at (10.2,-1.7) {$2$};
\node[gSUnode] (G6) at (11.2,0) {\scalebox{0.8}{$2N+1$}};
\node[fnode,gray] (F8) at (11.2,1.9) {$1$};
\node[fnode,Brown] (F9) at (12,-1.7) {$2$};
\node[gUSpnode] (G7) at (12.9,0) {$2N$};
\node[fnode,Orchid] (F10) at (14.4,0) {$3$};
\draw (F1) -- pic[pos=0.7,sloped]{arrow} (G1) -- pic[pos=0.7,sloped]{arrow} (G2) -- pic[pos=0.7,sloped]{arrow} (G3) -- pic[pos=0.7,sloped]{arrow} (G4) -- pic[pos=0.7,sloped]{arrow} (G5) -- pic[pos=0.7,sloped]{arrow} (G6) -- pic[pos=0.7,sloped]{arrow} (G7) -- pic[pos=0.7,sloped]{arrow} (F10);
\draw (G2) -- pic[pos=0.6,sloped,very thick]{arrow=latex reversed} (F3);
\draw (G6) -- pic[pos=0.7,sloped]{arrow} (F8);
\draw (G1) -- (F2);
\draw (G2) -- pic[pos=0.5,sloped,very thick]{arrow=latex reversed} (F2);
\draw (G2) -- pic[pos=0.5,sloped,very thick]{arrow=latex reversed} (F4);
\draw (G3) -- pic[pos=0.5,sloped,very thick]{arrow=latex reversed} (F4);
\draw (G3) -- pic[pos=0.4,sloped,very thick]{arrow=latex reversed} (F5);
\draw (G4) -- pic[pos=0.5,sloped,very thick]{arrow=latex reversed} (F5);
\draw (G4) -- pic[pos=0.5,sloped,very thick]{arrow=latex reversed} (F6);
\draw (G5) -- pic[pos=0.5,sloped,very thick]{arrow=latex reversed} (F6);
\draw (G5) -- pic[pos=0.5,sloped,very thick]{arrow=latex reversed} (F7);
\draw (G6) -- pic[pos=0.5,sloped,very thick]{arrow=latex reversed} (F7);
\draw (G6) -- pic[pos=0.5,sloped,very thick]{arrow=latex reversed} (F9);
\draw (G7) -- (F9);
\node at (0.8,0.4) {$\Lt$};
\node at (2.8,1.1) {$\Ft_l$};
\node at (2.3,0.4) {$\Ut_l$};
\node at (4.2,0.4) {$\Bt_1$};
\node at (6.2,0.4) {$\Bt_2$};
\node at (10.2,0.6) {$\Bt_{m-2}$};
\node at (10.8,1.1) {$\Ft_r$};
\node at (12.1,0.4) {$\Ut_r$};
\node at (13.7,0.4) {$\Rt$};
\node at (1.6,-1) {$\Vt_l$};
\node at (3,-1) {$\Dt_l$};
\node at (4.1,-1) {$\Vt_1$};
\node at (5,-1) {$\Dt_1$};
\node at (9.4,-1.1) {$\Vt_{m-2}$};
\node at (11.3,-1) {$\Vt_r$};
\node at (12.7,-1) {$\Dt_r$};
\end{scope}
\epic} \ee
Without the flippers, these two theories are \emph{not} dual to each other.
\subsubsection*{Strategy to get the set of flippers}\label{flippers}
In order to obtain the correct set of flippers to make $\tsup[1]{\star}_1)$ and $\tsup[1]{\star}_2)$ dual, we did the following procedure. 

\noindent Starting with $\tsup[1]{\star}_1)$ and do the following operations:
\begin{itemize}
\item deconfinement of the two antisymmetric\footnote{The deconfinement is a name of a technique that replaces an antisymmetric field by a confining $USp$ gauge node, see \cite{Bajeot:2022kwt, Bottini:2022vpy, Bajeot:2022lah} for details.}
\item m Seiberg dualities on the m SU nodes
\item CSST duality on the left SU(2)
\item m-2 confinements
\end{itemize}
\be \label{4dGeneralFamilyIIOdd3} \scalebox{0.85}{\bpic[node distance=2cm,gSUnode/.style={circle,red,draw,minimum size=8mm},gUSpnode/.style={circle,blue,draw,minimum size=8mm},fnode/.style={rectangle,red,draw,minimum size=8mm}]
\begin{scope}[shift={(0,0)}]
\node at (-6,3.5) {We get};
\node at (-3.8,3.5) {$\tsup[1]{\star}_1)$};
\node[gSUnode] (G1) at (0,0) {$2$};
\node[gSUnode] (G2) at (0,-2) {$2$};
\node[gUSpnode,MidnightBlue] (G3) at (-2,2) {\scalebox{0.8}{$2N-2$}};
\node[gUSpnode] (G4) at (2,2) {\scalebox{0.8}{$2N-2$}};
\node[fnode,Orchid] (F1) at (-2,0) {$3$};
\node[fnode] (F2) at (2,0) {$3$};
\node[fnode,gray] (F3) at (-3,-2) {$1$};
\node[fnode] (F4) at (-1.5,-2) {$1$};
\node[fnode,Brown] (F5) at (1.5,-2) {$1$};
\node[fnode,magenta] (F6) at (3,-2) {$1$};
\node[fnode,ForestGreen] (F7) at (-1,-4) {$2$};
\node[fnode,YellowGreen] (F8) at (1,-4) {$2$};
\node (F9) at (0,-4) {$\dots$};
\draw (G1) -- (G2);
\draw (G3) -- (G1) -- (G4);
\draw (G3) -- (G4);
\draw (G3) -- pic[pos=0.7,sloped]{arrow} (F1);
\draw (G4) -- pic[pos=0.7,sloped,very thick]{arrow=latex reversed} (F2);
\draw (F1) --pic[pos=0.7,sloped]{arrow} (G1) --pic[pos=0.7,sloped]{arrow} (F2); 
\draw (F4) -- (G2) -- (F5);
\draw (F7) -- (G2) -- (F8);
\draw (F1) --pic[pos=0.2,sloped]{arrow} (F3) -- (F4) --pic[pos=0.6,sloped,very thick]{arrow=latex reversed} (F1);
\draw (F2) --pic[pos=0.3,sloped,very thick]{arrow=latex reversed} (F5) -- (F6) --pic[pos=0.5,sloped]{arrow} (F2);
\draw (F3.south east) to[out=-45,in=180] (-1.5,-3) to[out=0,in=-135]  (G2);
\draw (F6.south west) to[out=-135,in=0] (1.5,-3) to[out=180,in=-45]  (G2);
\draw (F1.west) to[out=135,in=-90] (-3,2) to[out=90,in=180] pic[pos=1,sloped,very thick]{arrow=latex reversed} (0,3.5) to[out=0,in=90] (3,2) to[out=-90,in=45]  (F2.east);
\draw (G3) to[out=45,in=180] (-0.8,3) node{$+$};
\draw (F6.north east) to[out=45,in=-90] (3.5,-1) node{$+$};
\draw (G4) to[out=135,in=0] (0.8,3) node{$\bullet$};
\draw (F3.north west) to[out=135,in=-90] (-3.5,-1) node{$\bullet$};
\node[right] at (5,0) {$ \cW= 6 \, \quartic + 6 \, \triangles$};
\node[right] at (5.7,-1.5) {$+ \, \Flip[b \, b] + \displaystyle\sum_{i=1}^{m-1} \Flip[p_i \, p_i]$};
\draw[decorate,decoration={calligraphic brace,mirror,amplitude=7pt}] (-1.6,-4.6) -- (1.6,-4.6) node[pos=0.5,below=9pt,black] {$m-1$};
\node at (0.2,-1.1) {$b$};
\node at (-2.9,-1.1) {$\a_l$};
\node at (2.9,-1.1) {$\a_r$};
\node at (-2.2,-2.2) {$s_l$};
\node at (-1.4,-1.1) {$d_l$};
\node at (1.4,-1.1) {$d_r$};
\node at (2.2,-2.2) {$s_r$};
\node at (-0.2,-3.2) {$p_1$};
\node at (1.1,-3.2) {$p_{m-1}$};
\node at (0.2,3.7) {$\eta$};
\end{scope}
\epic} \ee
Then, we start with $\tsup[1]{\star}_2)$ and do the following operations:
\begin{itemize}
\item m-1 Seiberg dualities on the m-1 SU nodes
\item CSST duality on the left SU(2)
\item m-3 confinements
\end{itemize}
\be \label{4dGeneralFamilyIIOdd4} \scalebox{0.85}{\bpic[node distance=2cm,gSUnode/.style={circle,red,draw,minimum size=8mm},gUSpnode/.style={circle,blue,draw,minimum size=8mm},fnode/.style={rectangle,red,draw,minimum size=8mm}]
\begin{scope}[shift={(0,0)}]
\node at (-6,3) {We get};
\node at (-4.5,3) {$\tsup[1]{\star}_2)$};
\node[gSUnode] (G1) at (0,0) {$2$};
\node[gSUnode] (G2) at (0,-2) {$2$};
\node[gUSpnode,MidnightBlue] (G3) at (-1.7,2) {$2N$};
\node[gUSpnode] (G4) at (1.7,2) {$2N$};
\node[fnode] (F1) at (-3.5,2) {$3$};
\node[fnode,Orchid] (F2) at (3.5,2) {$3$};
\node[fnode,gray] (F3) at (-3,0) {$1$};
\node[fnode,magenta] (F4) at (3,0) {$1$};
\node[fnode,Violet] (F5) at (-2,-2) {$2$};
\node[fnode,Brown] (F6) at (2,-2) {$2$};
\node[fnode,ForestGreen] (F7) at (-1,-4) {$2$};
\node[fnode,YellowGreen] (F8) at (1,-4) {$2$};
\node (F9) at (0,-4) {$\dots$};
\draw (G1) -- (G2);
\draw (G3) -- (G1) -- (G4);
\draw (G3) -- (G4);
\draw (G3) -- pic[pos=0.5,sloped,very thick]{arrow=latex reversed} (F1);
\draw (G4) -- pic[pos=0.7,sloped,very thick]{arrow=latex reversed} (F2);
\draw (G3) -- (F3) -- (G1) -- (F4) -- (G4);
\draw (F3) -- (F5) -- (G2) -- (F6) -- (F4);
\draw (F7) -- (G2) -- (F8);
\draw (F3.west) -- (-4,0) node{$\bullet$};
\draw (F4.east) -- (4,0) node{$\bullet$};
\node[right] at (5,0) {$ \cW= 2 \, \quartic + 4 \, \triangles$};
\node[right] at (5.7,-1) {$+ \, \Flip[\bt \, \bt; \pt_l \, \pt_l; \pt_r \, \pt_r]$};
\node[right] at (5.7,-2.2) {$+ \, \displaystyle\sum_{i=1}^{m-2} \Flip[\pt_i \, \pt_i]$};
\node at (0.2,-1.1) {$\tilde{b}$};
\node at (-2.9,-1.1) {$\tilde{\a}_l$};
\node at (2.9,-1.1) {$\tilde{\a}_r$};
\node at (-1,-1.7) {$\tilde{p}_l$};
\node at (1,-1.7) {$\tilde{p}_r$};
\node at (-0.8,-3) {$\tilde{p}_1$};
\node at (1.2,-3) {$\tilde{p}_{m-2}$};
\node at (3.8,-0.3) {$\tilde{\eta}$};
\draw[decorate,decoration={calligraphic brace,mirror,amplitude=7pt}] (-1.6,-4.6) -- (1.6,-4.6) node[pos=0.5,below=9pt,black] {$m-2$};
\end{scope}
\epic} \ee
Then we can play with \eqref{4dGeneralFamilyIIOdd3} and \eqref{4dGeneralFamilyIIOdd4} to make manifest a bigger flavor symmetry group. Concretely, we flip the operators ($\a_l; \a_r; d_l; d_r; s_l; s_r; \eta; \Flipper[p_i \, p_i]$) in \eqref{4dGeneralFamilyIIOdd3}  and ($\tilde{\a}_l; \tilde{\a}_r; \tilde{\eta};  \Flipper[\pt_i \, \pt_i, \pt_l \, \pt_l, \pt_r \, \pt_r] $)  in \eqref{4dGeneralFamilyIIOdd4}. We therefore consider the following theories 
\be \label{4dGeneralFamilyIIOdd5} \scalebox{0.85}{\bpic[node distance=2cm,gSUnode/.style={circle,red,draw,minimum size=8mm},gUSpnode/.style={circle,blue,draw,minimum size=8mm},fnode/.style={rectangle,red,draw,minimum size=8mm}]
\begin{scope}[shift={(0,0)}]
\node at (-3.8,3) {$\tsup{\star}_1)$};
\node[gSUnode] (G1) at (0,0) {$2$};
\node[gSUnode] (G2) at (0,-2) {$2$};
\node[gUSpnode,MidnightBlue] (G3) at (-2,2) {\scalebox{0.8}{$2N-2$}};
\node[gUSpnode] (G4) at (2,2) {\scalebox{0.8}{$2N-2$}};
\node[fnode,Orchid] (F1) at (-2,0) {$3$};
\node[fnode] (F2) at (2,0) {$3$};
\node[fnode,magenta] (F3) at (-2,-2) {$1$};
\node[fnode,gray] (F4) at (2,-2) {$1$};
\node[fnode] (F5) at (0,-4) {$2m$};
\draw (G1) -- (G2) -- (F5);
\draw (G3) -- (G1) -- (G4);
\draw (G3) -- (G4);
\draw (G3) --pic[pos=0.7,sloped]{arrow} (F1) --pic[pos=0.7,sloped]{arrow} (G1) --pic[pos=0.7,sloped]{arrow} (F2) --pic[pos=0.7,sloped]{arrow} (G4); 
\draw (F3) -- (G2) -- (F4);
\draw (F3.west) to[out=135,in=-135] (G3.west);
\draw (F4.east) to[out=45,in=-45] (G4.east);
\node[right] at (-2.5,-5.5) {$ \cW= 2 \, \quartic + 3 \, \triangles$};
\node at (-1,-0.3) {$c$};
\node at (1,-0.3) {$\ct$};
\node at (-0.7,1) {$o$};
\node at (0.7,1) {$\ot$};
\node at (-1,-1.7) {$s$};
\node at (1,-1.7) {$\st$};
\node at (-3.7,0) {$u$};
\node at (3.7,0) {$\ut$};
\node at (-2.3,0.8) {$n$};
\node at (2.3,0.8) {$\nt$};
\node at (0,2.3) {$l$};
\node at (0.2,-1.1) {$b$};
\node at (0.2,-3) {$f$};
\end{scope}
\begin{scope}[shift={(9,0)}]
\node at (-4.5,3) {$\tsup{\star}_2)$};
\node[gSUnode] (G1) at (0,0) {$2$};
\node[gSUnode] (G2) at (0,-2) {$2$};
\node[gUSpnode,MidnightBlue] (G3) at (-1.5,2) {$2N$};
\node[gUSpnode] (G4) at (1.5,2) {$2N$};
\node[fnode] (F1) at (-3.2,2) {$3$};
\node[fnode,Orchid] (F2) at (3.2,2) {$3$};
\node[fnode,gray] (F3) at (-2,0) {$1$};
\node[fnode,magenta] (F4) at (2,0) {$1$};
\node[fnode] (F5) at (0,-4) {$2m$};
\draw (G1) -- (G2) -- (F5);
\draw (G3) -- (G1) -- (G4);
\draw (G3) -- (G4);
\draw (G3) -- pic[pos=0.5,sloped,very thick]{arrow=latex reversed} (F1);
\draw (G4) -- pic[pos=0.7,sloped,very thick]{arrow=latex reversed} (F2);
\draw (G3) -- (F3) -- (G1) -- (F4) -- (G4);
\node[right] at (-1.5,-5.5) {$ \cW= 3 \, \triangles$};
\node at (-2.4,2.4) {$\qt$};
\node at (2.4,2.3) {$q$};
\node at (-0.5,1) {$k$};
\node at (0.5,1) {$\kt$};
\node at (-1,-0.3) {$\ttilde$};
\node at (1,-0.3) {$t$};
\node at (-2,1) {$\dt$};
\node at (2,1) {$d$};
\node at (0,2.3) {$p$};
\node at (0.2,-1.1) {$w$};
\node at (0.2,-3) {$f$};
\end{scope}
\epic} \ee
Once again, at this stage $\tsup{\star}_1)$ and $\tsup{\star}_2)$ are \emph{not} dual. Now to make progress, we will focus on the case $N=1$ and generic $m$. Moreover we saw that in the $m=1$ case we had to flip the whole towers in the frame $\star_2$) \eqref{4dFamilyIIOdd}. Therefore we decide to study the following theory
\be \label{4dGeneralFamilyIIOdd6} \scalebox{0.85}{\bpic[node distance=2cm,gSUnode/.style={circle,red,draw,minimum size=8mm},gUSpnode/.style={circle,blue,draw,minimum size=8mm},fnode/.style={rectangle,red,draw,minimum size=8mm}]
\begin{scope}[shift={(0,0)}] 
\node at (-4.5,3) {$\star_2)$};
\node[gSUnode] (G1) at (0,0) {$2$};
\node[gSUnode] (G2) at (0,-2) {$2$};
\node[gUSpnode,MidnightBlue] (G3) at (-1.5,2) {$2$};
\node[gUSpnode] (G4) at (1.5,2) {$2$};
\node[fnode] (F1) at (-3.2,2) {$3$};
\node[fnode,Orchid] (F2) at (3.2,2) {$3$};
\node[fnode,gray] (F3) at (-2,0) {$1$};
\node[fnode,magenta] (F4) at (2,0) {$1$};
\node[fnode] (F5) at (0,-4) {$2m$};
\draw (G1) -- (G2) -- (F5);
\draw (G3) -- (G1) -- (G4);
\draw (G3) -- (G4);
\draw (G3) -- pic[pos=0.5,sloped,very thick]{arrow=latex reversed} (F1);
\draw (G4) -- pic[pos=0.7,sloped,very thick]{arrow=latex reversed} (F2);
\draw (G3) -- (F3) -- (G1) -- (F4) -- (G4);
\node[right] at (4.5,0) {$ \cW= 3 \, \triangles + \, \Flip[ww; qq; qd; \qt \qt; \qt \dt;$};
\node[right] at (8.5,-1) {$ q p \qt; q p \dt; \qt p d; d p \dt]$};
\node at (-2.4,2.4) {$\qt$};
\node at (2.4,2.3) {$q$};
\node at (-0.5,1) {$k$};
\node at (0.5,1) {$\kt$};
\node at (-1,-0.3) {$\ttilde$};
\node at (1,-0.3) {$t$};
\node at (-2,1) {$\dt$};
\node at (2,1) {$d$};
\node at (0,2.3) {$p$};
\node at (0.2,-1.1) {$w$};
\node at (0.2,-3) {$f$};
\end{scope}
\epic} \ee
Now we start by doing a CSST duality on the \textcolor{MidnightBlue}{$USp(2) \equiv SU(2)$}. We get 
\be \label{4dGeneralFamilyIIOdd7} \scalebox{0.85}{\bpic[node distance=2cm,gSUnode/.style={circle,red,draw,minimum size=8mm},gUSpnode/.style={circle,blue,draw,minimum size=8mm},fnode/.style={rectangle,red,draw,minimum size=8mm}]
\begin{scope}[shift={(0,0)}] 
\node[gSUnode] (G1) at (0,0) {$2$};
\node[gSUnode] (G2) at (0,-2) {$2$};
\node[gUSpnode,MidnightBlue] (G3) at (-1.5,2) {$2$};
\node[gUSpnode] (G4) at (1.5,2) {$2$};
\node[fnode] (F1) at (-3.2,2) {$3$};
\node[fnode,Orchid] (F2) at (3.2,2) {$3$};
\node[fnode,gray] (F3) at (-2,0) {$1$};
\node[fnode,magenta] (F4) at (2,0) {$1$};
\node[fnode] (F5) at (0,-4) {$2m$};
\draw (G1) -- (G2) -- (F5);
\draw (G3) -- (G1);
\draw (G3) -- (G4);
\draw (G3) -- pic[pos=0.5,sloped]{arrow} (F1);
\draw (G4) -- pic[pos=0.7,sloped,very thick]{arrow=latex reversed} (F2);
\draw (G3) -- (F3) -- (G1) -- (F4) -- (G4);
\node[right] at (4.5,0) {$ \cW= 1 \, \Triangle + 1 \, \quartic + \Flip[pp; kk; ww;$};
\node[right] at (8.5,-1) {$q q; q d; q p \qt; \qt p d; q p \dt; d p \dt]$};
\end{scope}
\epic} \ee
Then we use the IP confinement \cite{Intriligator:1995ne} on the \textcolor{blue}{$USp(2)$}. We obtain
\be \label{4dGeneralFamilyIIOdd8} \scalebox{0.85}{\bpic[node distance=2cm,gSUnode/.style={circle,red,draw,minimum size=8mm},gUSpnode/.style={circle,blue,draw,minimum size=8mm},fnode/.style={rectangle,red,draw,minimum size=8mm}]
\begin{scope}[shift={(0,0)}] 
\node[gSUnode] (G1) at (0,0) {$2$};
\node[gSUnode] (G2) at (0,-2) {$2$};
\node[gUSpnode,MidnightBlue] (G3) at (0,2) {$2$};
\node[fnode] (F1) at (-2.5,2) {$3$};
\node[fnode,Orchid] (F2) at (2.5,2) {$3$};
\node[fnode,gray] (F3) at (-2,0) {$1$};
\node[fnode,magenta] (F4) at (2,0) {$1$};
\node[fnode] (F5) at (0,-4) {$2m$};
\draw (G1) -- (G2) -- (F5);
\draw (G3) -- (G1);
\draw (F1) -- pic[pos=0.5,sloped]{arrow} (G3) -- pic[pos=0.7,sloped,very thick]{arrow=latex reversed} (F2);
\draw (G3) -- (F3) -- (G1) -- (F4) -- (G3);
\node[right] at (4.5,0) {$\cW= 2 \, \triangles + \Flip[mm; ww; \ct \dt;$};
\node[right] at (9.5,-1) {$c d; c \ct; d \dt]$};
\node at (-1,2.3) {$\ct$};
\node at (1,2.3) {$c$};
\node at (-1,-0.3) {$\ttilde$};
\node at (1,-0.3) {$t$};
\node at (-1.3,1) {$\dt$};
\node at (1.3,1) {$d$};
\node at (0.3,0.7) {$m$};
\node at (0.3,-1) {$w$};
\node at (0.3,-3) {$f$};
\end{scope}
\epic} \ee
The last step is the IP confinement on the middle \textcolor{red}{$SU(2)$} to get
\be \label{4dGeneralFamilyIIOdd9} \scalebox{0.85}{\bpic[node distance=2cm,gSUnode/.style={circle,red,draw,minimum size=8mm},gUSpnode/.style={circle,blue,draw,minimum size=8mm},fnode/.style={rectangle,red,draw,minimum size=8mm}]
\begin{scope}[shift={(0,0)}] 
\node[gUSpnode,MidnightBlue] (G1) at (0,0) {$2$};
\node[gSUnode] (G2) at (0,-2) {$2$};
\node[fnode] (F1) at (-2,0) {$3$};
\node[fnode,Orchid] (F2) at (2,0) {$3$};
\node[fnode,gray] (F3) at (-2,-2) {$1$};
\node[fnode,magenta] (F4) at (2,-2) {$1$};
\node[fnode] (F5) at (0,-4) {$2m$};
\draw (G1) -- (G2) -- (F5);
\draw (F1) -- (G1) -- (F2);
\draw (F3) -- (G2) -- (F4);
\node[right] at (4.5,-2) {$ \cW= \Flip[bb; c\ct; cb \st; \ct b s; b s b \st]$};
\node at (1,0.3) {$c$};
\node at (-1,0.3) {$\ct$};
\node at (1,-1.7) {$s$};
\node at (-1,-1.7) {$\st$};
\node at (0.2,-1) {$b$};
\node at (0.2,-3) {$f$};
\end{scope}
\epic} \ee
Which is of the form $\tsup{\star}_1)$ in \eqref{4dGeneralFamilyIIOdd5} specified to the case $N=1$. This result motivates the following educated guess for generic $N$:
\be \label{4dGeneralFamilyIIOdd10} \scalebox{0.85}{\bpic[node distance=2cm,gSUnode/.style={circle,red,draw,minimum size=8mm},gUSpnode/.style={circle,blue,draw,minimum size=8mm},fnode/.style={rectangle,red,draw,minimum size=8mm}]
\begin{scope}[shift={(0,0)}]
\node at (-3.8,3) {$\star_1)$};
\node[gSUnode] (G1) at (0,0) {$2$};
\node[gSUnode] (G2) at (0,-2) {$2$};
\node[gUSpnode,MidnightBlue] (G3) at (-2,2) {\scalebox{0.8}{$2N-2$}};
\node[gUSpnode] (G4) at (2,2) {\scalebox{0.8}{$2N-2$}};
\node[fnode,Orchid] (F1) at (-2,0) {$3$};
\node[fnode] (F2) at (2,0) {$3$};
\node[fnode,magenta] (F3) at (-2,-2) {$1$};
\node[fnode,gray] (F4) at (2,-2) {$1$};
\node[fnode] (F5) at (0,-4) {$2m$};
\draw (G1) -- (G2) -- (F5);
\draw (G3) -- (G1) -- (G4);
\draw (G3) -- (G4);
\draw (G3) --pic[pos=0.7,sloped]{arrow} (F1) --pic[pos=0.7,sloped]{arrow} (G1) --pic[pos=0.7,sloped]{arrow} (F2) --pic[pos=0.7,sloped]{arrow} (G4); 
\draw (F3) -- (G2) -- (F4);
\draw (F3.west) to[out=135,in=-135] (G3.west);
\draw (F4.east) to[out=45,in=-45] (G4.east);
\node[right] at (-2.5,-5.5) {$ \cW= 2 \, \quartic + 3 \, \triangles$};
\node[right] at (-2.5,-6.5) {$+ \, \Flip[bb, c \ct, cb \st, \ct b s, bsb \st]$};
\node at (-1,-0.3) {$c$};
\node at (1,-0.3) {$\ct$};
\node at (-0.7,1) {$o$};
\node at (0.7,1) {$\ot$};
\node at (-1,-1.7) {$s$};
\node at (1,-1.7) {$\st$};
\node at (-3.7,0) {$u$};
\node at (3.7,0) {$\ut$};
\node at (-2.3,0.8) {$n$};
\node at (2.3,0.8) {$\nt$};
\node at (0,2.3) {$l$};
\node at (0.3,-1) {$b$};
\node at (0.2,-3) {$f$};
\end{scope}
\begin{scope}[shift={(9,0)}]
\node at (-4.5,3) {$\star_2)$};
\node[gSUnode] (G1) at (0,0) {$2$};
\node[gSUnode] (G2) at (0,-2) {$2$};
\node[gUSpnode,MidnightBlue] (G3) at (-1.5,2) {$2N$};
\node[gUSpnode] (G4) at (1.5,2) {$2N$};
\node[fnode] (F1) at (-3.2,2) {$3$};
\node[fnode,Orchid] (F2) at (3.2,2) {$3$};
\node[fnode,gray] (F3) at (-2,0) {$1$};
\node[fnode,magenta] (F4) at (2,0) {$1$};
\node[fnode] (F5) at (0,-4) {$2m$};
\draw (G1) -- (G2) -- (F5);
\draw (G3) -- (G1) -- (G4);
\draw (G3) -- (G4);
\draw (G3) -- pic[pos=0.5,sloped,very thick]{arrow=latex reversed} (F1);
\draw (G4) -- pic[pos=0.7,sloped,very thick]{arrow=latex reversed} (F2);
\draw (G3) -- (F3) -- (G1) -- (F4) -- (G4);
\node[right] at (-2.5,-5.5) {$ \cW= 3 \, \triangles + \, \Flip[ww]$};
\node[right] at (-3.8,-6.7) {$+ \, \displaystyle \sum_{i=o}^{N-1} \Flip[q (p p)^i q; \qt (p p)^i \qt; q (p p)^i d; \qt (p p)^i \dt; $};
\node[right] at (-2.5,-7.9) {$q p (p p)^i \qt; q p (p p)^i \dt; \qt p (p p)^i d; d p (p p)^i \dt]$};
\node at (-2.4,2.4) {$\qt$};
\node at (2.4,2.3) {$q$};
\node at (-0.5,1) {$k$};
\node at (0.5,1) {$\kt$};
\node at (-1,-0.3) {$\ttilde$};
\node at (1,-0.3) {$t$};
\node at (-2,1) {$\dt$};
\node at (2,1) {$d$};
\node at (0,2.3) {$p$};
\node at (0.3,-1) {$w$};
\node at (0.2,-3) {$f$};
\end{scope}
\epic} \ee
We claim that $\star_1)$ and $\star_2)$ in \eqref{4dGeneralFamilyIIOdd10} are dual. For generic $N$ and $m$, we don't have a proof of this statement. However we provided a proof for the special case of $N=1$ and generic $m$. 
The first non-trivial test of this duality is the matching of the central charges for generic $N$ and $m$. Then we can match 't Hooft anomalies. We have reported the computation in the appendix \ref{MatchingAnomalies}. 

The mapping of the chiral ring generators is given by
\be \label{mappingGeneralFamilyIIOdd}
\scalebox{0.9}{$
\ba{c}\star_1) \\
n \, (l \, l)^i \, n \\
\nt \, (l \, l)^i \, \nt \\
n \, (l \, l)^i \, u \\
\nt \, (l \, l)^i \, \ut \\
n \, l \, (l \, l)^i \, \nt \\
n \, l \, (l \, l)^i \, \ut \\
\nt \, l \, (l \, l)^i \, u \\
u \, l \, (l \, l)^i \, \ut \\
(l \, l)^j \\
c b s \\
\ct b \st \\
c c \\
\ct \ct \\
\Flipper[c \ct] \\
\Flipper[c b \st] \\
\Flipper[\ct b s] \\
\Flipper[b s b \st] \\
s \st \\
c \ot \ut \\
\ct o u \\
c b f \\
\ct b f \\
o o \\
\ot \ot \\
\Flipper[b b] \\
s f \\
\st f \\
f f 
\ea
\quad
\ba{c} \\
\\
\\
\\
\\
\\
\\
\\
\\
\\
\\
\\
\\
\\
\Longleftrightarrow \\
\\
\\
\\
\\
\\
\\
\\
\\
\\
\\
\\
\\
\ea
\quad
\ba{c}\star_2) \\
\Flipper[q \, (p p)^{N-1-i} \, q] \\
\Flipper[\qt \, (p p)^{N-1-i} \, \qt] \\
\Flipper[q \, (p p)^{N-1-i} \, d] \\
\Flipper[\qt \, (p p)^{N-1-i} \, \dt] \\
\Flipper[q \, p \, (p p)^{N-2-i} \, \qt] \\
\Flipper[q \, p \, (p p)^{N-2-i} \, \dt] \\
\Flipper[\qt \, p \, (p p)^{N-2-i} \, d] \\
\Flipper[d \, p \, (p p)^{N-2-i} \, \dt] \\
(p p)^j \\
\Flipper[q \, q] \\
\Flipper[\qt \, \qt] \\
\Flipper[q \, d] \\
\Flipper[\qt \, \dt] \\
\Flipper[q \, p \, (p p)^{N-1} \, \qt] \\
\Flipper[q \, p \, (p p)^{N-1} \, \dt] \\
\Flipper[\qt \, p \, (p p)^{N-1} \, d] \\
\Flipper[d \, p \, (p p)^{N-1} \, \dt] \\
(p p)^N \\
q \kt \ttilde \\
\qt k t \\
q \kt w f \\
\qt k w f \\
k k \\
\kt \kt \\
t \ttilde \\
t w f \\
\ttilde w f \\
ff
\ea
\qquad
\ba{l} \\
i=0, \dots, N-2 \\
i=0, \dots, N-2 \\
i=0, \dots, N-2 \\
i=0, \dots, N-2 \\
i=0, \dots, N-2 \\
i=0, \dots, N-2 \\
i=0, \dots, N-2 \\
i=0, \dots, N-2 \\
j=1, \dots N-1 \\
\\
\\
\\
\\
\\
\\
\\
\\
\\
\\
\\
\\
\\
\\
\\
\\
\\
\\
\\
\ea
$}
\ee

\subsection{$n$ even: 5d duality $A_{2N,m} \lra U_{2N,m}$}
\begin{figure}[H]
\scalebox{0.9}{\bpic[node distance=2cm,gSUnode/.style={circle,red,draw,minimum size=8mm},gUSpnode/.style={circle,blue,draw,minimum size=8mm},fnode/.style={rectangle,draw,minimum size=8mm}]
\begin{scope}[shift={(-5,0)}] 
\node[inner sep=0pt] at (0,0) {\includegraphics[height=.25\textheight]{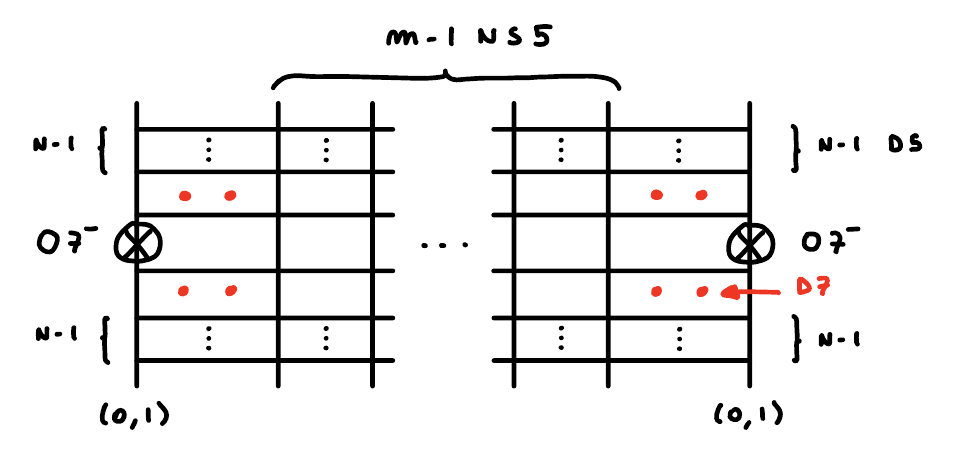}};
\end{scope}
\begin{scope}[shift={(4.5,0)}] 
\node[inner sep=0pt] at (0,0) {\includegraphics[height=.25\textheight]{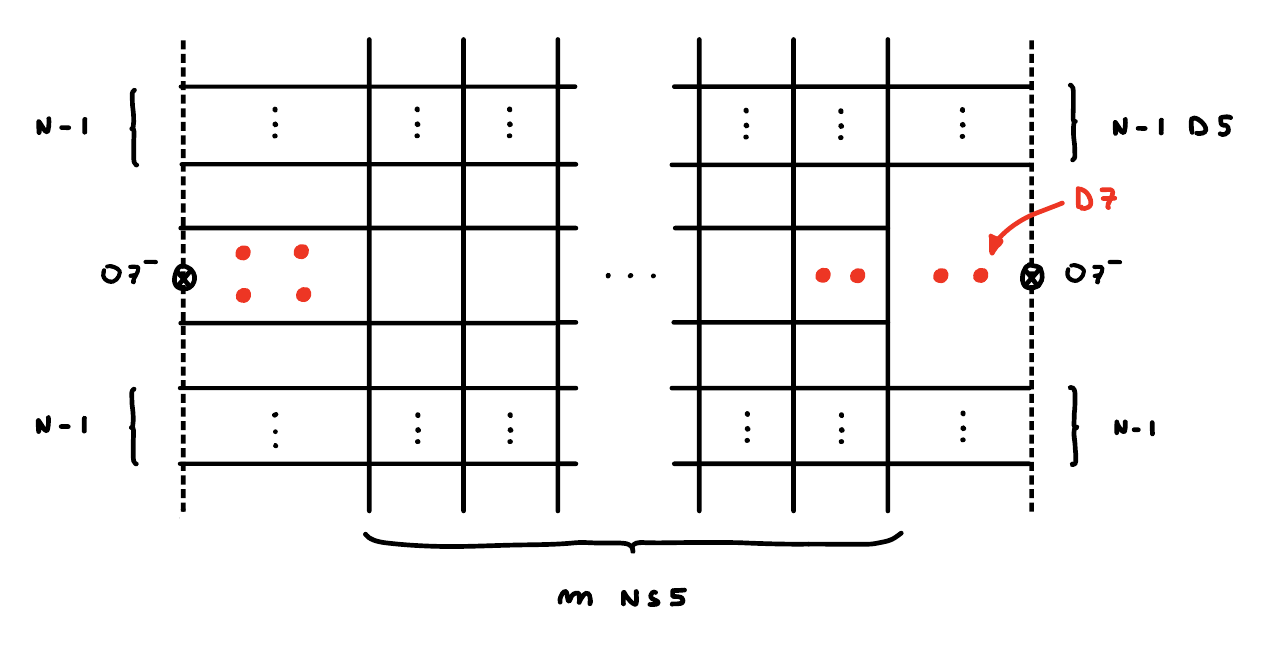}};
\end{scope}
\epic}
\caption{Brane setup for $A_{2N,m}$ on the left with an $NS5$ stuck on each $O7^-$ plane and for $U_{2N,m}$ on the right.}
\label{TypeIIB5dGeneralFamilyIIEven1}
\end{figure}

In this section, we generalize the duality \eqref{UVdualitiesFamilyIIEven}. We name  $U_{2N,m}$ the dual of $A_{2N,m}$. $A_{2N,m}$  ($U_{2N,m}$) contains a hyper in the antisymmetric representation of the gauge group (a $USp$ gauge node) at each end of the quiver. The quiver for $A_{2N,m}$/$U_{2N,m}$ is shown in \eqref{UVdualitiesGeneralFamilyIIEven1}/\eqref{UVdualitiesGeneralFamilyIIEven2}. We have also depicted the brane systems in Figure \ref{TypeIIB5dGeneralFamilyIIEven1}. The claim is that $A_{2N,m}$ and $U_{2N,m}$ are UV dual. The analysis of the brane systems that leads to this duality can be found in \cite{Hayashi:2015zka}.
\be \label{UVdualitiesGeneralFamilyIIEven1} \scalebox{0.9}{\bpic[node distance=2cm,gSUnode/.style={circle,red,draw,minimum size=8mm},gUSpnode/.style={circle,blue,draw,minimum size=8mm},fnode/.style={rectangle,red,draw,minimum size=8mm}]  
\begin{scope}[shift={(0,0)}]
\node at (-1.8,1) {$\star_1) \, \, A_{2N,m}:$};
\node[fnode] (F1) at (0,0) {$4$};
\node[gSUnode] (G1) at (1.5,0) {$2N$};
\node[gSUnode] (G2) at (3.2,0) {$2N$};
\node (G3) at (4.7,0) {$\dots$};
\node[gSUnode] (G4) at (6.2,0) {$2N$};
\node[fnode] (F2) at (7.7,0) {$4$};
\draw (F1) -- (G1) -- (G2) -- (G3) -- (G4) -- (F2);
\draw (1.9,0.3) to[out=90,in=0]  (1.5,0.9) to[out=180,in=90] (1.1,0.3);
\draw (6.6,0.3) to[out=90,in=0]  (6.2,0.9) to[out=180,in=90] (5.8,0.3);
\draw[decorate,decoration={calligraphic brace,mirror,amplitude=7pt}] (0.7,-0.8) -- (6.9,-0.8) node[pos=0.5,below=9pt,black] {$m$};
\end{scope}
\epic} \ee
\be \label{UVdualitiesGeneralFamilyIIEven2} \scalebox{0.9}{\bpic[node distance=2cm,gSUnode/.style={circle,red,draw,minimum size=8mm},gUSpnode/.style={circle,blue,draw,minimum size=8mm},fnode/.style={rectangle,red,draw,minimum size=8mm}]  
\begin{scope}[shift={(0,0)}]
\node at (-1.3,2) {$\star_2) \, \, U_{2N,m}:$};
\node[fnode] (F1) at (0,0) {$4$};
\node[gUSpnode] (G1) at (1.5,0) {$2N$};
\node[gSUnode] (G2) at (3.2,0) {$2N$};
\node[gSUnode] (G3) at (5.2,0) {$2N$};
\node (G4) at (7.2,0) {$\dots$};
\node[gSUnode] (G5) at (9.2,0) {$2N$};
\node[gSUnode] (G6) at (11.2,0) {$2N$};
\node[fnode] (F3) at (11.2,1.7) {$2$};
\node[gUSpnode] (G7) at (12.9,0) {\scalebox{0.8}{$2N-2$}};
\node[fnode] (F4) at (14.4,0) {$2$};
\draw (F1) -- (G1) -- (G2) -- (G3) -- (G4) -- (G5) -- (G6) -- (G7) -- (F4);
\draw (G6) -- (F3);
\draw[decorate,decoration={calligraphic brace,mirror,amplitude=7pt}] (1,-0.8) -- (13.5,-0.8) node[pos=0.5,below=9pt,black] {$m+1$};
\end{scope}
\epic} \ee 

\subsection{6d UV completion}
The UV completion of the $5d$ theories in \eqref{UVdualitiesGeneralFamilyIIEven1}-\eqref{UVdualitiesGeneralFamilyIIEven2} is a $6d$ given by the following Type IIA brane setup \cite{Hayashi:2015zka,Zafrir:2015rga}: 
\begin{figure}[H]
\centering
\includegraphics[height=0.25\textheight]{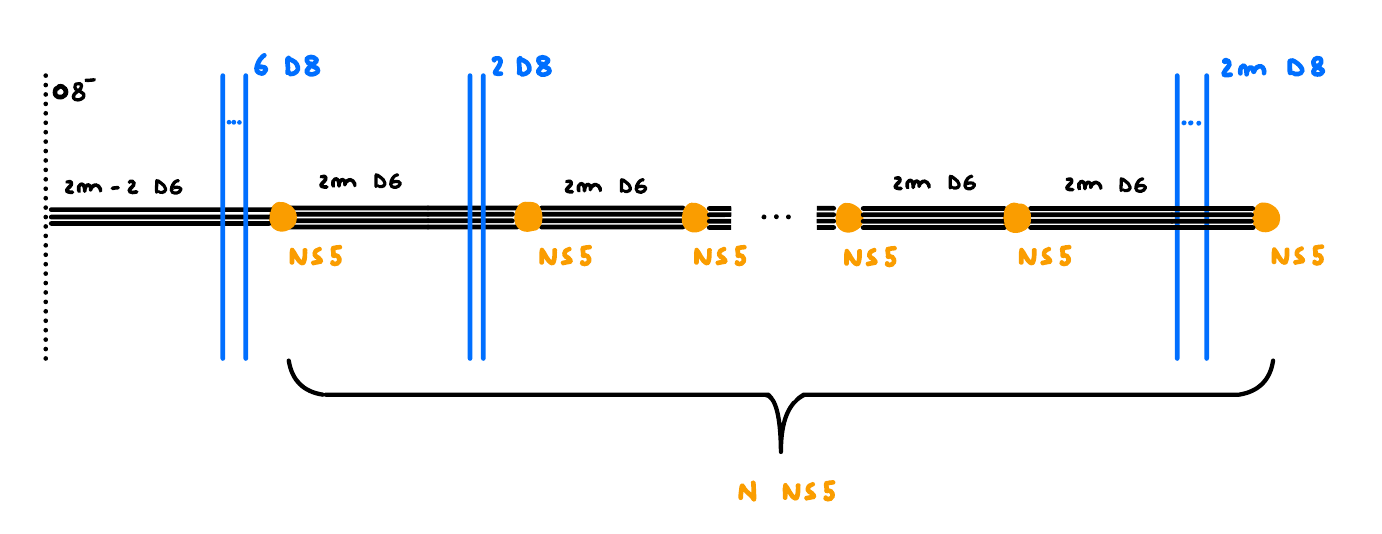}
\caption{Type IIA brane setup corresponding to the $6d$ UV completion of $A_{2N,m}$.}
\label{TypeIIA6dCompletionGeneralFamilyIIEvenFig}
\end{figure}
On the tensor branch, the system flows to the following gauge theory:

\be \label{TypeIIA6dCompletionGeneralFamilyIIEven} \scalebox{0.9}{\bpic[node distance=2cm,gSUnode/.style={circle,red,draw,minimum size=8mm},gUSpnode/.style={circle,blue,draw,minimum size=8mm},fnode/.style={rectangle,red,draw,minimum size=8mm}]   
\node[fnode] (F1) at (-6.6,0) {$6$};
\node[gUSpnode] (G0) at (-4.8,0) {\scalebox{0.8}{$2N-2$}};
\node[gSUnode] (G1) at (-3,0) {$2m$};
\node[fnode] (F2) at (-3,1.7) {$2$};
\node[gSUnode] (G2) at (-1.5,0) {$2m$};
\node (G3) at (0,0) {$\dots$};
\node[gSUnode] (G4) at (1.5,0) {$2m$};
\node[fnode] (F3) at (3,0) {$2m$};
\draw (F1) -- (G0) -- (G1) -- (G2) -- (G3) -- (G4) -- (F3);
\draw (G1) -- (F2);
\draw[decorate,decoration={calligraphic brace,mirror,amplitude=7pt}] (-5.2,-0.8) -- (2,-0.8) node[pos=0.5,below=9pt,black] {$N$};
\epic} \ee

\subsection{4d duality}
Our prescription of Sec. \ref{Algorithm} applied to the $5d$ duality \eqref{UVdualitiesGeneralFamilyIIEven1}-\eqref{UVdualitiesGeneralFamilyIIEven2} gives the following $4d$ duality
\be \label{4dGeneralFamilyIIEven1} \scalebox{0.9}{\bpic[node distance=2cm,gSUnode/.style={circle,red,draw,minimum size=8mm},gUSpnode/.style={circle,blue,draw,minimum size=8mm},fnode/.style={rectangle,red,draw,minimum size=8mm}]
\begin{scope}[shift={(0,0)}]
\node at (-1.5,1) {$\tsup[1]{\star}_1)$};
\node[fnode] (F1) at (-0.5,0) {$3$};
\node[fnode,magenta] (F2) at (0.2,-1.7) {$1$};
\node[gSUnode] (G1) at (1.5,0) {$2N$};
\node[fnode,ForestGreen] (F3) at (2.4,-1.7) {$2$};
\node[gSUnode] (G2) at (3.4,0) {$2N$};
\node[fnode] (F4) at (4.2,-1.7) {$2$};
\node (G3) at (5.1,0) {$\dots$};
\node[fnode,YellowGreen] (F5) at (5.8,-1.7) {$2$};
\node[gSUnode] (G4) at (6.8,0) {$2N$};
\node[fnode,gray] (F6) at (8.1,-1.7) {$1$};
\node[fnode,Orchid] (F7) at (8.8,0) {$3$};
\draw (F1) -- pic[pos=0.7,sloped]{arrow} (G1) -- pic[pos=0.7,sloped]{arrow} (G2) -- pic[pos=0.7,sloped]{arrow} (G3) -- pic[pos=0.7,sloped]{arrow} (G4) -- pic[pos=0.7,sloped]{arrow} (F7);
\draw (G1) -- pic[pos=0.4,sloped,very thick]{arrow=latex reversed} (F2);
\draw (G1) -- pic[pos=0.5,sloped,very thick]{arrow=latex reversed} (F3);
\draw (G2) -- pic[pos=0.5,sloped,very thick]{arrow=latex reversed} (F3);
\draw (G2) -- pic[pos=0.5,sloped,very thick]{arrow=latex reversed} (F4);
\draw (G3) -- pic[pos=0.4,sloped,very thick]{arrow=latex reversed} (F4);
\draw (G3) -- pic[pos=0.5,sloped,very thick]{arrow=latex reversed} (F5);
\draw (G4) -- pic[pos=0.5,sloped,very thick]{arrow=latex reversed} (F5);
\draw (G4) -- pic[pos=0.5,sloped,very thick]{arrow=latex reversed} (F6);
\draw (1.9,0.3) to[out=90,in=0] pic[pos=0.2,sloped]{arrow} (1.5,0.9) to[out=180,in=90] pic[pos=0.5,sloped,very thick]{arrow=latex reversed} (1.1,0.3);
\draw (7.2,0.3) to[out=90,in=0] pic[pos=0.4,sloped,very thick]{arrow=latex reversed} (6.8,0.9) to[out=180,in=90] pic[pos=0.3,sloped]{arrow} (6.4,0.3);
\node at (0.3,0.4) {$L$};
\node at (0.4,-0.8) {$F_l$};
\node at (2,1.1) {$A_l$};
\node at (2.5,0.4) {$B_1$};
\node at (4.4,0.4) {$B_2$};
\node at (5.6,0.4) {$B_{m-1}$};
\node at (7.3,1.1) {$A_r$};
\node at (8,0.4) {$R$};
\node at (7.9,-0.8) {$F_r$};
\node at (1.7,-1.1) {$V_1$};
\node at (3.2,-1.1) {$D_1$};
\node at (6.8,-1.1) {$D_{m-1}$};
\end{scope}
\epic} \ee
\be \label{4dGeneralFamilyIIEven2} \scalebox{0.9}{\bpic[node distance=2cm,gSUnode/.style={circle,red,draw,minimum size=8mm},gUSpnode/.style={circle,blue,draw,minimum size=8mm},fnode/.style={rectangle,red,draw,minimum size=8mm}]
\begin{scope}[shift={(0,0)}]
\node at (-0.5,2.2) {$\tsup[1]{\star}_2)$};
\node[fnode] (F1) at (0,0) {$4$};
\node[gUSpnode] (G1) at (1.5,0) {$2N$};
\node[fnode,Violet] (F2) at (2.3,-1.7) {$2$};
\node[gSUnode] (G2) at (3.2,0) {$2N$};
\node[fnode,ForestGreen] (F4) at (4.2,-1.7) {$2$};
\node[gSUnode] (G3) at (5.2,0) {$2N$};
\node[fnode] (F5) at (6.2,-1.7) {$2$};
\node (G4) at (7.2,0) {$\dots$};
\node[fnode] (F6) at (8.2,-1.7) {$2$};
\node[gSUnode] (G5) at (9.2,0) {$2N$};
\node[fnode,YellowGreen] (F7) at (10.2,-1.7) {$2$};
\node[gSUnode] (G6) at (11.2,0) {$2N$};
\node[fnode,gray] (F8) at (11.2,1.9) {$2$};
\node[fnode,Brown] (F9) at (12,-1.7) {$2$};
\node[gUSpnode] (G7) at (13.2,0) {\scalebox{0.8}{$2N-2$}};
\node[fnode,Orchid] (F10) at (14.9,0) {$2$};
\draw (F1) -- pic[pos=0.7,sloped]{arrow} (G1) -- pic[pos=0.7,sloped]{arrow} (G2) -- pic[pos=0.7,sloped]{arrow} (G3) -- pic[pos=0.7,sloped]{arrow} (G4) -- pic[pos=0.7,sloped]{arrow} (G5) -- pic[pos=0.7,sloped]{arrow} (G6) -- pic[pos=0.7,sloped]{arrow} (G7) -- pic[pos=0.7,sloped]{arrow} (F10);
\draw (G6) -- pic[pos=0.7,sloped]{arrow} (F8);
\draw (G1) -- (F2);
\draw (G2) -- pic[pos=0.5,sloped,very thick]{arrow=latex reversed} (F2);
\draw (G2) -- pic[pos=0.5,sloped,very thick]{arrow=latex reversed} (F4);
\draw (G3) -- pic[pos=0.5,sloped,very thick]{arrow=latex reversed} (F4);
\draw (G3) -- pic[pos=0.4,sloped,very thick]{arrow=latex reversed} (F5);
\draw (G4) -- pic[pos=0.5,sloped,very thick]{arrow=latex reversed} (F5);
\draw (G4) -- pic[pos=0.5,sloped,very thick]{arrow=latex reversed} (F6);
\draw (G5) -- pic[pos=0.5,sloped,very thick]{arrow=latex reversed} (F6);
\draw (G5) -- pic[pos=0.5,sloped,very thick]{arrow=latex reversed} (F7);
\draw (G6) -- pic[pos=0.5,sloped,very thick]{arrow=latex reversed} (F7);
\draw (G6) -- pic[pos=0.5,sloped,very thick]{arrow=latex reversed} (F9);
\draw (G7) -- (F9);
\node at (0.8,0.4) {$\Lt$};
\node at (2.3,0.4) {$\Ut_l$};
\node at (4.2,0.4) {$\Bt_1$};
\node at (6.2,0.4) {$\Bt_2$};
\node at (10.2,0.6) {$\Bt_{m-2}$};
\node at (10.8,1.1) {$\Ft_r$};
\node at (12.1,0.4) {$\Ut_r$};
\node at (14.2,0.4) {$\Rt$};
\node at (1.6,-1) {$\Vt_l$};
\node at (3.1,-1) {$\Dt_l$};
\node at (4.1,-1) {$\Vt_1$};
\node at (5,-1) {$\Dt_1$};
\node at (9.4,-1.1) {$\Vt_{m-2}$};
\node at (11.3,-1) {$\Vt_r$};
\node at (12.9,-1) {$\Dt_r$};
\end{scope}
\epic} \ee
Without the flippers, these two theories are \emph{not} dual to each other.
\subsubsection*{Strategy to get the set of flippers}\label{flippers2}
Once again in order to find the correct set of flippers we do a similar procedure as in the odd case. We first put the two theories $\tsup[1]{\star}_1)$ and $\tsup[1]{\star}_2)$ in a simpler form. It means that we do to each theories the following set of manipulations.

\noindent Starting with $\tsup[1]{\star}_1)$:
\begin{itemize}
\item deconfinement of the two antisymmetric
\item m Seiberg dualities on the m SU nodes
\item CSST duality on the left SU(2)
\item m-2 confinements
\end{itemize}
We end up with a frame similar to \eqref{4dGeneralFamilyIIOdd3}. 

\noindent Starting with $\tsup[1]{\star}_2)$:
\begin{itemize}
\item m-1 Seiberg dualities on the m-1 SU nodes
\item CSST duality on the left SU(2)
\item m-3 confinements
\end{itemize}
We end up with a frame similar to \eqref{4dGeneralFamilyIIOdd4}. 

Then we arrange the two resulting theories by a flipping procedure equivalent to the one after \eqref{4dGeneralFamilyIIOdd4}. We are lead to consider the following theories
\be \label{4dGeneralFamilyIIEven3} \scalebox{0.85}{\bpic[node distance=2cm,gSUnode/.style={circle,red,draw,minimum size=8mm},gUSpnode/.style={circle,blue,draw,minimum size=8mm},fnode/.style={rectangle,red,draw,minimum size=8mm}]
\begin{scope}[shift={(0,0)}]
\node at (-3.8,5) {$\tsup{\star}_1)$};
\node[gSUnode] (G1) at (0,0) {$2$};
\node[gSUnode] (G2) at (0,-2) {$2$};
\node[gUSpnode,MidnightBlue] (G3) at (-2,2) {\scalebox{0.8}{$2N-2$}};
\node[gUSpnode] (G4) at (2,2) {\scalebox{0.8}{$2N-2$}};
\node[fnode,Orchid] (F1) at (-2,0) {$2$};
\node[fnode] (F2) at (2,0) {$2$};
\node[fnode,magenta] (F3) at (-2,-2) {$1$};
\node[fnode,gray] (F4) at (2,-2) {$1$};
\node[fnode] (F5) at (0,-4) {$2m$};
\node[fnode] (F6) at (-2,4.2) {$1$};
\node[fnode] (F7) at (2,4.2) {$1$};
\draw (G1) -- (G2) -- (F5);
\draw (G3) -- (G1) -- (G4);
\draw (G3) -- (G4);
\draw (G3) -- (F1) -- (G1) -- (F2) -- (G4); 
\draw (F3) -- (G2) -- (F4);
\draw (G3) -- (F6);
\draw (G4) -- (F7);
\draw (F3.west) to[out=135,in=-135] (G3.west);
\draw (F4.east) to[out=45,in=-45] (G4.east);
\node[right] at (-2.7,-5.5) {$ \cW= 2 \, \quartic + 3 \, \triangles$};
\node at (-1,-0.3) {$\rt$};
\node at (1,-0.3) {$r$};
\node at (-0.7,1) {$\ct$};
\node at (0.7,1) {$c$};
\node at (-1,-1.7) {$s$};
\node at (1,-1.7) {$\tilde{s}$};
\node at (-3.8,0) {$\nt^{(2)}$};
\node at (3.9,0) {$n^{(2)}$};
\node at (-2.4,0.9) {$\nt^{(1)}$};
\node at (2.5,0.9) {$n^{(1)}$};
\node at (0,2.3) {$M$};
\node at (0.2,-1.1) {$b$};
\node at (0.2,-3) {$f$};
\end{scope}
\begin{scope}[shift={(9,0)}]
\node at (-3.5,5) {$\tsup{\star}_2)$};
\node[gSUnode] (G1) at (0,0) {$2$};
\node[gSUnode] (G2) at (0,-2) {$2$};
\node[gUSpnode] (G3) at (-1.5,2) {\scalebox{0.8}{$2N-2$}};
\node[gUSpnode] (G4) at (1.5,2) {$2N$};
\node[fnode,Violet] (F1) at (-1.5,4.2) {$2$};
\node[fnode] (F2) at (1.5,4.2) {$4$};
\node[fnode,Brown] (F4) at (2,0) {$2$};
\node[fnode] (F5) at (0,-4) {$2m$};
\draw (G1) -- (G2) -- (F5);
\draw (G3) -- (G1) -- (G4);
\draw (G3) -- (G4);
\draw (G3) -- pic[pos=0.5,sloped,very thick]{arrow=latex reversed} (F1);
\draw (G4) -- pic[pos=0.7,sloped,very thick]{arrow=latex reversed} (F2);
\draw (G1) -- (F4) -- (G4);
\node[right] at (-1.3,-5.5) {$ \cW= 3 \, \triangles $};
\node at (-2,3.2) {$R$};
\node at (2.1,3.2) {$L^{(1)}$};
\node at (-0.4,1) {$c$};
\node at (0.5,1) {$r$};
\node at (1,-0.3) {$t$};
\node at (2.3,1) {$L^{(2)}$};
\node at (0,2.3) {$B$};
\node at (0.2,-1.1) {$\tilde{f}$};
\node at (0.2,-3) {$f$};
\end{scope}
\epic} \ee
Once again at this stage $\tsup{\star}_1)$ and $\tsup{\star}_2)$ are \emph{not} dual, it misses the set of flippers in both sides. In the odd case, in order to make progress at this stage we studied the $N=1$ case. It allowed us to come up with the educated guess \eqref{4dGeneralFamilyIIOdd10} for generic $N$. This educated guess turned out to be correct because it passes the non-trivial checks of matching the central charges as well as 't Hooft anomalies. Now, for the even case we consider a different procedure to obtain an educated guess. We do the following steps:
\begin{itemize}
\item Start with the theory with no flipper
\item Compute the R-charges of all the chiral ring generators
\item Flip all operators with R-charge \emph{less} than $1$
\item Compute again all R-charges
\item Flip additional chiral ring generators with R-charge \emph{less} than $1$ if present
\item Repeat this procedure until reaching a frame with only chiral ring generators with R-charge \emph{bigger} than $1$
\end{itemize}
After applying these algorithm to \eqref{4dGeneralFamilyIIEven3} we obtain
\be \label{4dGeneralFamilyIIEven5} \scalebox{0.85}{\bpic[node distance=2cm,gSUnode/.style={circle,red,draw,minimum size=8mm},gUSpnode/.style={circle,blue,draw,minimum size=8mm},fnode/.style={rectangle,red,draw,minimum size=8mm}]
\begin{scope}[shift={(0,0)}]
\node at (-3.8,5) {$\star_1)$};
\node[gSUnode] (G1) at (0,0) {$2$};
\node[gSUnode] (G2) at (0,-2) {$2$};
\node[gUSpnode,MidnightBlue] (G3) at (-2,2) {\scalebox{0.8}{$2N-2$}};
\node[gUSpnode] (G4) at (2,2) {\scalebox{0.8}{$2N-2$}};
\node[fnode,Orchid] (F1) at (-2,0) {$2$};
\node[fnode] (F2) at (2,0) {$2$};
\node[fnode,magenta] (F3) at (-2,-2) {$1$};
\node[fnode,gray] (F4) at (2,-2) {$1$};
\node[fnode] (F5) at (0,-4) {$2m$};
\node[fnode] (F6) at (-2,4.2) {$1$};
\node[fnode] (F7) at (2,4.2) {$1$};
\draw (G1) -- (G2) -- (F5);
\draw (G3) -- (G1) -- (G4);
\draw (G3) -- (G4);
\draw (G3) -- (F1) -- (G1) -- (F2) -- (G4); 
\draw (F3) -- (G2) -- (F4);
\draw (G3) -- (F6);
\draw (G4) -- (F7);
\draw (F3.west) to[out=135,in=-135] (G3.west);
\draw (F4.east) to[out=45,in=-45] (G4.east);
\node[right] at (-4.2,-5.5) {$ \cW= 2 \, \quartic + 3 \, \triangles + \Flip[b \, b]$};
\node[right] at (-4.2,-6.8) {$+ \, \displaystyle\sum_{a=0}^{\left\lfloor \frac{N-4}{2} \right\rfloor} \, \Flip[n^{(1)} \, (M \, M)^a \, \nt^{(1)}]$};
\node[right] at (-4.2,-8.6) {$+ \, \displaystyle\sum_{b=0}^{\left\lfloor \frac{N-3}{2} \right\rfloor} \, \Flip[n^{(1)} \, (M \, M)^b \, n^{(1)}; \nt^{(1)} \, (M \, M)^b \, \nt^{(1)};$};
\node[right] at (-4.2,-9.8) {$l \, M (M \, M)^b \, n^{(1)}; \lt \, M (M \, M)^b \, \nt^{(1)}; n^{(1)} \, M (M \, M)^b \, \nt^{(2)};$};
\node[right] at (-4.2,-11.2) {$\nt^{(1)} \, M (M \, M)^b \, n^{(2)}] + \displaystyle\sum_{c=0}^{\left\lfloor \frac{N-2}{2} \right\rfloor} \, \Flip[n^{(1)} \, (M \, M)^c \, n^{(2)};$};
\node[right] at (-4.2,-12.6) {$\nt^{(1)} \, (M \, M)^c \, \nt^{(2)}; l \, M (M \, M)^c \, \nt^{(2)}; \lt \, M (M \, M)^c \, n^{(2)};$};
\node[right] at (-4.2,-13.8) {$n^{(1)} \, (M \, M)^c \, \lt; \nt^{(1)} \, (M \, M)^c \, l; l \, M (M \, M)^c \, \lt;$};
\node[right] at (-4.2,-15.2) {$n^{(2)} \, M (M \, M)^c \, \nt^{(2)}] + \displaystyle\sum_{d=0}^{\left\lfloor \frac{N-1}{2} \right\rfloor} \, \Flip[n^{(2)} \, (M \, M)^c \, \lt;$};
\node[right] at (-4.2,-16.5) {$\nt^{(2)} \, (M \, M)^c \, l]$};
\node at (-1,-0.3) {$\rt$};
\node at (1,-0.3) {$r$};
\node at (-0.7,1) {$\ct$};
\node at (0.7,1) {$c$};
\node at (-1,-1.7) {$s$};
\node at (1,-1.7) {$\tilde{s}$};
\node at (-3.8,0) {$\nt^{(2)}$};
\node at (3.9,0) {$n^{(2)}$};
\node at (-2.4,0.9) {$\nt^{(1)}$};
\node at (2.5,0.9) {$n^{(1)}$};
\node at (0,2.3) {$M$};
\node at (0.2,-1.1) {$b$};
\node at (0.2,-3) {$f$};
\end{scope}
\begin{scope}[shift={(9,0)}]
\node at (-3.5,5) {$\star_2)$};
\node[gSUnode] (G1) at (0,0) {$2$};
\node[gSUnode] (G2) at (0,-2) {$2$};
\node[gUSpnode] (G3) at (-1.5,2) {\scalebox{0.8}{$2N-2$}};
\node[gUSpnode] (G4) at (1.5,2) {$2N$};
\node[fnode,Violet] (F1) at (-1.5,4.2) {$2$};
\node[fnode] (F2) at (1.5,4.2) {$4$};
\node[fnode,Brown] (F4) at (2,0) {$2$};
\node[fnode] (F5) at (0,-4) {$2m$};
\draw (G1) -- (G2) -- (F5);
\draw (G3) -- (G1) -- (G4);
\draw (G3) -- (G4);
\draw (G3) -- pic[pos=0.5,sloped,very thick]{arrow=latex reversed} (F1);
\draw (G4) -- pic[pos=0.7,sloped,very thick]{arrow=latex reversed} (F2);
\draw (G1) -- (F4) -- (G4);
\node[right] at (-2,-5.5) {$ \cW= 3 \, \triangles + \Flip[\tilde{f} \, \tilde{f}]$};
\node[right] at (-2,-7) {$+ \, \displaystyle\sum_{a=0}^{\left\lfloor \frac{N-3}{2} \right\rfloor} \, \Flip[L^{(2)} \, (B \, B)^a \, L^{(2)};$};
\node[right] at (-0.8,-8.3) {$L^{(2)} \, B (B \, B)^a \, R] +$};
\node[right] at (-2,-9.6) {$\displaystyle\sum_{b=0}^{\left\lfloor \frac{N-2}{2} \right\rfloor} \, \Flip[L^{(1)} \, (B \, B)^b \, L^{(2)};$};
\node[right] at (-0.8,-11) {$L^{(1)} \, B (B \, B)^b \, R] +$};
\node[right] at (-2,-12.3) {$\displaystyle\sum_{c=0}^{\left\lfloor \frac{N-1}{2} \right\rfloor} \, \Flip[L^{(1)} \, (B \, B)^c \, L^{(1)};$};
\node[right] at (-0.8,-13.6) {$R \, (B \, B)^c \, R]$};
\node at (-2,3.2) {$R$};
\node at (2.1,3.2) {$L^{(1)}$};
\node at (-0.4,1) {$c$};
\node at (0.5,1) {$r$};
\node at (1,-0.3) {$t$};
\node at (2.3,1) {$L^{(2)}$};
\node at (0,2.3) {$B$};
\node at (0.2,-1.1) {$\tilde{f}$};
\node at (0.2,-3) {$f$};
\end{scope}
\epic} \ee
We claim that $\star_1)$ and $\star_2)$ in \eqref{4dGeneralFamilyIIEven5} are dual. For generic $N$ and $m$, we do not have a proof of this statement. The non-trivial check of the claim is the matching of the central charges with $a$-maximization.

\section{Higgsing $R_{N,k}$}\label{higgsing}
In the last section, we start the study of Higgsing of the $5d$ $R_{N,k}$ theories \eqref{RNk}.
\subsection{5d UV duality}
\begin{figure}[H]
\scalebox{0.9}{\bpic[node distance=2cm,gSUnode/.style={circle,red,draw,minimum size=8mm},gUSpnode/.style={circle,blue,draw,minimum size=8mm},fnode/.style={rectangle,draw,minimum size=8mm}]
\begin{scope}[shift={(-5,0)}] 
\node[inner sep=0pt] at (0,0) {\includegraphics[height=.25\textheight]{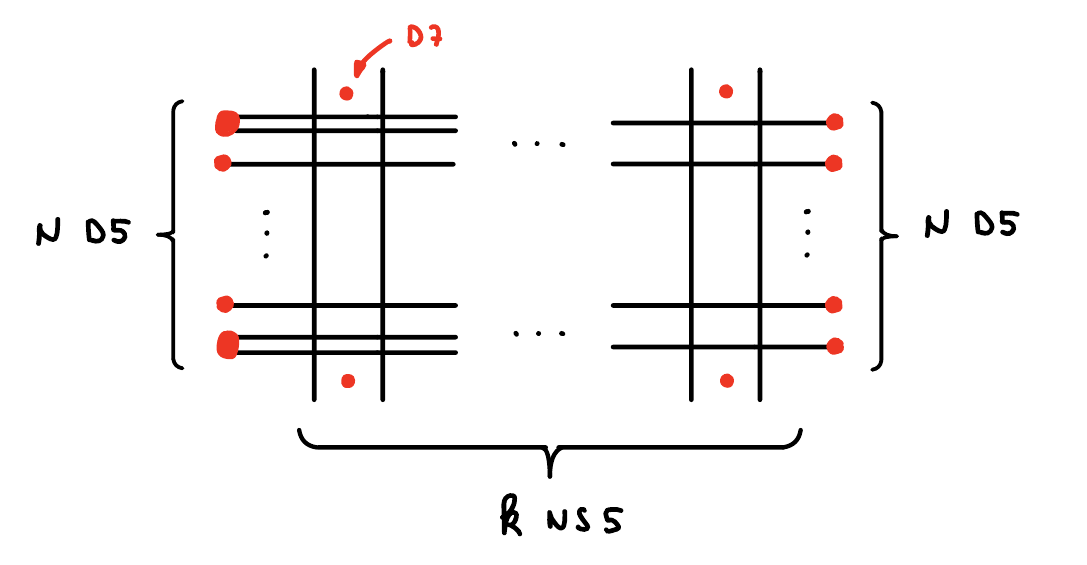}};
\end{scope}
\begin{scope}[shift={(4.5,0)}] 
\node[inner sep=0pt] at (0,0) {\includegraphics[height=.25\textheight]{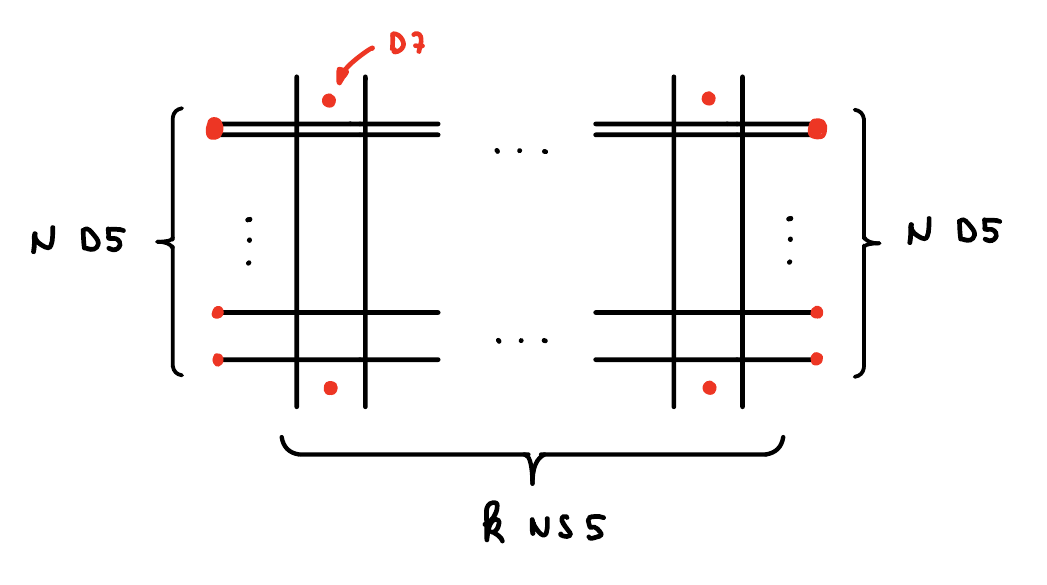}};
\end{scope}
\epic}
\caption{Brane setup for \eqref{HiggsingRNk1} on the left and for \eqref{HiggsingRNk2} on the right.}
\label{HiggsingRNk}
\end{figure}

Concretely we will study two different Higgsing in $5d$ that is mapped to the same deformation of the $6d$ SCFT. Therefore we are left with another example of $5d$ UV duality. The question that we can ask: does the $5d$ UV duality that we obtain after the Higgsing procedure leads to another $4d$ IR duality? We do not have a general answer to this question but we will study the simplest Higgsing and the answer will turn out to be positive. At the level of the brane systems, the Higgsing is manifested by breaking 5-branes on the same 7-brane \cite{Benini:2009gi}. The example of Higgsing that we consider is the following. We start with the brane web on the left of Figure \ref{TypeIIB5dGeneralFamilyI1} and force two pairs of 5-branes to end on the same 7-brane. We have the choice to take the two pairs either on the same side of the brane web or the opposite side. We obtain the brane systems of Figure \ref{HiggsingRNk} and the gauge theories associated are shown in \eqref{HiggsingRNk1}-\eqref{HiggsingRNk2}. The details of this example can be found in \cite{Zafrir:2015rga}. 

\be \label{HiggsingRNk1} \scalebox{0.9}{\bpic[node distance=2cm,gSUnode/.style={circle,red,draw,minimum size=8mm},gUSpnode/.style={circle,blue,draw,minimum size=8mm},fnode/.style={rectangle,red,draw,minimum size=8mm}]  
\begin{scope}[shift={(0,0)}]
\node at (-0.5,2) {$\star_1)$};
\node[fnode] (F1) at (-0.3,0) {$N-2$};
\node[gSUnode] (G1) at (1.5,0) {\scalebox{0.8}{$N-2$}};
\node[gSUnode] (G2) at (3.2,0) {$N$};
\node[fnode] (F2) at (3.2,1.7) {$2$};
\node[gSUnode] (G3) at (5.2,0) {$N$};
\node (G4) at (7.2,0) {$\dots$};
\node[gSUnode] (G5) at (9.2,0) {$N$};
\node[gSUnode] (G6) at (11.2,0) {$N$};
\node[fnode] (F3) at (12.9,0) {$N+2$};
\draw (F1) -- (G1) -- (G2) -- (G3) -- (G4) -- (G5) -- (G6) -- (F3);
\draw (G2) -- (F2);
\draw[decorate,decoration={calligraphic brace,mirror,amplitude=7pt}] (0.7,-0.8) -- (11.7,-0.8) node[pos=0.5,below=9pt,black] {$k-1$};
\end{scope}
\epic} \ee
\be \label{HiggsingRNk2} \scalebox{0.9}{\bpic[node distance=2cm,gSUnode/.style={circle,red,draw,minimum size=8mm},gUSpnode/.style={circle,blue,draw,minimum size=8mm},fnode/.style={rectangle,red,draw,minimum size=8mm}]  
\begin{scope}[shift={(0,0)}]
\node at (-0.5,2) {$\star_2)$};
\node[fnode] (F1) at (0,0) {$N$};
\node[gSUnode] (G1) at (1.5,0) {\scalebox{0.8}{$N-1$}};
\node[gSUnode] (G2) at (3.2,0) {$N$};
\node[fnode] (F2) at (3.2,1.7) {$1$};
\node[gSUnode] (G3) at (5.2,0) {$N$};
\node (G4) at (7.2,0) {$\dots$};
\node[gSUnode] (G5) at (9.2,0) {$N$};
\node[gSUnode] (G6) at (11.2,0) {$N$};
\node[fnode] (F3) at (11.2,1.7) {$1$};
\node[gSUnode] (G7) at (12.9,0) {\scalebox{0.8}{$N-1$}};
\node[fnode] (F4) at (14.6,0) {$N$};
\draw (F1) -- (G1) -- (G2) -- (G3) -- (G4) -- (G5) -- (G6) -- (G7) -- (F4);
\draw (G2) -- (F2);
\draw (G6) -- (F3);
\draw[decorate,decoration={calligraphic brace,mirror,amplitude=7pt}] (0.8,-0.8) -- (13.6,-0.8) node[pos=0.5,below=9pt,black] {$k-1$};
\end{scope}
\epic} \ee

\subsection{6d UV completion}
The $6d$ UV completion of the theories \eqref{HiggsingRNk1}-\eqref{HiggsingRNk2} depends on the parity of $k$ and can be obtain by doing the Higgsing at the level of the Type IIA brane setup corresponding to the $6d$ UV completion of $R_{N,k}$.

\subsection*{$k$ even: $k=2l$}
In this case, the $6d$ completion is given by the following Type IIA brane setup \cite{Hayashi:2015zka,Zafrir:2015rga}: 
\begin{figure}[H]
\centering
\includegraphics[height=0.25\textheight]{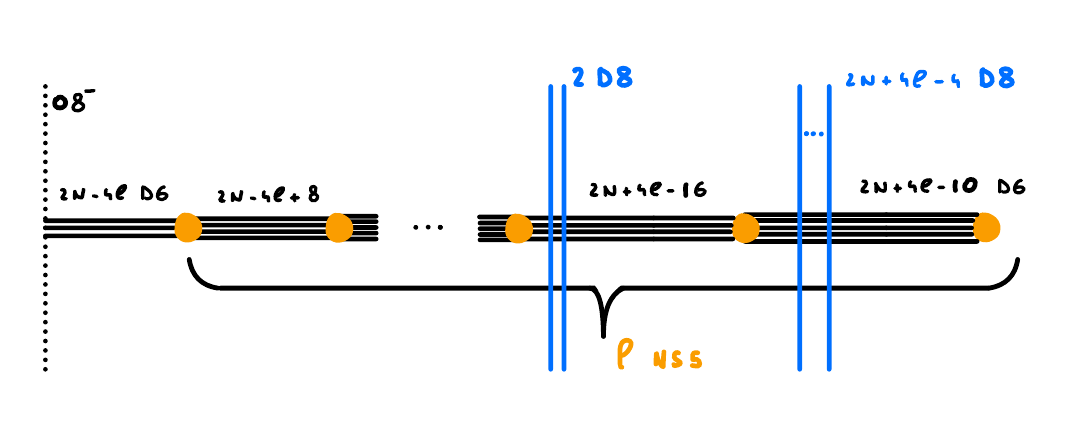}
\caption{Type IIA brane setup corresponding to the $6d$ UV completion of \eqref{HiggsingRNk1}-\eqref{HiggsingRNk2} for $k=2l$. It is obtained by Higgsing the Type IIA brane system \eqref{TypeIIA6dCompletionGeneralFamilyIkEvenFig} corresponding to the UV completion of $R_{N,2l}$.}
\label{TypeIIA6dCompletionHiggsingEvenFig}
\end{figure}
The gauge theory corresponding to this brane system is a linear quiver with one $USp$ gauge node and $l-1$ $SU$ gauge nodes: 
\be \label{TypeIIA6dCompletionHiggsingEven} \scalebox{0.9}{\bpic[node distance=2cm,gSUnode/.style={circle,red,draw,minimum size=8mm},gUSpnode/.style={circle,blue,draw,minimum size=8mm},fnode/.style={rectangle,red,draw,minimum size=8mm}]   
\node[gUSpnode] (G1) at (-4,0) {\scalebox{0.9}{$2N-4l$}};
\node[gSUnode] (G2) at (-1.5,0) {\scalebox{0.8}{$2N-4l+8$}};
\node[gSUnode] (G3) at (1.5,0) {\scalebox{0.8}{$2N-4l+16$}};
\node (G4) at (3.5,0) {$\dots$};
\node[gSUnode] (G5) at (5.5,0) {\scalebox{0.8}{$2N+4l-16$}};
\node[gSUnode] (G6) at (8.5,0) {\scalebox{0.8}{$2N+4l-10$}};
\node[fnode] (F1) at (11,0) {\scalebox{0.8}{$2N+4l-4$}};
\node[fnode] (F2) at (5.5,2) {$2$};
\draw (G1) -- (G2) -- (G3) -- (G4) -- (G5) -- (G6) -- (F1);
\draw (G5) -- (F2);
\epic} \ee
\subsection*{$k$ odd: $k=2l+1$}
The $6d$ completion is given by the following Type IIA brane setup \cite{Hayashi:2015zka,Zafrir:2015rga}: 
\begin{figure}[H]
\centering
\includegraphics[height=0.25\textheight]{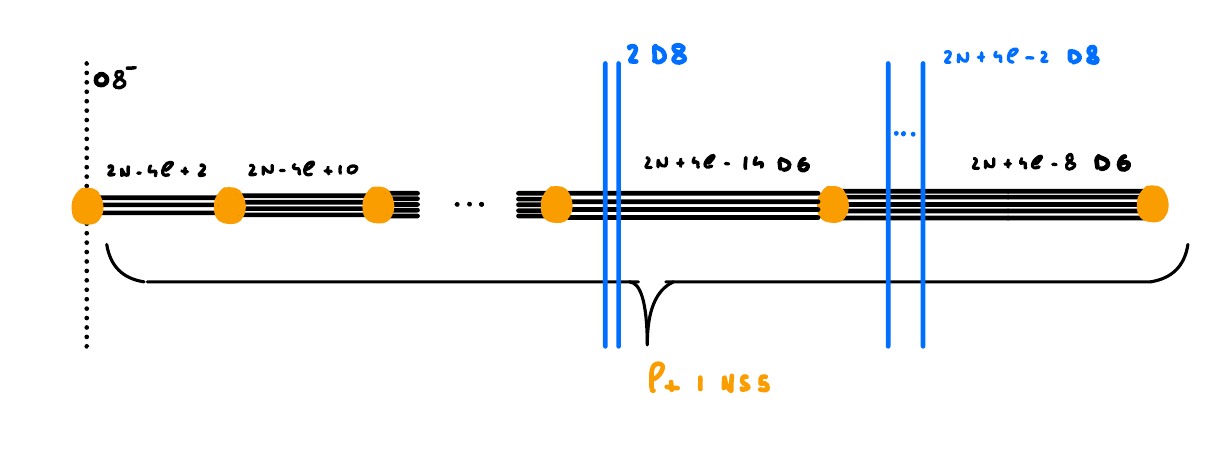}
\caption{Type IIA brane setup corresponding to the $6d$ UV completion of \eqref{HiggsingRNk1}-\eqref{HiggsingRNk2} for $k=2l+1$. It is obtained by Higgsing the Type IIA brane system \eqref{TypeIIA6dCompletionGeneralFamilyIkOddFig} corresponding to the UV completion of $R_{N,2l+1}$.}
\label{TypeIIA6dCompletionHiggsingOddFig}
\end{figure}
The gauge theory corresponding to this brane system is a linear quiver with $l$ $SU$ gauge nodes and an antisymmetric hyper attached to the first node: 
\be \label{TypeIIA6dCompletionHiggsingOdd} \scalebox{0.9}{\bpic[node distance=2cm,gSUnode/.style={circle,red,draw,minimum size=8mm},gUSpnode/.style={circle,blue,draw,minimum size=8mm},fnode/.style={rectangle,red,draw,minimum size=8mm}]    
\node[gSUnode] (G1) at (-4,0) {\scalebox{0.8}{$2N-4l+2$}};
\node[gSUnode] (G2) at (-1.5,0) {\scalebox{0.8}{$2N-4l+10$}};
\node[gSUnode] (G3) at (1.5,0) {\scalebox{0.8}{$2N-4l+18$}};
\node (G4) at (3.5,0) {$\dots$};
\node[gSUnode] (G5) at (5.5,0) {\scalebox{0.8}{$2N+4l-14$}};
\node[gSUnode] (G6) at (8.5,0) {\scalebox{0.8}{$2N+4l-8$}};
\node[fnode] (F1) at (11,0) {\scalebox{0.8}{$2N+4l-2$}};
\node[fnode] (F2) at (5.5,2) {$2$};
\draw (G1) -- (G2) -- (G3) -- (G4) -- (G5) -- (G6) -- (F1);
\draw (G5) -- (F2);
\draw (-3.5,0.9) to[out=90,in=0]  (-4,1.5) to[out=180,in=90] (-4.5,0.9);
\epic} \ee

\subsection{4d duality}
Applying our procedure to the $5d$ UV duality \eqref{HiggsingRNk1}-\eqref{HiggsingRNk1} we produce the following $4d$ $\cN=1$ theories
\be \label{4dHiggsingRNk1} \scalebox{0.9}{\bpic[node distance=2cm,gSUnode/.style={circle,red,draw,minimum size=8mm},gUSpnode/.style={circle,blue,draw,minimum size=8mm},fnode/.style={rectangle,red,draw,minimum size=8mm}]  
\begin{scope}[shift={(0,0)}]
\node at (-0.5,1.5) {$\star_1)$};
\node[fnode] (F1) at (-0.3,0) {$N-2$};
\node[gSUnode] (G1) at (1.5,0) {\scalebox{0.8}{$N-2$}};
\node[fnode] (F2) at (2.4,-1.7) {$2$};
\node[gSUnode] (G2) at (3.2,0) {$N$};
\node[fnode] (F3) at (3.2,1.7) {$2$};
\node[fnode] (F4) at (4.2,-1.7) {$2$};
\node[gSUnode] (G3) at (5.2,0) {$N$};
\node[fnode] (F5) at (6.2,-1.7) {$2$};
\node (G4) at (7.2,0) {$\dots$};
\node[fnode] (F6) at (8.2,-1.7) {$2$};
\node[gSUnode] (G5) at (9.2,0) {$N$};
\node[fnode] (F7) at (10.2,-1.7) {$2$};
\node[gSUnode] (G6) at (11.2,0) {$N$};
\node[fnode] (F8) at (11.2,1.7) {$2$};
\node[fnode] (F9) at (12.9,0) {$N$};
\draw (F1) --pic[pos=0.7,sloped]{arrow} (G1) --pic[pos=0.7,sloped]{arrow} (G2) --pic[pos=0.7,sloped]{arrow} (G3) --pic[pos=0.7,sloped]{arrow} (G4) --pic[pos=0.7,sloped]{arrow} (G5) --pic[pos=0.7,sloped]{arrow} (G6) --pic[pos=0.7,sloped]{arrow} (F9);
\draw (G2) --pic[pos=0.5,sloped,very thick]{arrow=latex reversed} (F3);
\draw (G6) --pic[pos=0.5,sloped,very thick]{arrow=latex reversed} (F8);
\draw (G1) -- pic[pos=0.5,sloped,very thick]{arrow=latex reversed} (F2);
\draw (G2) -- pic[pos=0.5,sloped,very thick]{arrow=latex reversed} (F2);
\draw (G2) -- pic[pos=0.5,sloped,very thick]{arrow=latex reversed} (F4);
\draw (G3) -- pic[pos=0.5,sloped,very thick]{arrow=latex reversed} (F4);
\draw (G3) -- pic[pos=0.5,sloped,very thick]{arrow=latex reversed} (F5);
\draw (G4) -- pic[pos=0.5,sloped,very thick]{arrow=latex reversed} (F5);
\draw (G4) -- pic[pos=0.5,sloped,very thick]{arrow=latex reversed} (F6);
\draw (G5) -- pic[pos=0.5,sloped,very thick]{arrow=latex reversed} (F6);
\draw (G5) -- pic[pos=0.5,sloped,very thick]{arrow=latex reversed} (F7);
\draw (G6) -- pic[pos=0.5,sloped,very thick]{arrow=latex reversed} (F7);
\end{scope}
\epic} \ee
\be \label{4dHiggsingRNk2} \scalebox{0.9}{\bpic[node distance=2cm,gSUnode/.style={circle,red,draw,minimum size=8mm},gUSpnode/.style={circle,blue,draw,minimum size=8mm},fnode/.style={rectangle,red,draw,minimum size=8mm}]  
\begin{scope}[shift={(0,0)}]
\node at (-0.5,1.5) {$\star_2)$};
\node[fnode] (F1) at (-0.3,0) {$N-1$};
\node[gSUnode] (G1) at (1.5,0) {\scalebox{0.8}{$N-1$}};
\node[fnode] (F2) at (1.5,1.7) {$1$};
\node[fnode] (F3) at (2.4,-1.7) {$2$};
\node[gSUnode] (G2) at (3.2,0) {$N$};
\node[fnode] (F4) at (3.2,1.7) {$1$};
\node[fnode] (F5) at (4.2,-1.7) {$2$};
\node[gSUnode] (G3) at (5.2,0) {$N$};
\node[fnode] (F6) at (6.2,-1.7) {$2$};
\node (G4) at (7.2,0) {$\dots$};
\node[fnode] (F7) at (8.2,-1.7) {$2$};
\node[gSUnode] (G5) at (9.2,0) {$N$};
\node[fnode] (F8) at (10.2,-1.7) {$2$};
\node[gSUnode] (G6) at (11.2,0) {$N$};
\node[fnode] (F9) at (11.2,1.7) {$1$};
\node[fnode] (F10) at (12,-1.7) {$2$};
\node[gSUnode] (G7) at (12.9,0) {\scalebox{0.8}{$N-1$}};
\node[fnode] (F11) at (12.9,1.7) {$1$};
\node[fnode] (F12) at (14.7,0) {$N-1$};
\draw (F1) --pic[pos=0.7,sloped]{arrow} (G1) --pic[pos=0.7,sloped]{arrow} (G2) --pic[pos=0.7,sloped]{arrow} (G3) --pic[pos=0.7,sloped]{arrow} (G4) --pic[pos=0.7,sloped]{arrow} (G5) --pic[pos=0.7,sloped]{arrow} (G6) --pic[pos=0.7,sloped]{arrow} (G7) --pic[pos=0.7,sloped]{arrow} (F12);
\draw (G1) --pic[pos=0.7,sloped]{arrow} (F2);
\draw (G2) --pic[pos=0.6,sloped,very thick]{arrow=latex reversed} (F4);
\draw (G6) --pic[pos=0.7,sloped]{arrow} (F9);
\draw (G7) --pic[pos=0.6,sloped,very thick]{arrow=latex reversed} (F11);
\draw (G1) -- pic[pos=0.5,sloped,very thick]{arrow=latex reversed} (F3);
\draw (G2) -- pic[pos=0.5,sloped,very thick]{arrow=latex reversed} (F3);
\draw (G2) -- pic[pos=0.5,sloped,very thick]{arrow=latex reversed} (F5);
\draw (G3) -- pic[pos=0.5,sloped,very thick]{arrow=latex reversed} (F5);
\draw (G3) -- pic[pos=0.5,sloped,very thick]{arrow=latex reversed} (F6);
\draw (G4) -- pic[pos=0.5,sloped,very thick]{arrow=latex reversed} (F6);
\draw (G4) -- pic[pos=0.5,sloped,very thick]{arrow=latex reversed} (F7);
\draw (G5) -- pic[pos=0.5,sloped,very thick]{arrow=latex reversed} (F7);
\draw (G5) -- pic[pos=0.5,sloped,very thick]{arrow=latex reversed} (F8);
\draw (G6) -- pic[pos=0.5,sloped,very thick]{arrow=latex reversed} (F8);
\draw (G6) -- pic[pos=0.5,sloped,very thick]{arrow=latex reversed} (F10);
\draw (G7) -- pic[pos=0.5,sloped,very thick]{arrow=latex reversed} (F10);
\end{scope}
\epic} \ee
Let us remark that in order to get non-anomalous theories we were forced to split the flavor nodes at the edge of the quiver compared to the 5d avatars.

\subsection{Proof of the 4d duality}  
The proof is really similar to the section \ref{Proof4dGeneralFamilyI} so we will be brief. We start from either $\star_1)$ or $\star_2)$, and do the following operations:
\begin{itemize}
\item $k-1$ Seiberg dualities on the SU nodes from left to right
\item CSST duality on the left SU(2)
\item $k-3$ confinements
\end{itemize}
Finally, we introduce some flippers and both $\star_1)$ and $\star_2)$ take the same following form
\be \label{4dHiggsingRNk3} \scalebox{0.9}{\bpic[node distance=2cm,gSUnode/.style={circle,red,draw,minimum size=8mm},gUSpnode/.style={circle,blue,draw,minimum size=8mm},fnode/.style={rectangle,red,draw,minimum size=8mm}]    
\node[gSUnode] (G1) at (-2,0) {$2$};
\node[gSUnode] (G2) at (0,0) {$2$};
\node[fnode] (F1) at (-4,0) {$2k-2$};
\node[fnode] (F2) at (2,0) {$2N-2$};
\node[fnode] (F3) at (-1,-1.5) {$2$};
\draw (F1) -- (G1) -- (G2) -- (F2);
\draw (G1) -- (F3) -- (G2);
\node[right] at (-2.5,-2.5) {$ \cW= \Triangle$};
\epic} \ee
The fact that both $\star_1)$ and $\star_2)$ are dual (modulo flips) to the same theory \eqref{4dHiggsingRNk3} proves the duality.

\acknowledgments
We are grateful to Oren Bergman, Shlomo Razamat and Gabi Zafrir for various useful discussions.\\
Stephane Bajeot is partially supported by the INFN Research Project STEFI. Sergio Benvenuti is partially supported by the INFN Research Project GAST.

\appendix
\section{'t Hooft anomaly matching for the duality of section \ref{4ddualityToCheck}} \label{MatchingAnomalies}
In this appendix we present the matching of the t'Hooft anomalies for the duality \eqref{4dGeneralFamilyIIOdd10} that we report in \ref{4ddualityAppendix} for convenience. On both LHS and RHS of the duality, the theories have 4 $U(1)$'s global symmetries. The following charges assignments respect the constraints coming from ABJ anomalies and the superpotential terms. The matching of the t'Hooft anomalies are really non-trivial, especially the ones involving the $U(1)$'s symmetries, and relies on having the correct set of flipper fields.

\be \label{4ddualityAppendix} \scalebox{0.85}{\bpic[node distance=2cm,gSUnode/.style={circle,red,draw,minimum size=8mm},gUSpnode/.style={circle,blue,draw,minimum size=8mm},fnode/.style={rectangle,red,draw,minimum size=8mm}]
\begin{scope}[shift={(0,0)}]
\node at (-3.8,3) {$\star_1)$};
\node[gSUnode] (G1) at (0,0) {$2$};
\node[gSUnode] (G2) at (0,-2) {$2$};
\node[gUSpnode,MidnightBlue] (G3) at (-2,2) {\scalebox{0.8}{$2N-2$}};
\node[gUSpnode] (G4) at (2,2) {\scalebox{0.8}{$2N-2$}};
\node[fnode,Orchid] (F1) at (-2,0) {$3$};
\node[fnode] (F2) at (2,0) {$3$};
\node[fnode,magenta] (F3) at (-2,-2) {$1$};
\node[fnode,gray] (F4) at (2,-2) {$1$};
\node[fnode] (F5) at (0,-4) {$2m$};
\draw (G1) -- (G2) -- (F5);
\draw (G3) -- (G1) -- (G4);
\draw (G3) -- (G4);
\draw (G3) --pic[pos=0.7,sloped]{arrow} (F1) --pic[pos=0.7,sloped]{arrow} (G1) --pic[pos=0.7,sloped]{arrow} (F2) --pic[pos=0.7,sloped]{arrow} (G4); 
\draw (F3) -- (G2) -- (F4);
\draw (F3.west) to[out=135,in=-135] (G3.west);
\draw (F4.east) to[out=45,in=-45] (G4.east);
\node[right] at (-2.5,-5.5) {$ \cW= 2 \, \quartic + 3 \, \triangles$};
\node[right] at (-2.5,-6.5) {$+ \, \Flip[bb, c \ct, cb \st, \ct b s, bsb \st]$};
\node at (-1,-0.3) {$c$};
\node at (1,-0.3) {$\ct$};
\node at (-0.7,1) {$o$};
\node at (0.7,1) {$\ot$};
\node at (-1,-1.7) {$s$};
\node at (1,-1.7) {$\st$};
\node at (-3.7,0) {$u$};
\node at (3.7,0) {$\ut$};
\node at (-2.3,0.8) {$n$};
\node at (2.3,0.8) {$\nt$};
\node at (0,2.3) {$l$};
\node at (0.3,-1) {$b$};
\node at (0.2,-3) {$f$};
\end{scope}
\begin{scope}[shift={(9,0)}]
\node at (-4.5,3) {$\star_2)$};
\node[gSUnode] (G1) at (0,0) {$2$};
\node[gSUnode] (G2) at (0,-2) {$2$};
\node[gUSpnode,MidnightBlue] (G3) at (-1.5,2) {$2N$};
\node[gUSpnode] (G4) at (1.5,2) {$2N$};
\node[fnode] (F1) at (-3.2,2) {$3$};
\node[fnode,Orchid] (F2) at (3.2,2) {$3$};
\node[fnode,gray] (F3) at (-2,0) {$1$};
\node[fnode,magenta] (F4) at (2,0) {$1$};
\node[fnode] (F5) at (0,-4) {$2m$};
\draw (G1) -- (G2) -- (F5);
\draw (G3) -- (G1) -- (G4);
\draw (G3) -- (G4);
\draw (G3) -- pic[pos=0.5,sloped,very thick]{arrow=latex reversed} (F1);
\draw (G4) -- pic[pos=0.7,sloped,very thick]{arrow=latex reversed} (F2);
\draw (G3) -- (F3) -- (G1) -- (F4) -- (G4);
\node[right] at (-2.5,-5.5) {$ \cW= 3 \, \triangles + \, \Flip[ww]$};
\node[right] at (-3.8,-6.7) {$+ \, \displaystyle \sum_{i=o}^{N-1} \Flip[q (p p)^i q; \qt (p p)^i \qt; q (p p)^i d; \qt (p p)^i \dt; $};
\node[right] at (-2.5,-7.9) {$q p (p p)^i \qt; q p (p p)^i \dt; \qt p (p p)^i d; d p (p p)^i \dt]$};
\node at (-2.4,2.4) {$\qt$};
\node at (2.4,2.3) {$q$};
\node at (-0.5,1) {$k$};
\node at (0.5,1) {$\kt$};
\node at (-1,-0.3) {$\ttilde$};
\node at (1,-0.3) {$t$};
\node at (-2,1) {$\dt$};
\node at (2,1) {$d$};
\node at (0,2.3) {$p$};
\node at (0.3,-1) {$w$};
\node at (0.2,-3) {$f$};
\end{scope}
\epic} \ee

\begin{table}[H]
\centering
\begin{tabular}{|c|c|c|c|c|}
\hline
Fields LHS & $U(1)_1$ & $U(1)_2$ & $U(1)_3$ & $U(1)_4$ \\
\hline
$f$ & $0$ & $0$ & $\frac{1}{2m}$ & $0$ \\
\hline
$s$ & $0$ & $1$ & $-\frac{1}{2}$ & $-\frac{1}{2}$ \\
\hline
$\st$ & $0$ & $-1$ & $-\frac{1}{2}$ & $-\frac{1}{2}$ \\
\hline
$b$ & $0$ & $0$ & $0$ & $\frac{1}{2}$ \\
\hline
$u$ & $-1$ & $-2$ & $\frac{2N-1}{4N}$ & $-\frac{1}{4N}$ \\
\hline
$\ut$ & $1$ & $2$ & $\frac{2N-1}{4N}$ & $-\frac{1}{4N}$ \\
\hline
$c$ & $-\frac{2}{3}$ & $-1$ & $-\frac{N-1}{6N}$ & $-\frac{2N-1}{6N}$ \\
\hline
$n$ & $-\frac{1}{3}$ & $0$ & $\frac{1}{3} \frac{2N-5}{4N}$ & $\frac{4N-5}{12N}$ \\
\hline
$\nt$ & $\frac{1}{3}$ & $0$ & $\frac{1}{3} \frac{2N-5}{4N}$ & $\frac{4N-5}{12N}$ \\
\hline
$o$ & $1$ & $1$ & $\frac{1}{4N}$ & $\frac{1}{4N}$ \\
\hline
$\ot$ & $-1$ & $-1$ & $\frac{1}{4N}$ & $\frac{1}{4N}$ \\
\hline
$l$ & $0$ & $0$ & $-\frac{1}{2N}$ & $-\frac{1}{2N}$ \\
\hline
\end{tabular}
\caption{$U(1)$'s charges of the fields in LHS of \eqref{4ddualityAppendix}.}
\label{chargesLHS}
\end{table}

\begin{table}[H]
\centering
\begin{tabular}{|c|c|c|c|c|}
\hline
Fields RHS & $U(1)_1$ & $U(1)_2$ & $U(1)_3$ & $U(1)_4$ \\
\hline
$f$ & $0$ & $0$ & $\frac{1}{2m}$ & $0$ \\
\hline
$w$ & $0$ & $0$ & $-\frac{1}{2}$ & $0$ \\
\hline
$t$ & $0$ & $-1$ & $0$ & $-\frac{1}{2}$ \\
\hline
$\ttilde$ & $0$ & $1$ & $0$ & $-\frac{1}{2}$ \\
\hline
$d$ & $1$ & $2$ & $-\frac{1}{4N}$ & $\frac{2N-1}{4N}$ \\
\hline
$\dt$ & $-1$ & $-2$ & $-\frac{1}{4N}$ & $\frac{2N-1}{4N}$ \\
\hline
$k$ & $1$ & $1$ & $\frac{1}{4N}$ & $\frac{1}{4N}$ \\
\hline
$\kt$ & $-1$ & $-1$ & $\frac{1}{4N}$ & $\frac{1}{4N}$ \\
\hline
$q$ & $\frac{1}{3}$ & $0$ & $\frac{4N-1}{12N}$ & $\frac{2N-1}{12N}$ \\
\hline
$\qt$ & $-\frac{1}{3}$ & $0$ & $\frac{4N-1}{12N}$ & $\frac{2N-1}{12N}$ \\
\hline
$p$ & $0$ & $0$ & $-\frac{1}{2N}$ & $-\frac{1}{2N}$ \\
\hline
\end{tabular}
\caption{$U(1)$'s charges of the fields in RHS of \eqref{4ddualityAppendix}.}
\label{chargesRHS}
\end{table}

\noindent \textbf{'t Hooft anomalies involving non-abelian symmetries:}\\
\noindent $\bullet$ \textbf{LHS:}
\begin{align}
\tr(\textcolor{Orchid}{SU(3)^3}) &= (2N-2) A(\scalebox{0.4}{\ydiagram{1}}) + 2 A(\bar{\scalebox{0.4}{\ydiagram{1}}}) + 3 A(\scalebox{0.4}{\ydiagram{1}}) + A(\scalebox{0.4}{\ydiagram{1}}) = 2N \\
\tr(\textcolor{red}{SU(3)^3}) &= -2N \\
\tr(\textcolor{red}{SU(2m)^3}) &= 2
\end{align}

\noindent $\bullet$ \textbf{RHS:}
\begin{align}
\tr(\textcolor{Orchid}{SU(3)^3}) &= 2N A(\bar{\scalebox{0.4}{\ydiagram{1}}}) + N \left( A(\scalebox{0.4}{\ydiagram{1,1}}) + A(\scalebox{0.4}{\ydiagram{1}}) + 3 A(\scalebox{0.4}{\ydiagram{1}}) + A(\scalebox{0.4}{\ydiagram{1}}) \right) = 2N \\
\tr(\textcolor{red}{SU(3)^3}) &= -2N \\
\tr(\textcolor{red}{SU(2m)^3}) &= 2
\end{align}

In the previous equations, $A$ corresponds to the anomaly coefficient. In our normalization, it takes the following value for $SU(N)$: $A(\scalebox{0.4}{\ydiagram{1}}) = 1 = -A(\bar{\scalebox{0.4}{\ydiagram{1}}})$ and $A(\scalebox{0.4}{\ydiagram{1,1}}) = N-4$.

\noindent \textbf{'t Hooft anomalies involving abelian symmetries:}\\
\noindent $\bullet$ \textbf{LHS:}
\begin{align*}
\tr(\textcolor{Orchid}{SU(3)^2} U(1)_i) &= (2N-2) \, q^i_{n} \, \mu(\bar{\scalebox{0.4}{\ydiagram{1}}}) + 2 \, q^i_{c} \, \mu(\scalebox{0.4}{\ydiagram{1}}) - 3( q^i_c + q^i_{\ct}) \, \mu(\bar{\scalebox{0.4}{\ydiagram{1}}}) - (q^i_{c} + q^i_b + q^i_{\st}) \, \mu(\bar{\scalebox{0.4}{\ydiagram{1}}}) \\
i=1: &= -\frac{2N}{3} \\
i=2: &= 0 \\
i=3: &= \frac{2N+1}{6} \\
i=4: &= \frac{4N+1}{6}
\end{align*} 
\begin{align*}
\tr(\textcolor{red}{SU(3)^2} U(1)_i) &= (2N-2) \, q^i_{\nt} \, \mu(\bar{\scalebox{0.4}{\ydiagram{1}}}) + 2 \, q^i_{\ct} \, \mu(\scalebox{0.4}{\ydiagram{1}}) - 3( q^i_c + q^i_{\ct}) \, \mu(\bar{\scalebox{0.4}{\ydiagram{1}}}) - (q^i_{\ct} + q^i_b + q^i_s) \, \mu(\bar{\scalebox{0.4}{\ydiagram{1}}}) \\
i=1: &= \frac{2N}{3} \\
i=2: &= 0 \\
i=3: &= \frac{2N+1}{6} \\
i=4: &= \frac{4N+1}{6} 
\end{align*} 
In the previous equations, $\mu$ corresponds to the Dynkin index of the representation. In our normalization, it takes the following value for $SU(N)$: $\mu(\scalebox{0.4}{\ydiagram{1}}) = 1 = \mu(\bar{\scalebox{0.4}{\ydiagram{1}}})$ and $\mu(\scalebox{0.4}{\ydiagram{1,1}}) = N-2 = \mu(\bar{\scalebox{0.4}{\ydiagram{1,1}}})$.\\

Same kind of computations give for the linear anomalies:\\
\noindent $\tr(U(1)_1) = 0$ \\
\noindent $\tr(U(1)_2) = 0$ \\
\noindent $\tr(U(1)_3) = 2N+2$ \\
\noindent $\tr(U(1)_4) = 2N-1$ 

\noindent $\bullet$ \textbf{RHS:}
\begin{align*}
\tr(\textcolor{Orchid}{SU(3)^2} U(1)_i) &= (2N) \, q^i_{q} \, \mu(\bar{\scalebox{0.4}{\ydiagram{1}}}) - \sum_{j=0}^{N-1} \left[ (2 \, q^j_{q} + 2j \, q^j_p) \, \mu(\scalebox{0.4}{\ydiagram{1,1}}) + (q^j_q + 2j \, q^j_p + q^j_{d}) \, \mu(\scalebox{0.4}{\ydiagram{1}}) \right. \\
&+ \left. 3(q^j_q + q^j_{\qt} + (2j+1) \, q^j_p) \, \mu(\scalebox{0.4}{\ydiagram{1}}) + (q^j_q + q^j_{\dt} + (2j+1) \, q^j_p) \, \mu(\scalebox{0.4}{\ydiagram{1}}) \right] \\
i=1: &= -\frac{2N}{3} \\
i=2: &= 0 \\
i=3: &= \frac{2N+1}{6} \\
i=4: &= \frac{4N+1}{6}
\end{align*} 
\begin{align*}
\tr(\textcolor{red}{SU(3)^2} U(1)_i) &= (2N) \, q^i_{q} \, \mu(\scalebox{0.4}{\ydiagram{1}}) - \sum_{j=0}^{N-1} \left[ (2 \, q^j_{\qt} + 2j \, q^j_p) \, \mu(\bar{\scalebox{0.4}{\ydiagram{1,1}}}) + (q^j_{\qt} + 2j \, q^j_p + q^j_{\dt}) \, \mu(\bar{\scalebox{0.4}{\ydiagram{1}}}) \right. \\
&+ \left. 3(q^j_q + q^j_{\qt} + (2j+1) \, q^j_p) \, \mu(\bar{\scalebox{0.4}{\ydiagram{1}}}) + (q^j_{\qt} + q^j_{d} + (2j+1) \, q^j_p) \, \mu(\bar{\scalebox{0.4}{\ydiagram{1}}}) \right] \\
i=1: &= \frac{2N}{3} \\
i=2: &= 0 \\
i=3: &= \frac{2N+1}{6} \\
i=4: &= \frac{4N+1}{6}
\end{align*} 
\noindent $\tr(U(1)_1) = 0$ \\
\noindent $\tr(U(1)_2) = 0$ \\
\noindent $\tr(U(1)_3) = 2N+2$ \\
\noindent $\tr(U(1)_4) = 2N-1$ 

We can indeed see the matching of the anomalies.

\bibliographystyle{ytphys}
\bibliography{refs}
\end{document}